\RequirePackage[british]{babel}
\documentclass[reqno,a4paper,12pt,final]{amsart}
\usepackage[a4paper,hmargin=2cm,tmargin=3cm,bmargin=3cm]{geometry}
\usepackage[utf8x]{inputenc}
\usepackage{amsmath,amssymb,amstext,amsthm,amscd,mathrsfs,eucal}
\usepackage{graphicx,color}
\usepackage{array}
\usepackage{cite}
\usepackage[all]{xy}
\usepackage[vcentermath]{youngtab}
\Yboxdim{4pt}
\usepackage{hyperref}
\hypersetup{%
  pdftitle   = {Superpotentials for superconformal Chern--Simons theories from representation theory},
  pdfkeywords = {M2, M Theory, Bagger-Lambert, Lie algebra, Filippov, 3-algebra, Lie triple system, Chern-Simons, Faulkner, superpotential},
  pdfauthor  = {Paul de Medeiros, José Figueroa-O'Farrill, Elena Méndez-Escobar},
  pdfcreator = {\LaTeX\ with package \flqq hyperref\frqq}
}
\PrerenderUnicode{éÉ}
\let\into\hookrightarrow
\newcommand{\half}{\tfrac12}
\newcommand{\fg}{\mathfrak{g}}

\newcommand{\fB}{\mathfrak{B}}

\newcommand{\fF}{\mathfrak{F}}

\newcommand{\fM}{\mathfrak{M}}
\newcommand{\fV}{\mathfrak{V}}
\newcommand{\fW}{\mathfrak{W}}
\newcommand{\fgl}{\mathfrak{gl}}

\newcommand{\fk}{\mathfrak{k}}

\newcommand{\fso}{\mathfrak{so}}
\newcommand{\fosp}{\mathfrak{osp}}
\newcommand{\fspin}{\mathfrak{spin}}
\newcommand{\fusp}{\mathfrak{usp}}

\newcommand{\fsp}{\mathfrak{sp}}
\newcommand{\fsu}{\mathfrak{su}}
\newcommand{\fu}{\mathfrak{u}}

\newcommand{\Cl}{\mathrm{C}\ell}

\newcommand{\TT}{\mathbb{T}}
\newcommand{\RR}{\mathbb{R}}
\newcommand{\CC}{\mathbb{C}}
\newcommand{\HH}{\mathbb{H}}
\newcommand{\KK}{\mathbb{K}}
\newcommand{\ZZ}{\mathbb{Z}}
\newcommand{\XX}{\mathbb{X}}

\newcommand{\eL}{\mathscr{L}}

\newcommand{\eR}{\mathscr{R}}

\newcommand{\be}{\boldsymbol{e}}

\newcommand{\sW}{\mathsf{W}}

\newcommand{\Vbar}{\overline{V}}

\renewcommand{\Re}{\mathrm{Re}}
\newcommand{\AdS}{\mathrm{AdS}}

\DeclareMathOperator{\Dar}{Rep}
\DeclareMathOperator{\Irr}{Irr}

\newcommand{\rf}[1]{[\![#1]\!]}
\newcommand{\rh}[1]{(\!(#1)\!)}
\theoremstyle{plain}
\newtheorem{lemma}{Lemma}
\newtheorem{proposition}[lemma]{Proposition}
\newtheorem{theorem}[lemma]{Theorem}
\newtheorem{corollary}[lemma]{Corollary}
\theoremstyle{definition}

\newtheorem{example}[lemma]{Example}
\newcommand{\MUNCH}[1]{\relax}

\allowdisplaybreaks
\begin{document}
\title[Chern--Simons superpotentials from representation theory]{Superpotentials for superconformal Chern--Simons theories from representation theory}
\author[de Medeiros]{Paul de Medeiros}
\author[Figueroa-O'Farrill]{José Figueroa-O'Farrill}
\author[Méndez-Escobar]{Elena Méndez-Escobar}
\address{School of Mathematics and Maxwell Institute for Mathematical Sciences, University of Edinburgh, James Clerk Maxwell Building, King's Buildings,
  Edinburgh EH9 3JZ, UK}
\address[JMF also]{Institute for the Physics and Mathematics of the Universe, University of Tokyo, Kashiwa, Chiba 277-8586, Japan}
\email{\{P.deMedeiros,J.M.Figueroa,E.Mendez\}@ed.ac.uk}
\date{\today}
\begin{abstract}
   These notes provide a detailed account of the universal structure of superpotentials defining a large class of superconformal Chern--Simons theories with matter, many of which appear as the low-energy descriptions of multiple M2-brane configurations.  The amount of superconformal symmetry in the Chern--Simons-matter theory determines the minimum amount of global symmetry that the associated quartic superpotential must realise, which in turn restricts the matter superfield representations.  Our analysis clarifies the necessary representation-theoretic data which guarantees a particular amount of superconformal symmetry.  Thereby we shall recover all the examples of M2-brane effective field theories that have appeared in the recent literature.  The results are based on a refinement of the unitary representation theory of Lie algebras to the case when the Lie algebra admits an ad-invariant inner product.  The types of representation singled out by the superconformal symmetry turn out to be intimately associated with triple systems admitting embedding Lie (super)algebras and we obtain a number of new results about these triple systems which might be of independent interest.  In particular, we prove that any metric 3-Lie algebra embeds into a real metric 3-graded Lie superalgebra in such a way that the 3-bracket is given by a nested Lie bracket.
\end{abstract}
\maketitle
\tableofcontents

\section{Introduction and summary}
\label{sec:introduction}

The past couple of years have witnessed a remarkable amount of progress in our ability to understand configurations of coincident M2-branes in M-theory preserving large amounts of supersymmetry.  This progress followed the seminal work of Bagger and Lambert \cite{BL1,BL2} and Gustavsson \cite{GustavssonAlgM2} who managed to construct a lagrangian field theory in three-dimensional Minkowski spacetime that is invariant under the same maximal $N{=}8$ superconformal algebra as that of the near-horizon geometry of the M2-brane solution of eleven-dimensional supergravity.  This theory is now understood to arise as a special case of a more general type of $N{=}6$ superconformal field theory in three dimensions discovered by Aharony, Bergman, Jafferis and Maldacena in \cite{MaldacenaBL} (see also \cite{KlebanovBL}).  These $N{=}6$ theories are thought to describe the low-energy dynamics of multiple coincident M2-brane configurations whose near-horizon geometries are of the form $\AdS_4 \times S^7 / \ZZ_k$, for some positive integer $k$, with maximal $N{=}8$ supersymmetry recovered only for $k=1,2$.  New $N{=}5$ superconformal field theories were obtained in \cite{3Lee,BHRSS} and the regular ones which cannot enhance to $N=6$ are thought \cite{ABJ} to describe near-horizon geometries of the form $\AdS_4 \times S^7 / {\hat D}_k$ (${\hat D}_k$ being the binary dihedral group of order $4k$).  Moreover they can be considered as special cases of the $N{=}4$ superconformal field theories found in \cite{pre3Lee} generalising the $N{=}4$ theories obtained first by Gaiotto and Witten in \cite{GaiottoWitten}.

The lagrangian for all these new superconformal field theories can be written in terms of a Chern--Simons term for a non-dynamical gauge field which is coupled to matter fields in a particular unitary representation of the Lie algebra which describes the gauge symmetry.  In addition to the standard kinetic terms for the bosonic scalar and fermionic spinor matter fields, there are quartic scalar-fermion Yukawa couplings and a sextic scalar potential.  The order of these matter couplings is precisely as expected for a generic on-shell superconformal Chern--Simons-matter theory and is just a consequence of superconformal symmetry.  What is novel, at least relative to known superconformal gauge theories in other dimensions, is that the representation and indeed the Lie algebra describing the gauge-matter couplings is restricted by the amount of superconformal symmetry to be realised.  As noted in \cite{VanRaamsdonkBL}, for the $N{=}8$ theory of \cite{BL1,BL2,GustavssonAlgM2} the unitary representation for the matter fields corresponds to the bifundamental of $\fsu(2) \oplus \fsu(2) \cong \fso(4)$.  The fact that this is the only possibility follows from the classifications in \cite{NagykLie,GP3Lie,GG3Lie} of euclidean 3-Lie algebras, in terms of which the original description in \cite{BL1,BL2,GustavssonAlgM2} was based.  A classification of the $N{=}6$ theories of \cite{MaldacenaBL} has been provided by \cite{SchnablTachikawa} (see also \cite{Lie3Algs,JMFSimplicity}) and the unitary representation for the matter fields must correspond to the bifundamental of either $\fu(m)\oplus \fu(n)$ or $\fsp(m) \oplus \fu(1)$.  The $N{=}5$ theories obtained in \cite{3Lee,BHRSS} can have either $\fso(m) \oplus \fsp(n)$, $\fspin(7) \oplus \fsp(1)$, $\fg_2 \oplus \fsp(1)$ or $\fso(4) \oplus \fsp(1)$ (the second and fourth being distinguished from the first of these possibilities due to the different representation inhabited by the matter fields in these cases).  The $N{=}4$ theories in \cite{GaiottoWitten,pre3Lee} can have matter fields in any of the representations noted above.

The goal of this paper, which is to be viewed as a companion to \cite{Lie3Algs}, is to survey this emerging landscape of superconformal Chern--Simons theories and establish precisely what representation-theoretic conditions must be obeyed by the matter couplings in order to realise a particular amount of superconformal symmetry.  Our starting point will be the generic off-shell $N{=}1$ superconformal Chern--Simons-matter theory.  To define this (at least as a classical field theory) requires two sets of ingredients: a unitary representation $\fM$ (which the matter fields will inhabit) of a metric Lie algebra $\fg$ and a real, quartic, $\fg$-invariant function on $\fM$ (that defines a superpotential $\sW$ from which the quartic Yukawa couplings and sextic scalar potential will be obtained in the on-shell theory after integrating out all the auxiliary fields).  The construction of $N{=}2$ superconformal Chern--Simons-matter theories is somewhat canonical given the existence of an off-shell $N{=}2$ superspace formalism in three dimensions and the generic $N{=}2$ theory can be obtained from the $N{=}1$ construction whenever $\fM$ is complex unitary (provided one fixes certain components of the $N{=}1$ superpotential appropriately so as to reproduce the correct $N{=}2$ gauge-matter couplings).  Prior to the recent progress described above, the maximum amount of superconformal symmetry that could be realised in the lagrangian for a Chern--Simons-matter theory was suspected to be $N{=}3$ \cite{GYCS}.  The construction of the $N{=}3$ superconformal Chern--Simons-matter theory not only requires $\fM$ to be quaternionic unitary but also fixes uniquely the superpotential and so indeed one should expect no generic possibilities for obtaining $N > 3$ superconformal symmetry.  Recalling that $N{=}3$ theories should describe M2-brane near-horizon geometries of the form $\AdS_4 \times X_7$, where $X_7$ is a 3-Sasakian 7-manifold, a nice consistency check is that there exists a corresponding infinitesimal rigidity theorem \cite[Theorem~13.3.24]{MR2382957} for such geometries.

The way we shall proceed to $N>3$ is by examining certain special classes of unitary representations for which this same superpotential realises an enhanced $\fso(N-1)$ global symmetry.  For a given value of $N>3$, this property, albeit non-trivial, need not guarantee that the resulting on-shell Chern--Simons-matter theory realises an enhanced $N$-extended superconformal symmetry.  Clearly though for such an on-shell theory to be derivable from an $N{=}1$ superpotential requires that superpotential to be invariant under those R-symmetries preserving the choice of $N{=}1$ superspace parameter, which is an $\fso(N-1)$ subalgebra of the $\fso(N)$ R-symmetry.  Nevertheless in all the cases we consider we will find that this property does give rise to superconformal symmetry enhancement and thereby this construction will recover all the known examples of $N \geq 4$ superconformal Chern--Simons-matter theories described above.

These notes are organised as follows.  In the rest of this introductory section we outline the representation-theoretic structure of three-dimensional superconformal Chern--Simons theories with matter.  We begin with a cursory glance at some of the essential notation to be used for the generic types of unitary representations we shall encounter throughout the paper.  The representation theory of $N$-extended superconformal algebras in three dimensions is then reviewed focusing on the specific types of unitary representations of $\fso(N) \oplus \fg$ which the matter fields must inhabit for each value of $N \leq 8$.  We then summarise for $N \geq 4$ the representation-theoretic criteria which determine the indecomposable $N$-extended superconformal Chern--Simons theories.  Section~\ref{sec:introduction} is concluded with a summary of the functional forms of the quartic superpotentials that we will find give rise to all the $N$-extended superconformal theories in the rest of the paper. 

In Section~\ref{sec:scft-ternary} we will review the generic $N \leq 3$ superconformal Chern--Simons-matter theories, describing the lagrangians and supersymmetry transformations both off- and on-shell.  Starting from the generic $N=1$ theory, we will describe in detail how one obtains the generic $N{=}2,3$ theories by taking the ground field for the matter representation to be respectively $\KK = \CC,\HH$ and then choosing an appropriate superpotential $\sW_{\CC,\HH}$ in terms of this data.  The realisation of the non-abelian $\fusp(2)$ R-symmetry in the $N{=}3$ theory is also detailed.  In Section~\ref{sec:scft-ternary-3plus}, for incremental values of $N > 3$, we proceed to describe how one obtains $N$-extended superconformal Chern--Simons-matter theories from the generic $N{=}1$ superspace formalism by taking the matter representations to be of the types established in Section~\ref{sec:introduction} and using the corresponding $\fso(N-1)$-invariant forms of the rigid superpotential $\sW_\HH$ describing the $N{=}3$ theory.  For $N{=}4$ we will thereby recover the theories of \cite{GaiottoWitten,pre3Lee} and determine the conditions for indecomposability of these theories.  For the theories of \cite{GaiottoWitten} this condition is just irreducibility of the matter representation, while for the more general theories in \cite{pre3Lee} indecomposability is the connectedness of the corresponding quiver.  Similarly for $N{=}5,6,8$ we recover the theories of \cite{3Lee,BHRSS}, \cite{MaldacenaBL} and \cite{BL1,BL2,GustavssonAlgM2} respectively.  We find that indecomposability in all the $N>4$ theories is equivalent to irreducibility of the associated matter representations, thus allowing their classification in terms of certain simple Lie superalgebras into which the matter fields embed.  In the process of obtaining the $N{=}8$ theory we rule out the possibility of a Chern--Simons-matter theory with precisely $N{=}7$ superconformal symmetry in the sense that we find such a theory must automatically enhance to $N{=}8$.

In the interest of accessibility we have collected many crucial technical results concerning unitary representations of metric Lie algebras into a comprehensive Appendix~\ref{sec:rep-theory}.  This appendix contains precise definitions of all the basic representation-theoretic notions employed in this paper.  We begin by reviewing the notions of real, complex and quaternionic unitary representations of a Lie algebra $\fg$.  We then describe the canonical functors mapping between these three categories of representations as we change the ground field.  We also describe the various useful identities which these functors satisfy under composition and finally the effect of these functors on the irreducible representations.  When the Lie algebra $\fg$ admits an ad-invariant inner product, it is possible to refine the standard representation theory and this allows us to distinguish among the generic unitary representation some which \emph{a posteriori} can be characterised by the existence of a certain embedding Lie (super)algebra structure.  We discuss how to each of these ``Lie-embeddable'' representations one can attach a triple system of the type which appears naturally in the 3-algebraic description \cite{BL1,BL2,GustavssonAlgM2,BL4} of the $N \geq 4$ theories related to M2-branes.  In particular, a new characterisation (Theorem~\ref{th:3LAchar}) is established of the metric 3-Lie algebras of the $N{=}8$ theory in terms of certain real metric 3-graded Lie superalgebras.  We then determine the action of the aforementioned functors on the Lie-embeddable representations and in this way shed some light on the representation-theoretic underpinning of supersymmetry enhancement in these theories.  Finally, for completeness, we include a derivation of the superpotentials describing the generic $N{=}2,3$ theories in Appendix~\ref{sec:der-superpotentials}.

\subsection*{Matters of notation}

In this paper we will make much use of certain basic concepts in the theory of unitary representations of a Lie algebra.  There is a great deal of bookkeeping for which perhaps there is no universally agreed notation.  Although the notation is presented as the definitions are introduced, it may benefit the reader to see the notation at glance in one place.  This small section aims to be that place.

Following \cite{Adams, MR781344}, we have used the letters $U$, $V$ and $W$, also with decorations, to refer to real, complex and quaternionic representations, respectively.  The notation $\Dar(\fg,\KK)$ (resp. $\Irr(\fg,\KK)$), where $\fg$ is a Lie algebra, denotes the (resp. irreducible) unitary representations over $\KK = \RR,\CC,\HH$.  We will also let $\Dar(\fg,\KK)_{\text{C}}$ (resp. $\Irr(\fg,\KK)_{\text{C}}$) denote the subset of those (resp. irreducible) $\KK$-representations of $\fg$ of class $\text{C}$, where these are defined in Appendix \ref{sec:special-reps}.  There are natural operations on representations which act on the ground field $\KK$:
\begin{itemize}
\item \emph{extending from $\RR$ to $\CC$}: $U \mapsto V = U_\CC$ (complexification)
\item \emph{extending from $\CC$ to $\HH$}: $V \mapsto W = V_\HH$ (quaternionification)
\item \emph{conjugation}: $V \mapsto \Vbar$
\item \emph{restricting from $\HH$ to $\CC$}: $W \mapsto V = \rh{W}$
\item \emph{restricting from $\CC$ to $\RR$}: $V \mapsto U = \rf{V}$
\end{itemize}
These operations obey relations which are reviewed in Proposition~\ref{prop:diamond} in Appendix~\ref{sec:relat-betw-real} and whose effect on the irreducibles is described by Proposition~\ref{prop:irreducible} in the same appendix, and on the special classes of representations by Proposition~\ref{prop:relations} in Appendix~\ref{sec:some-relat-betw}.

\subsection{Matter content of three-dimensional superconformal theories}
\label{sec:matter-reps}

Superconformal field theories in three-dimensional Minkowski spacetime $\RR^{1,2}$ are invariant under a conformal superalgebra.  These algebras are denoted type VII in \cite[Proposition~2.2]{Nahm} and are indexed by a positive integer $N$.  The even Lie algebra is isomorphic to $\fso(N) \oplus \fso(2,3)$, where $\fso(N)$ is the R-symmetry of the field theory and $\fso(2,3)$ is the three-dimensional conformal algebra.  The odd subspace is in the tensor product representation of the vector of $\fso(N)$ and the spinor of $\fso(2,3)$.  The spinor representation of $\fso(2,3)$ defines an isomorphism $\fso(2,3) \cong \fsp(4,\RR)$, which means that the spin representation is real, four-dimensional and symplectic.  In other words, the conformal superalgebra is isomorphic to the orthosymplectic Lie superalgebra $\fosp(N|4)$.  This Lie superalgebra can be understood as endomorphisms of the vector superspace $\RR^{N|4}$ of $N$ even and $4$ odd dimensions preserving a euclidean structure on the even subspace and a symplectic structure on the odd subspace.  The supercharges in the superalgebra are the odd endomorphisms which map the even and odd subspaces to each other.   Standard arguments restricting unitary interacting field theories to admit at most 32 supercharges impose an upper bound $N\leq 8$ for the interesting theories.  Having said that, such arguments must be taken with a pinch of salt for theories in less than 4 dimensions which do not admit a dimensional oxidation to four dimensions, as is the case with the three-dimensional superconformal Chern--Simons theories which are the subject of this paper.  Nevertheless we will restrict attention to $N\leq 8$; although see \cite{Bergshoeff:2008ta} for a topologically massive Chern--Simons theory admitting $N>8$.

As we will discuss in more detail in Section~\ref{sec:scft-ternary}, the lagrangian for superconformal Chern--Simons theories with matter consists of two types of terms: a Chern--Simons lagrangian for a gauge field taking values in a metric Lie algebra $\fg$ and a lagrangian involving matter fields in a faithful unitary representation $\fM$ of $\fg$.  Unitarity of the theory requires the inner product on $\fM$ to be positive-definite, but since the Chern--Simons gauge fields are non-propagating, the inner product on $\fg$ can, and in some cases must, have indefinite signature.  Indeed, one of the remarkable features of the theories with $N\geq 4$ supersymmetry is that the Lie algebra $\fg$ cannot be taken to be simple.  The assumption that the representation be faithful is done for convenience and we believe that there is minimal (if any) loss in generality.  We recall that a unitary representation $\rho: \fg \to \fu(\fM)$ is faithful if $\rho$ has trivial kernel.  In most other contexts, this assumption can be made without any loss of generality: the kernel $\fk$ of $\rho$ is an ideal and the quotient $\fg/\fk$ is a Lie algebra which does act faithfully.  However here we have gauge fields taking values in $\fg$, hence in particular also in $\fk$ and one would like to show that these fields somehow decouple.  The gauge fields $A$ enter the matter lagrangian via covariant derivatives of the form $d + \rho(A)$, whence any gauge field in $\fk$ appears only in the Chern--Simons lagrangian, which depends not just on $\fg$ as a Lie algebra but as a metric Lie algebra.  In other words, it depends on a choice of ad-invariant inner product on $\fg$.  If $\fk$ is a nondegenerate ideal, so that $\fg = \fk \oplus \fk^\perp$, then it is not hard to see that since both $\fk$ and $\fk^\perp$ are ideals and are orthogonal, the theory decouples into a Chern--Simons term for $\fk$ and a Chern--Simons-matter term for $\fk^\perp$.  In the case when $\fk$ is not nondegenerate, preliminary results with abelian quiver theories suggest that one ends up with a Chern--Simons-matter theory for $\fk^\perp/(\fk \cap\fk^\perp)$ which is a metric Lie algebra which does act faithfully.

The matter fields in three dimensions arrange themselves into supermultiplets which, ignoring auxiliary fields, consist of a bosonic scalar $X$ and a fermionic Majorana spinor $\Psi$ on $\RR^{1,2}$ in representations which we will denote schematically by $\fB \otimes \fM_1$ and $\fF\otimes \fM_2$, respectively, where $\fB$ and $\fF$ are the bosonic and fermionic R-symmetry representations and $\fM_1,\fM_2$ are representations of $\fg$.  The supersymmetry transformations take the generic form \begin{equation}
  \label{eq:susy-sketch}
  \delta_\epsilon X = \overline{\epsilon}\Psi \qquad\text{and}\qquad \delta_\epsilon \Psi = dX \cdot \epsilon + \cdots,
\end{equation}
where $\epsilon$ is the supersymmetry parameter which is a vector under the R-symmetry, but inert under $\fg$, reflecting the fact that for a rigidly supersymmetric theory, supersymmetry and gauge transformations commute.  It then follows that $\fM_1 = \fM_2$, whence we will drop the subscript, and focusing on the R-symmetry, we see that letting $\fV$ denote the vector representation of the R-symmetry,
\begin{equation*}
  \fB \subset \fV \otimes \fF \qquad\text{and}\qquad   \fF \subset \fV \otimes \fB.
\end{equation*}
This suggests taking $\fB$ and $\fF$ to be spinor representations in such a way that the above inclusions are induced from the Clifford actions $\fV \otimes \fF \to \fB$ and $\fV \otimes \fB \to \fF$, respectively.  We will do so.  This means that when $N$ is odd, bosons and fermions will be in the same representation, whereas if $N$ is even, since Clifford multiplication by vectors reverses chirality, the fermionic representation will be obtained from the bosonic one by changing the chirality of the spinor representations.

\begin{table}[h!]
  \centering
  \begin{tabular}{|>{$}c<{$}|>{$}c<{$}|>{$}c<{$}|}
    \hline
    N & \fso(N) & \text{spinor irreps}\\\hline
    2 &  \fu(1) & \CC\\
    3 &  \fsp(1) & \HH\\
    4 &  \fsp(1) \oplus \fsp(1) & \HH \oplus \HH\\
    5 &  \fsp(2) & \HH^2\\
    6 &  \fsu(4) & \CC^4\\
    7 &  \fso(7) & \RR^8\\
    8 &  \fso(8) & \RR^8 \oplus \RR^8\\\hline
  \end{tabular}
  \vspace{8pt}
  \caption{Spinor representations of $\fso(N)$ for $N\leq 8$}
  \label{tab:spinors}
\end{table}

Table~\ref{tab:spinors} summarises the spinor representations for $N\leq 8$.  It lists the exceptional low-dimensional isomorphisms which are induced by the spinor representations and lists the types of representation with their dimension.  For $N$ odd there is a unique irreducible spinor representation (up to isomorphism) which is real for $N\equiv \pm 1 \pmod{8}$ and quaternionic for $N\equiv \pm 3 \pmod{8}$.  For $N$ even there are two, distinguished by chirality.  They are complex for $N\equiv\pm 2 \pmod{8}$, with opposite chiralities being related by complex conjugation, real for $N\equiv 0 \pmod 8$ and quaternionic for $N\equiv 4 \pmod 8$.  It will be convenient to introduce the following notation for the spinor representations: for $N$ odd, we let $\Delta^{(N)}$ denote the unique irreducible spinor representation of $\fso(N)$, whereas for $N$ even, we let $\Delta^{(N)}_\pm$ denote the unique irreducible spinor representation of $\fso(N)$ with positive/negative chirality, with the understanding that for $N{=}2,6$, $\overline{\Delta^{(N)}_\pm} = \Delta^{(N)}_\mp$.

The degrees of freedom described by the matter fields are fundamentally real and hence this fact determines the type of the representation $\fM$ in terms of the type of the relevant spinor representation.  This means that if the spinor representation is real then so must $\fM$, whereas if the spinor representation is quaternionic then so must $\fM$, but we are then supposed to take the fields to be in the underlying real representation of the tensor product of the two quaternionic representations.  In practical terms, this means imposing a reality condition on the fields which involves the symplectic structure of both the spinor representation and $\fM$, as described for example in Appendix~\ref{sec:reality}.  Finally, if the spinor representation is complex, we can take $\fM$ to be complex without loss of generality, with the understanding that we may think of both reals and quaternionic representations as special types of complex representations.  In this case, the matter fields take values in the real representation given by their real and imaginary parts.  In conclusion, for $N{=}1,7,8$ the representations $\fM$ are real, for $N{=}3,4,5$ quaternionic and for $N{=}2,6$ complex.

Summarising thus far, for odd $N$ the bosonic and fermionic matter fields both take values in the representation $\Delta^{(N)} \otimes \fM$, with the proviso that for $N{=}3,5$, when $\Delta^{(N)}$ is quaternionic, fields must obey the natural reality condition.  For even $N$ the bosonic matter fields can take values in the representation $\Delta^{(N)}_+ \otimes \fM_1 \oplus \Delta^{(N)}_- \otimes \fM_2$, whereas the fermionic matter fields take values in $\Delta^{(N)}_- \otimes \fM_1 \oplus \Delta^{(N)}_+ \otimes \fM_2$, where a priori both representations $\fM_1$ and $\fM_2$ can be different.  Again, if $N{=}4$, then all representations are quaternionic, so that we must impose the natural symplectic reality condition on the fields.  If $N{=}2,6$ then all representations are complex and we must consider the representation made up of by the real and imaginary parts of the fields or, said differently, to consider both the fields and their complex conjugates.  In this case one may ignore the distinction between $\fM_1$ and $\fM_2$ because taking real and imaginary parts of $\Delta^{(N)}_+ \otimes \fM_1 \oplus \Delta^{(N)}_- \otimes \fM_2$ is the same as taking real and imaginary parts of $\Delta^{(N)}_+ \otimes (\fM_1\oplus\overline\fM_2)$, so that we can always take the matter fields to be in (the underlying real form of) a particular chiral spinor representation of $\fso(N)$.  Of course, for $N{=}4$ the symplectic reality condition for the matter fields does not eliminate the distinction and we will see in later sections how this can lead to the notion of ``twisted'' and ``untwisted'' $N{=}4$ hypermultiplets according to the relative chiralities of the spinor representation of $\fso(4)$ they transform under.  Similarly for $N{=}8$ the matter representations are real and one might expect to be able to distinguish between different types of matter in $\Delta^{(8)}_\pm$.  However, this case of maximal supersymmetry will turn out to be rather special in that we will find one can obtain any $N{=}8$ supermultiplet from an $N{=}7$ one wherein both $\Delta^{(8)}_\pm$ are identified with $\Delta^{(7)}$ under the embedding $\fso(7) \into \fso(8)$ thus eliminating the apparent distinction between the two possible types of matter.

\subsection{Constraints from supersymmetry}
\label{sec:constr-from-supersym}

We will see that for the $N{=}1,2,3$ theories we can take the matter to be in any real, complex or quaternionic unitary representations, respectively, whilst for $N>3$ the allowed representations are subject to further restrictions.  In Appendix~\ref{sec:special-reps} we introduce several classes of ``Lie-embeddable'' unitary representations of a metric Lie algebra, which are summarised in Table~\ref{tab:special} below.  The notation is explained in Appendix~\ref{sec:special-reps}.  For now suffice it to say that $\eR$ is a natural fourth-rank tensor constructed canonically from the representation data.  The names of the classes stand for the corresponding triple system naturally attached to such representations: Lie (LTS), anti-Lie (aLTS), Jordan (JTS), anti-Jordan (aJTS), 3-Lie (3LA) and quaternionic (QTS), this last one being a non-standard nomenclature introduced in this paper.

\begin{table}[h!]
  \centering
  \begin{tabular}{|c|>{$}c<{$}|>{$}c<{$}|l|}
    \hline
    Class & \text{type} & \eR & \multicolumn{1}{|c|}{Embedding Lie (super)algebra}\\\hline
    LTS & \RR & U^{\yng(2,2)} & 2-graded metric Lie algebra\\
    JTS & \CC & S^2V \otimes S^2\Vbar & 3-graded complex metric Lie algebra\\
    QTS & \HH & S^4W & 3-graded complex metric Lie algebra\\
    aLTS & \HH & W^{\yng(2,2)} & complex metric Lie superalgebra\\
    aJTS & \CC & \Lambda^2V \otimes \Lambda^2\Vbar & 3-graded complex metric Lie superalgebra\\
    3LA & \RR & \Lambda^4 U & 3-graded metric Lie superalgebra\\
    \hline
  \end{tabular}
  \vspace{8pt}
  \caption{Lie-embeddable unitary representations of a metric Lie algebra}
  \label{tab:special}
\end{table}

As discussed in Section~\ref{sec:scft-ternary-3plus}, the matter representations for $N>3$ supersymmetry are forced to belong to some of these special classes.  Table \ref{tab:matter-reps} summarises the situation, where the notation $\Dar(\fg,\KK)_{\text{C}}$ denotes the unitary representations of $\fg$ of type $\KK$ and class ${\text{C}}$, where $\KK = \RR, \CC, \HH$ and ${\text{C}}$ can be either 3LA, aJTS or aLTS, and similarly $\Irr(\fg,\KK)_{\text{C}}$ for the irreducibles.  If the class ${\text{C}}$ is absent, then we mean any representation of type $\KK$.  The irreducibility of the representation for $N\geq 5$  theories is imposed by the requirement that the theory should not decouple into two or more nontrivial theories.  For $N$ even, the bosonic and fermionic matter representations are different, related by changing the chirality of the R-symmetry spinor representations.  In those cases we list them both in one line, with the top reading for bosons and the bottom reading for fermions.  In the case of representations which are not manifestly real, one is instructed to take the underlying real representation by imposing the appropriate reality conditions on the fields, which follow from the discussion in Appendix \ref{sec:reality}.  This applies to $2\leq N \leq 6$.

\begin{table}[h!]
  \centering
  \begin{tabular}{|>{$}c<{$}|>{$}c<{$}|>{$}c<{$}|}
    \hline
    N & \text{Matter representation} & \text{Remarks}\\\hline
    1 & U & U \in \Dar(\fg,\RR)\\
    2 & \Delta^{(2)}_\pm \otimes V \oplus  \Delta^{(2)}_\mp \otimes \Vbar & V \in \Dar(\fg,\CC)\\
    3 & \Delta^{(3)} \otimes W & W \in \Dar(\fg,\HH)\\
    4 & \Delta^{(4)}_\pm \otimes W_1 \oplus  \Delta^{(4)}_\mp \otimes W_2 & W_{1,2} \in \Dar(\fg,\HH)_{\text{aLTS}}\\
    5 & \Delta^{(5)} \otimes W & W \in \Irr(\fg,\HH)_{\text{aLTS}}\\
    6 & \Delta^{(6)}_\pm \otimes V \oplus  \Delta^{(6)}_\mp \otimes \Vbar & V \in \Irr(\fg,\CC)_{\text{aJTS}}\\
    7 & \Delta^{(7)} \otimes U & U \in \Irr(\fg,\RR)_{\text{3LA}}\\
    8 & \Delta^{(8)}_\pm \otimes U & U \in \Irr(\fg,\RR)_{\text{3LA}}\\
    \hline
  \end{tabular}
  \vspace{8pt}
  \caption{Matter representations for $N$-extended supersymmetry}
  \label{tab:matter-reps}
\end{table}

The representation theory also helps to explain the conditions for supersymmetry enhancement.  Table \ref{tab:spinor-decomp} summarises how the spinor representations decompose as a result of the embedding of the R-symmetry Lie algebras $\fso(N-1) \into \fso(N)$.  The notation $\rf{V}$, introduced in Appendix \ref{sec:relat-betw-real}, means the real representation obtained from the complex representation $V$ by restricting scalars to $\RR$.

\begin{table}[h!]
  \centering
  \begin{tabular}{|>{$}c<{$}|>{$}r<{$}>{$}c<{$}>{$}l<{$}|}
    \hline
    N & \fso(N) & \supset & \fso(N-1)\\\hline
    8 & \Delta^{(8)}_\pm &\cong & \Delta^{(7)}\\
    7 & \Delta^{(7)} &\cong & \rf{\Delta_+^{(6)}}\\
    6 & \Delta_\pm^{(6)} &\cong & \Delta^{(5)}\\
    5 & \Delta^{(5)} &\cong & \Delta_+^{(4)} \oplus \Delta_-^{(4)}\\
    4 & \Delta_\pm^{(4)} &\cong & \Delta^{(3)}\\
    3 & \Delta^{(3)} &\cong & \Delta_+^{(2)} \oplus \Delta_-^{(2)}\\\hline
  \end{tabular}
  \vspace{8pt}
  \caption{Spinor representations under $\fso(N-1) \into \fso(N)$}
  \label{tab:spinor-decomp}
\end{table}

This then implies the decomposition of the matter representations from $N$- to ($N-1$)-extended supersymmetry which is summarised in Table~\ref{tab:susy-enhancement}.  In that table we use notation introduced in Appendix~\ref{sec:relat-betw-real}.  In particular, $U_\CC$ is the complexification of a real representation $U$, whereas $V_\HH$ is the quaternionification of a complex representation $V$ and $\rh{W}$ is a complex representation obtained from a quaternionic representation $W$ by forgetting the quaternionic structure.  As usual, square brackets denote the underlying real representation, so that if $V$ is a complex representation with a real structure, then $[V]_\CC \cong V$.

\begin{table}[h!]
  \centering
  \begin{tabular}{|>{$}c<{$}|>{$}c<{$}|>{$}c<{$}|}
    \hline
    N & N-\text{matter representation} & (N-1)-\text{matter representation}\\\hline
    8 & \Delta^{(8)}_+ \otimes U & \Delta^{(7)} \otimes U\\
    7 & \Delta^{(7)}\otimes U & [ (\Delta_+^{(6)} \oplus \Delta_-^{(6)}) \otimes U_\CC] \\
    6 & \Delta_+^{(6)} \otimes V \oplus \Delta_-^{(6)} \otimes \Vbar & \Delta^{(5)} \otimes V_\HH\\
    5 & \Delta^{(5)} \otimes W & \Delta_+^{(4)} \otimes W \oplus \Delta_-^{(4)} \otimes W\\
    4 & \Delta_+^{(4)} \otimes W_1 \oplus \Delta_-^{(4)} \otimes W_2 & \Delta^{(3)} \otimes (W_1 \oplus W_2)\\
    3 & \Delta^{(3)} \otimes W & (\Delta_+^{(2)} \oplus \Delta_-^{(2)})\otimes \rh{W}\\\hline
  \end{tabular}
  \vspace{8pt}
  \caption{Decomposition of matter representations}
  \label{tab:susy-enhancement}
\end{table}

We may understand the following supersymmetry enhancements, by looking at the $N$-extended matter representation in terms of the ($N-1$)-extended representation and then comparing with the generic ($N-1$)-extended representation.  In practice one finds the $N$-extended matter representation in the second column of Table~\ref{tab:susy-enhancement}, then moves over to the third column which shows this representation in terms of ($N-1$)-extended supersymmetry and then moves back to the second column but one row below to compare with the generic ($N-1$)-extended representations.   This allows us to understand the enhancements $N{=}4\to N{=}5$, $N{=}5 \to N{=}6$ and $N{=}6 \to N>6$, as follows.
\begin{itemize}
\item In $N{=}4$, $W_1,W_2 \in \Dar(\fg,\HH)_{\text{aLTS}}$ and the enhancement to $N{=}5$ occurs precisely when $W_1 = W_2$:
  \begin{equation}
    \label{eq:4to5}
    \xymatrix{\Delta^{(5)} \otimes W \ar@{.>}[r] & \Delta_+^{(4)} \otimes W \oplus \Delta_-^{(4)} \otimes W \ar@{.>}[dl]\\
      \Delta_+^{(4)} \otimes W_1 \oplus \Delta_-^{(4)} \otimes W_2 & \\}
  \end{equation}
\item In $N{=}5$, $W \in \Irr(\fg,\HH)_{\text{aLTS}}$ and the enhancement to $N{=}6$ occurs when $W = V_\HH$, for $V \in \Irr(\fg,\CC)_{\text{aJTS}}$:
  \begin{equation}
    \label{eq:5to6}
    \xymatrix{\Delta_+^{(6)} \otimes V \oplus \Delta_-^{(6)} \otimes \Vbar \ar@{.>}[r] &  \Delta^{(5)} \otimes V_\HH\ar@{.>}[dl]\\
      \Delta^{(5)} \otimes W & \\}
  \end{equation}
\item Finally, in $N{=}6$, $V \in \Irr(\fg,\CC)_{\text{aJTS}}$ and enhancement to $N{=}7$ occurs when $V=U_\CC$ for $U \in \Irr(\fg,\RR)_{\text{3LA}}$:
  \begin{equation}
    \label{eq:6to7}
    \xymatrix{\Delta_+^{(7)} \otimes U \ar@{.>}[r] &  [\![\Delta_+^{(6)} \otimes U_\CC]\!] \ar@{.>}[dl]\\
      [\![\Delta_+^{(6)} \otimes V]\!] & \\}
  \end{equation}
\end{itemize}

We also see from Table~\ref{tab:susy-enhancement} that enhancement from $N{=}7$ to $N{=}8$ does not constrain the representation further.  This suggests that $N{=}7$ implies $N{=}8$ and we will show in Section~\ref{sec:nequal8} that this is indeed the case.

\subsection{Indecomposability and irreducibility}
\label{sec:indecomposability}

Given two $N$-extended superconformal Chern--Simons theories with matter with data $(\fg_1,\fM_1)$ and $(\fg_2,\fM_2)$ one can add their lagrangians to obtain a theory with the same amount of supersymmetry and with data $(\fg_1 \oplus \fg_2, (\fM_1\otimes \KK) \oplus (\KK \otimes \fM_2))$, where $\KK = \RR,\CC$ denotes the relevant trivial one-dimensional representation.  In other words, superconformal Chern--Simons theories admit direct sums and hence there is a notion of indecomposability; namely, an indecomposable theory is one which cannot be decoupled as a direct sum of two nontrivial theories.

For $N<4$ indecomposability places very weak constraints on the allowed representations.  For example, if the Chern--Simons Lie algebra $\fg$ is simple, then any direct sum of nontrivial irreducible unitary representations of the right type will give rise to an indecomposable theory, the Chern--Simons terms acting as the ``glue'' binding the matter together.

For the $N{=}4$ theories of the type discussed by Gaiotto and Witten \cite{GaiottoWitten}, where the bosonic matter lives in $\Delta^{(4)}_+ \otimes W$, for $W \in \Dar(\fg,\HH)_{\text{aLTS}}$, indecomposability forces $W$ to be irreducible.  For the general $N{=}4$ theories with twisted matter, indecomposability implies the connectedness of the corresponding quiver \cite{pre3Lee}, which imposes conditions --- albeit not irreducibility --- on the allowed representations.

Finally for $N>4$ indecomposability coincides with irreducibility of the matter representation.  The matter representations for $N>4$ are Lie-embeddable, which means that one can attach a Lie superalgebra to them and hence a triple system by nesting the Lie bracket: the Lie bracket of two odd elements is even and its Lie bracket with a third odd element will again be odd.  As discussed in Appendix \ref{sec:simplicity}, which is based on \cite{JMFSimplicity}, the notions of irreducibility of the representation agrees with the simplicity of the embedding Lie superalgebra and with that of the triple system, provided the representation is positive-definite, which is a tacit assumption in this paper.  This allows a classification of positive-definite irreducible matter representations for $N>4$ in terms of Lie superalgebras, which is summarised in Table \ref{tab:irreducible-LE} below, where $\fu(1)$ charges are denoted by subscripts where appropriate.  It should be noted that simplicity of the embedding Lie superalgebra does not imply the simplicity of the gauge Lie algebra $\fg$ and indeed in none of the cases in the table is $\fg$ allowed to be simple.  This fact may explain why these theories took a relatively long time to be discovered.

\begin{table}[h!]
  \centering
  \begin{tabular}{|>{$}c<{$}|c|>{$}c<{$}|>{$}c<{$}|>{$}c<{$}|}
    \hline
    N & Class & \text{Representation} & \fg & \text{Lie superalgebra} \\\hline
    8 & 3LA & (\boldsymbol{2},\boldsymbol{2}) & \fsu(2) \oplus \fsu(2) &  A(1,1) \\\hline
    5, 6 & aLTS, aJTS & (\boldsymbol{m+1},\overline{\boldsymbol{n+1}})_{m-n} & \fsu(m+1) \oplus \fsu(n+1) \oplus \fu(1) &  A(m,n), m\neq n\\
    5, 6 & aLTS, aJTS & (\boldsymbol{n+1},\overline{\boldsymbol{n+1}}) & \fsu(n+1) \oplus \fsu(n+1) &  A(n,n) \\
    5, 6 & aLTS, aJTS & (\boldsymbol{2n})_{+1} & \fusp(2n) \oplus \fu(1) &  C(n+1) \\\hline
    5 & aLTS & (\boldsymbol{2m+1},\boldsymbol{2n}) & \fso(2m+1) \oplus \fusp(2n) &  B(m,n) \\
    5 & aLTS & (\boldsymbol{2m},\boldsymbol{2n})  & \fso(2m) \oplus \fusp(2n) &  D(m,n) \\
    5 & aLTS & (\boldsymbol{2},\boldsymbol{2},\boldsymbol{2}) & \fsu(2) \oplus \fsu(2) \oplus \fsu(2) &  D(2,1;\alpha) \\
    5 & aLTS & (\boldsymbol{2},\boldsymbol{8}) & \fsu(2) \oplus \fspin(7) &  F(4) \\
    5 & aLTS & (\boldsymbol{2},\boldsymbol{7}) & \fsu(2) \oplus \fg_2 &  G(3) \\\hline
  \end{tabular}
  \vspace{8pt}
  \caption{Irreducible, positive-definite Lie-embeddable representations}
  \label{tab:irreducible-LE}
\end{table}

It must be remarked that all the matter representations for $N>4$ superconformal Chern--Simons theories in our Table~\ref{tab:irreducible-LE} have been found already in \cite{BHRSS} via a certain global limit of conformally gauged supergravities in three dimensions.  The gauging can be most conveniently described in terms of a so-called embedding tensor and it is from the linear constraint imposed on this object by supersymmetry that allows one to identify the different classes of representations in Table~\ref{tab:irreducible-LE} with those in Table 3 of \cite{BHRSS}. In each case, it is the tensor $\eR$ defined in Appendix~\ref{sec:rep-theory} that corresponds to an R-symmetry-singlet of the embedding tensor in the aforementioned global limit. Thus our classification from first principles establishes that there exist no other indecomposable $N>4$ theories.

\subsection{Superpotentials}
\label{sec:superpotentials}

The on-shell superconformal Chern--Simons-matter theories we will consider can all be derived in terms of an off-shell $N{=}1$ superspace formalism from a particular choice of quartic, gauge-invariant superpotential $\sW$.  For an $N$-extended superconformal symmetry to be realised in the on-shell theory requires $\sW$ to be invariant under a global $\fso(N-1)$ symmetry.  This can be understood as the subalgebra of the $\fso(N)$ R-symmetry in the on-shell theory which preserves the choice of $N{=}1$ superspace parameter in the off-shell theory.

The off-shell superfield that will describe the matter content in the on-shell $N$-extended superconformal Chern--Simons-matter theory can always be assembled into the representation of $\fso(N-1) \oplus \fg$ appearing in the third column of Table~\ref{tab:susy-enhancement}.  The $\fg$-modules $\fM$ for $2 \leq N \leq 6$ are always of complex or quaternionic type while only for $N{=}8$ is $\fM$ of real type.  As described in Appendix~\ref{sec:spec-compl-unit}, associated with any complex or quaternionic unitary representation $\fM$ there is canonical complex-sesquilinear map $\TT : \fM \times \fM \rightarrow \fg_\CC$ defined by the transpose of the action of $\fg$ with respect to the hermitian inner product.  When $\fM$ is real, the canonical real skewsymmetric map $T : \fM \times \fM \rightarrow \fg$ is defined in Appendix~\ref{sec:spec-real-orth} and, as explained in Appendix~\ref{sec:spec-complex-as-real}, can be thought of as the real part of $\TT$ by thinking of the real $\fM$ as a special kind of complex representation.  Using the inner product $(-,-)$ on $\fg$ one can thus define the canonical quartic, $\fg$-invariant tensor $\eR = ( \TT (-,-) , \TT (-,-) )$ on $\fM$ when it is of complex or quaternionic type and $\eR = (T(-,-),T(-,-))$ when $\fM$ is real.  All the superpotentials we shall consider can be expressed in terms of these tensors.  However, the fact that we are dealing with superfields in representations of $\fso(N-1) \oplus \fg$ rather than just $\fg$ means that for each $2 \leq N \leq 8$, $\fso(N-1)$-invariance of the superpotential is achieved only after using a particular quartic tensor on the appropriate spinor representation of $\fso(N-1)$ appearing in the third column of Table~\ref{tab:susy-enhancement} to create a singlet.

This structure will be made much more explicit in the rest of the paper but a schematic picture of what these superpotentials look like can be achieved by writing the superfield $\Xi = \Xi^a \, \be_a$ in terms of the basis $\{ \be_a \}$ for the spinor representation $\Delta^{(N-1)}$ of $\fso(N-1)$. (For $N$ odd, $\Delta^{(N-1)} = \Delta^{(N-1)}_+ \oplus \Delta^{(N-1)}_-$ and this basis is further decomposed in terms of the bases $\{ \be_\alpha \}$ on $\Delta^{(N-1)}_+$ and $\{ \be_{\dot \alpha} \}$ on $\Delta^{(N-1)}_-$.) For $N{=}4,5,6,8$, the component superfields $\Xi^a$ are valued respectively in unitary representations $\fM = W_1 \oplus W_2 , W , V , U$ where $W_1 , W_2, W \in \Dar(\fg,\HH)_{\text{aLTS}}$, $V \in \Dar(\fg,\CC)_{\text{aJTS}}$ and $U \in \Dar(\fg,\RR)_{\text{3LA}}$.  For $N{=}2$ the single component superfield $\Xi$ is valued in a generic complex unitary representation whereas for $N{=}3$ the single component superfield $\Xi$ is valued in a generic quaternionic unitary representation and taken to have charge $\half$ corresponding to chiral representation $\Delta^{(2)}_+$ of $\fso(2) \cong \fu(1)$.

The superpotentials can all be expressed as $\sW = \tfrac{1}{16} \int d^2 \theta \, \fW ( \Xi )$, where the measure is for the $N{=}1$ superspace parameter $\theta$, in terms of a real, quartic, $\fso(N-1) \oplus \fg$-invariant function $\fW$.  Table~\ref{tab:superpotentials} defines what this function is for all $N >1$.
\begin{table}[h!]
  \centering
  \begin{tabular}{|>{$}c<{$}|>{$}c<{$}|}
    \hline
    N & \fW ( \Xi ) \\\hline
    8 & \tfrac{1}{3} \, \Omega_{abcd} \, ( T ( \Xi^a , \Xi^b ) , T ( \Xi^c , \Xi^d ) ) \\
    6 & ( \TT ( \Xi^a , \Xi^b ) , \TT ( \Xi^b , \Xi^a ) ) + \Omega_{ab} \, \Omega^{cd} \, ( \TT ( \Xi^a , \Xi^c ) , \TT ( \Xi^b , \Xi^d ) ) \\
    5 & -\tfrac{1}{6} \, ( \TT ( \Xi^\alpha , \Xi^\beta ) , \TT ( \Xi^\beta , \Xi^\alpha ) ) -\tfrac{1}{6} \, ( \TT ( \Xi^{\dot \alpha} , \Xi^{\dot \beta} ) , \TT ( \Xi^{\dot \beta} , \Xi^{\dot \alpha} ) ) + ( \TT ( \Xi^\alpha , \Xi^{\dot \beta} ) , \TT ( \Xi^{\dot \beta} , \Xi^\alpha ) ) \\
    4 & \tfrac{1}{6} \, ( \TT_1 ( \Xi^a , \Xi^b ) , \TT_1 ( \Xi^b , \Xi^a ) ) + \tfrac{1}{6} \, ( \TT_2 ( \Xi^a , \Xi^b ) , \TT_2 ( \Xi^b , \Xi^a ) ) - ( \TT_1 ( \Xi^a , \Xi^b ) , \TT_2 ( \Xi^b , \Xi^a ) ) \\
    3 & ( \TT ( \Xi , \Xi ) , \TT ( \Xi , \Xi ) ) + {\mathrm{Re}} \, ( \TT ( \Xi , J \Xi ) , \TT ( \Xi , J \Xi ) )\\
    2 & ( \TT ( \Xi , \Xi ) , \TT ( \Xi , \Xi ) ) + {\mathrm{Re}} \, \fW_F ( \Xi ) \\\hline
  \end{tabular}
  \vspace{8pt}
  \caption{Superpotentials}
  \label{tab:superpotentials}
\end{table}
The tensor $\Omega$ appearing in the $N{=}6$ row is the $\fso(5) \cong \fusp(4)$-invariant symplectic form on $\Delta^{(5)}$ while in the $N{=}8$ row it denotes the $\fso(7)$-invariant self-dual Cayley 4-form on $\Delta^{(7)}$.  Repeated indices are contracted with respect to the hermitian inner product on $\Delta^{(N-1)}$.  The function $\fW_F$ in the $N{=}2$ row denotes an arbitrary quartic, $\fg$-invariant, holomorphic F-term superpotential that is compatible with $N{=}2$ supersymmetry.  The superpotential giving rise to $N{=}3$ superconformal symmetry is rigid.  In the course of the paper, we will show how one ascends Table~\ref{tab:superpotentials} realising more global symmetry for this same superpotential purely as a consequence of restricting attention to increasingly specialised types of unitary representations for the matter fields.

\section{Generic superconformal Chern--Simons-matter theories}
\label{sec:scft-ternary}

Having summarised the necessary data associated with matter representations of the superconformal algebra in three dimensions, we are now ready to describe some of the physical properties of the associated superconformal Chern--Simons-matter theories.  This section will focus on the structure of the generic theories with $N \leq 3$ superconformal symmetry.  This will pave the way for our consideration of further supersymmetry enhancement in subsequent sections.  We will begin by describing the generic $N{=}1$ superconformal Chern--Simons-matter theory that is built from any faithful unitary representation of a metric Lie algebra.  We then describe how supersymmetry is enhanced to $N{=}2$ and $N{=}3$ when the representation is assumed to be respectively of a generic complex and quaternionic type.  Some useful background references on this material are \cite{SchwarzCS,GYCS}.

\subsection{N{=}1 supersymmetry in three dimensions}
\label{sec:3dsusy}

$N{=}1$ supermultiplets in three dimensions are classed as either {\emph{gauge}} or {\emph{matter}}.

A gauge supermultiplet consists of a bosonic gauge field $A_\mu$ and a fermionic Majorana spinor $\chi$.  We will assume that both fields take values in a Lie algebra $\fg$ that is equipped with an ad-invariant inner product $(-,-)$.  It is sometimes assumed that these fields are valued in the adjoint representation of a semisimple Lie algebra.  This ensures that one can construct a supersymmetric lagrangian for the gauge supermultiplet in terms of the Killing form on the Lie algebra.  This is sufficient but not necessary for the construction of a lagrangian and in several important examples we will see that it is necessary for the inner product $(-,-)$ we have assumed on $\fg$ to not be the Killing form.

A matter supermultiplet consists of a bosonic scalar field $X$ and an auxiliary field $C$ plus a fermionic Majorana spinor $\Psi$.  We will assume that all the matter fields take values in a faithful unitary representation $\fM$ of $\fg$ and can be collected into a superfield $\Xi = X + {\bar \theta} \Psi + \half {\bar \theta} \theta C$ where the superspace coordinate $\theta$ is a fermionic Majorana spinor.

Our spinor conventions in three dimensions are as follows.  We take the Minkowski metric $\eta_{\mu\nu}$ on $\RR^{1,2}$ to have mostly plus signature and the orientation tensor $\varepsilon_{\mu\nu\rho}$ such that $\varepsilon_{012} = 1$.  The Clifford algebra $\Cl(1,2)$ has two inequivalent representations, both of which are real and two-dimensional.  Having chosen one of these representations, the Clifford algebra acts via $2\times2$ real matrices $\gamma_\mu$ which obey $\gamma_\mu \gamma_\nu + \gamma_\nu \gamma_\mu = 2 \eta_{\mu\nu} 1$ --- a suitable choice being $\gamma_0 = i \sigma_2$, $\gamma_1 = \sigma_1$ and $\gamma_2 = \sigma_3$.  A Majorana spinor $\xi$ has two real components and we define ${\bar \xi} := \xi^t \gamma^0$.  This implies ${\bar \chi} \xi = {\bar \xi} \chi$ and ${\bar \chi} \gamma_\mu \xi = - {\bar \xi} \gamma_\mu \chi$ for any fermionic Majorana spinors $\chi$ and $\xi$.  Finally, it will be useful to recall that $\gamma_{\mu\nu} = \varepsilon_{\mu\nu\rho} \gamma^\rho$, $\gamma_{\mu\nu\rho} = \varepsilon_{\mu\nu\rho} 1$ and the Fierz identity $\xi {\bar \chi} = -\half \left[ ( {\bar \chi} \xi ) 1 + ( {\bar \chi} \gamma^\mu \xi ) \gamma_\mu \right]$.

The coupling of gauge and matter supermultiplets is achieved using the action $\cdot$ of the $\fg$-module on $\fM$.  The $N{=}1$ supersymmetry transformations for the matter and gauge fields are
\begin{equation}
  \label{eq:susy}
  \begin{aligned}[m]
    \delta X &= {\bar \epsilon} \Psi\\
    \delta \Psi &= - (D_\mu X) \gamma^\mu \epsilon + C \epsilon\\
    \delta C &= - {\bar \epsilon} \gamma^\mu ( D_\mu \Psi ) - {\bar \epsilon} \chi \cdot X\\
    \delta A_\mu &= {\bar \epsilon} \gamma_\mu \chi\\
    \delta \chi &= \half F_{\mu\nu} \gamma^{\mu\nu} \epsilon,
  \end{aligned}
\end{equation}
where the parameter $\epsilon$ is a fermionic Majorana spinor and $D_\mu \phi = \partial_\mu \phi + A_\mu \cdot \phi$ for any field $\phi$ valued in $\fM$.  The derivative $D_\mu$ is covariant with respect to the gauge transformations $\delta \phi = - \Lambda \cdot \phi$ and $\delta A_\mu = \partial_\mu \Lambda + [ A_\mu , \Lambda ]$, for any gauge parameter $\Lambda$ valued in $\fg$ and where $[-,-]$ denotes the Lie bracket on $\fg$.  The curvature of this covariant derivative is $\fg$-valued and defined by $F_{\mu\nu} = [ D_\mu , D_\nu ] = \partial_\mu A_\nu - \partial_\nu A_\mu + [ A_\mu , A_\nu ]$.  It is easy to check that the commutator of two supersymmetry transformations in \eqref{eq:susy} closes off-shell giving a translation on $\RR^{1,2}$ plus a gauge transformation in $\fg$.  It is worth emphasising that up to this point we need not have assumed that $\fg$ is metric nor that the representation $\fM$ be unitary since neither of the inner products appear in \eqref{eq:susy} nor are they required for closure of the supersymmetry algebra.  It is in order to construct a lagrangian that is invariant under these supersymmetry transformations that necessitates this extra data.

\subsection{N{=}1 supersymmetric lagrangians in three dimensions}
\label{sec:3dsusylag}

There are three distinct contributions making up the most general Chern--Simons-matter lagrangian that is invariant under \eqref{eq:susy}.  They will be referred to as the supersymmetric Chern-Simons term $\eL_{CS}$, the supersymmetric matter term $\eL_{M}$ and the superpotential $\sW$ and will now be discussed in turn.

\subsubsection{Supersymmetric Chern-Simons term}
\label{sec:susy-chern-simons}

Given a gauge supermultiplet $( A_\mu , \chi )$ valued in the Lie algebra $\fg$ with ad-invariant inner product $(-,-)$, the integral of
\begin{equation}\label{eq:susy-cs}
  \eL_{CS} = -\varepsilon^{\mu\nu\rho} \left( A_\mu , \partial_\nu A_\rho + \tfrac{1}{3} [ A_\nu , A_\rho ] \right)  - ( {\bar \chi} , \chi ),
\end{equation}
is invariant under the last two supersymmetry transformations in \eqref{eq:susy}.  It is, of course, also manifestly gauge-invariant as a consequence of the ad-invariance of $(-,-)$.

\subsubsection{Supersymmetric matter term}
\label{sec:susy-matter}

Given a matter supermultiplet $(X,\Psi ,C)$ valued in a real representation $\fM$ of $\fg$ with an invariant positive-definite symmetric inner product $\left< -,- \right>$, that is coupled to the aforementioned gauge supermultiplet $(A_\mu , \chi )$, the supersymmetric matter term is
\begin{equation}\label{eq:susy-m}
  \eL_{M} = -\half \left< D_\mu X , D^\mu X \right> + \half \left< {\bar \Psi} , \gamma^\mu D_\mu \Psi \right> + \half \left< C , C \right> - \left< X ,  {\bar \chi} \cdot \Psi \right>.
\end{equation}
Replacing covariant with partial derivatives in the first two terms and dropping the fourth term would describe the supersymmetric lagrangian for the matter fields in the ungauged theory.  As it is, the integral of \eqref{eq:susy-m} is invariant under \eqref{eq:susy} with the fourth term describing an additional gauge-matter coupling that is required to cancel the supersymmetry variation of the first three terms in the gauged theory.

\subsubsection{Superpotential}
\label{sec:superpotential}

The integral of $\eL_{CS} + \eL_{M}$ from \eqref{eq:susy-cs} and \eqref{eq:susy-m} is classically scale-invariant with respect to the fields $(X,\Psi ,C,A_\mu , \chi )$ being assigned weights $( \half ,1, \tfrac{3}{2} ,1, \tfrac{3}{2} )$.  One could add a mass term of the form $\half \int d^2 \theta \left< \Xi , \Xi \right> = \left< X , C \right> - \half \left< {\bar \Psi} , \Psi \right>$ which is manifestly supersymmetric but of course breaks the classical scale invariance.

More generally, one could consider a superpotential
\begin{equation}\label{eq:superpotential}
  \sW = \int d^2 \theta \; \fW ( \Xi ) = C^a \frac{\partial}{\partial X^a} \, \fW (X) - \tfrac{1}{2} {\bar \Psi}^a \Psi^b \,
  \frac{\partial^2}{\partial X^a \partial X^b} \, \fW (X),
\end{equation}
where $\fW$ is an arbitrary polynomial function on $\fM$ and the matter superfield components have been written relative to a basis $\{ e_a \}$ for $\fM$.  Scale-invariance requires that $\fW$ must in fact be a quartic function.  Notice that the expression \eqref{eq:superpotential} does not require an inner product on $\fM$.  However, given the gauge couplings to the matter fields in \eqref{eq:susy}, invariance of the integral of the superpotential under these supersymmetry transformations requires $\fW$ to be an $\fg$-invariant function. 

Given a $\fg$-invariant inner product on $\fM$, as was required for the existence of $\eL_M$ in \eqref{eq:susy-m}, one has a generic type of quartic $\fg$-invariant superpotential proportional to $\int d^2 \theta \left< \Xi , \Xi \right>^2$.  However, this can be thought of as arising as a marginal deformation by the square of the $\fg$-invariant operator $\left< \Xi , \Xi \right>$ of an existing classically superconformal $N{=}1$ Chern--Simons-matter theory.  Such operators are of course generically unprotected from quantum corrections given only $N{=}1$ supersymmetry.  We shall not consider superpotentials which take the form of such deformations in our subsequent analysis.

\subsubsection{On-shell N{=}1 supersymmetric lagrangians}
\label{sec:on-shell-3dsusy}

Before discussing the superpotentials leading to increased amounts of supersymmetry, let us conclude this subsection by noting the on-shell form of the generic $N{=}1$ supersymmetric Chern-Simons-matter lagrangian $\eL_{CS} + \eL_{M} + \sW$, after integrating out the auxiliary fields $\chi$ and $C$.  Their respective equations of motion are $\chi = \half T (X, \Psi )$ and $\left< C,- \right> = - d \fW (X)$ (the terms in the equation for $C$ being thought of as $\fM^*$-valued).  Substituting these expressions into the lagrangian gives
\begin{equation}
  \label{eq:susy-lag-on-shell}
  \begin{aligned}[m]
    \eL_{CS} + \eL_{M} + \sW =&  -\varepsilon^{\mu\nu\rho} \left( A_\mu , \partial_\nu A_\rho + \tfrac{1}{3} [ A_\nu , A_\rho ] \right) -\half \left< D_\mu X , D^\mu X \right> -\half \left< d \fW (X) , d \fW(X) \right> \\
    &+ \half \left< {\bar \Psi} , \gamma^\mu D_\mu \Psi \right> - \half {\bar \Psi}^a \Psi^b \partial_a \partial_b \fW (X) + \tfrac{1}{4} ( T (X, {\bar \Psi} ) , T (X, \Psi )).
  \end{aligned}
\end{equation}
In a slight abuse of notation, the expression $-\half \left< d \fW (X) , d \fW(X) \right>$ for the scalar potential is shorthand for $-\half g^{ab} \partial_a \fW(X) \partial_b \fW(X)$ with $g^{ab}$ denoting components of the matrix inverse of $\left< e_a , e_b \right>$ on $\fM$.  It is straightforward to check that \eqref{eq:susy-lag-on-shell} is invariant under the $N{=}1$ supersymmetry transformations \eqref{eq:susy}, upon substituting into their expressions the field equations for the auxiliary fields $\chi$ and $C$.  These supersymmetry transformations are
\begin{equation}
  \label{eq:susy-onshell}
  \begin{aligned}[m]
    \delta X &= {\bar \epsilon} \Psi\\
    \delta \Psi &= - (D_\mu X) \gamma^\mu \epsilon - d \fW (X) \epsilon\\
    \delta A_\mu &= \half T \left( X , {\bar \epsilon} \gamma_\mu \Psi \right) ,
  \end{aligned}
\end{equation}
and close up to a translation on $\RR^{1,2}$ plus a gauge transformation, using the equations of motion from \eqref{eq:susy-lag-on-shell}.  Notice the effect of integrating out the auxiliary fields has been to generate a sextic potential for the scalar fields and various scalar-fermion Yukawa couplings that appear in the second line of \eqref{eq:susy-lag-on-shell}.

\subsection{N{=}2 supersymmetry for $\fM$ complex}
\label{sec:nequal2}

When the matter representation $\fM =V \in \Dar(\fg,\CC)$, there exists a superpotential
\begin{equation}\label{eq:superpotentialcomplex}
  \sW_{\CC} = \tfrac{1}{16}\int d^2 \theta \; (\TT (\Xi ,\Xi ),\TT (\Xi ,\Xi ) ).
\end{equation}
Proposition \ref{pr:T-map-C} implies that $\TT (\Xi ,\Xi )$ is pure imaginary (indeed $\TT (\Xi ,\Xi ) = i T(\Xi , I \Xi )$ from Lemma~\ref{le:T-map-C}), whence \eqref{eq:superpotentialcomplex} is real.  Notice that $T(\Xi , \Xi ) \equiv 0$ and so there is no possibility to build an alternative superpotential based on the real part $T$ of $\TT$ here.  The value of the coefficient is fixed uniquely by the requirement that this superpotential gives rise to an enhanced $N{=}2$ supersymmetry, when added to $\eL_{CS} + \eL_{M}$.  That is, it provides precisely the additional gauge-matter couplings that are required for $N{=}2$ supersymmetry.

To understand why this superpotential gives rise to an enhancement in supersymmetry, it will be useful to note that one can obtain \eqref{eq:superpotentialcomplex} via integrating out an auxiliary matter supermultiplet.  The auxiliary matter superfield $\Pi = \sigma - {\bar \theta} {\hat \chi} + \half {\bar \theta} \theta D$ here is just an $N{=}1$ matter superfield that is valued in $\fg$ instead of $V$ (the supersymmetry transformations for these fields just follow from \eqref{eq:susy} by taking the action $\cdot$ of $\fg$ to be the adjoint action of $\fg$ on itself).  The real superpotential
\begin{equation}\label{eq:superpotentialcomplexaux}
  \int d^2 \theta \; ( \Pi , \Pi ) + \tfrac{i}{2} ( \Pi , \TT (\Xi ,\Xi ) ) ,
\end{equation}
then gives precisely \eqref{eq:superpotentialcomplex} after integrating out $\Pi$.  Classical scale invariance here follows from the auxiliary components $(\sigma ,{\hat \chi},D)$ being assigned weights $( 1, \tfrac{3}{2} ,2)$.

However, before integrating out $\Pi$, notice that adding the first term in \eqref{eq:superpotentialcomplexaux} to the supersymmetric Chern-Simons term \eqref{eq:susy-cs} gives
\begin{equation}\label{eq:susy-cs2}
  \eL_{CS} + \int d^2 \theta \; ( \Pi , \Pi ) = -\varepsilon^{\mu\nu\rho} \left( A_\mu , \partial_\nu A_\rho + \tfrac{1}{3} [ A_\nu , A_\rho ] \right)  - ( {\bar \chi} , \chi ) - ( {\bar {\hat \chi}} , {\hat \chi} ) +2\, (\sigma ,D) ,
\end{equation}
which is precisely the lagrangian $\eL_{CS}^{N{=}2}$ for $N{=}2$ supersymmetric Chern-Simons theory.  The supersymmetry enhancement here can be seen to arise from the choice of taking either $\chi$ or ${\hat \chi}$ to describe the superpartner of the gauge field $A_\mu$ in the $N{=}1$ gauge supermultiplet.  It will be convenient to assemble the fermions into a complex spinor $\chi_{\CC} = \chi + i {\hat \chi}$, whereby $( {\bar \chi} , \chi ) + ( {\bar {\hat \chi}} , {\hat \chi} ) = ( {\bar \chi_{\CC}}^* , \chi_{\CC} )$, with $\chi_{\CC}^* = \chi - i {\hat \chi}$ here denoting the complex conjugate of $\chi_{\CC}$.  The integral of $\eL_{CS}^{N{=}2}$ in \eqref{eq:susy-cs2} is invariant under the $N{=}2$ supersymmetry transformations
\begin{equation}
  \label{eq:susy2g}
  \begin{aligned}[m]
    \delta A_\mu &= {\mathrm{Re}} \left( {\bar \epsilon_{\CC}}^* \gamma_\mu \chi_{\CC} \right) \\
    \delta \chi_{\CC} &= \half F_{\mu\nu} \gamma^{\mu\nu} \epsilon_{\CC} + i ( D_\mu \sigma ) \gamma^\mu \epsilon_{\CC} - iD\, \epsilon_{\CC} \\
    \delta \sigma &= - {\mathrm{Im}} \left( {\bar \epsilon_{\CC}}^* \chi_{\CC} \right) \\
    \delta D &= {\mathrm{Im}} \left( {\bar \epsilon_{\CC}}^* \left( \gamma^\mu D_\mu \chi_{\CC} +i [ \sigma , \chi_{\CC} ] \right) \right) ,
  \end{aligned}
\end{equation}
where the parameter $\epsilon_{\CC}$ is a complex spinor on $\RR^{1,2}$.  Associated with this enhanced $N{=}2$ supersymmetry there is a $\fu (1)$ R-symmetry under which $\chi_{\CC}$ and $\epsilon_{\CC}$ have charge -1 (their complex conjugates having charge +1) while $A_\mu$, $\sigma$ and $D$ are uncharged.

Combining the remaining term in \eqref{eq:superpotentialcomplexaux} with the supersymmetric matter term \eqref{eq:susy-m} gives
\begin{equation}\label{eq:susy-m2}
  \begin{aligned}[m]
  \eL_{M} +\half \int d^2 \theta \, \left< \Xi , i \Pi \cdot \Xi \right> =& -\half \left< D_\mu X , D^\mu X \right> + \half \left< {\bar \Psi} , \gamma^\mu D_\mu \Psi \right> + \half \left< C , C \right> - \left< X , {\bar \chi_{\CC}}^* \cdot \Psi \right> \\
  &+\half \left< X , i D \cdot X \right> - \half \left< {\bar \Psi} , i \sigma \cdot \Psi \right> + \left< X , i \sigma \cdot C \right>.
  \end{aligned}
\end{equation}
In terms of the redefined auxiliary matter field $F := C + i\sigma \cdot X$ one finally recovers from \eqref{eq:susy-m2} the standard off-shell gauged $N{=}2$ supersymmetric matter lagrangian
\begin{equation}\label{eq:susy-m2a}
  \begin{aligned}[m]
  \eL_{M}^{N{=}2} =&-\half \left< D_\mu X , D^\mu X \right> + \half \left< {\bar \Psi} , \gamma^\mu D_\mu \Psi \right> + \half \left< F , F \right> - \left< X , {\bar \chi_{\CC}}^* \cdot \Psi \right> \\
  &+\half \left< X , i D \cdot X \right> - \half \left< {\bar \Psi} , i \sigma \cdot \Psi \right> -\half \left< \sigma \cdot X , \sigma \cdot X \right>,
  \end{aligned}
\end{equation}
whose integral is invariant under the $N{=}2$ supersymmetry transformations
\begin{equation}
  \label{eq:susy2m}
  \begin{aligned}[m]
    \delta X &= {\bar \epsilon_{\CC}}^* \Psi \\
    \delta \Psi &= - ( D_\mu X ) \gamma^\mu \epsilon_{\CC} + F\, \epsilon_{\CC}^* -i\sigma \cdot X \, \epsilon_{\CC} \\
    \delta F &= - {\bar \epsilon_{\CC}} \left(  \gamma^\mu D_\mu \Psi + \chi_{\CC} \cdot X - i\sigma \cdot \Psi \right) ,
  \end{aligned}
\end{equation}
for the matter supermultiplet, combined with the transformations \eqref{eq:susy2g} for the gauge supermultiplet.  The matter fields $(X,\Psi ,F)$ have charges $(\half , -\half , -\tfrac{3}{2} )$ under the $\fu (1)$ R-symmetry.  The transformations \eqref{eq:susy2m} with $\epsilon_\CC = \epsilon$ real generate an $N{=}1$ subalgebra whose transformations are recovered from \eqref{eq:susy} in terms of the matter superfields $\Pi \in \fg$ and $\Xi \in V$ defined above precisely as a consequence of the identification $F = C + i\sigma \cdot X$.  The off-shell closure of the $N{=}2$ supersymmetry algebra from \eqref{eq:susy2g}, \eqref{eq:susy2m} is straightforward to check.

A rather more economical expression for $\eL_{M}^{N{=}2}$ can be obtained by collecting the fields into $N{=}2$ superfields involving a complex superspace coordinate $\theta_{\CC}$.  The matter fields $(X, \Psi , F)$ can be assembled into the chiral $N{=}2$ superfield $\Xi_{\CC} = X + {\bar \theta_{\CC}^*} \Psi + \half {\bar \theta_{\CC}^*} \theta_{\CC}^* F$.  The $N{=}2$ gauge supermultiplet fields $( A_\mu , \chi_{\CC} , \sigma , D)$ can be assembled into the $N{=}2$ superfield operator ${\sf V} = {\bar \theta_{\CC}^*} \gamma^\mu \theta_{\CC} D_\mu + i {\bar \theta_{\CC}^*} \theta_{\CC} \sigma - \tfrac{i}{4} ( {\bar \theta_{\CC}^*} \theta_{\CC}^* )( {\bar \theta_{\CC}} \theta_{\CC} ) D + \half ( {\bar \theta_{\CC}^*} \theta_{\CC}^* ) {\bar \theta_{\CC}} \chi_{\CC} - \half ( {\bar \theta_{\CC}} \theta_{\CC} ) {\bar \theta_{\CC}^*} \chi_{\CC}^*$.  Thus ${\sf V}$ is pure imaginary.  The superspace coordinate $\theta_{\CC}$ has $\fu (1)$ R-charge -1 and so $\Xi_{\CC}$ has R-charge $\half$ while ${\sf V}$ is neutral.  In terms of these quantities, one can write
\begin{equation}\label{eq:susy-m2b}
\eL_{M}^{N{=}2} = \int d^4 \theta_{\CC} \; \half \left< \Xi_{\CC} , e^{- {\sf V}} \cdot \Xi_{\CC} \right> ,
\end{equation}
up to total derivatives and where $d^4 \theta_{\CC} = d^2 \theta_{\CC} d^2 \theta_{\CC}^*$ is shorthand for the full $N{=}2$ superspace measure.

\subsubsection{N{=}2 F-term superpotential}
\label{sec:ftermsuperpotential}

To the $N{=}2$ supersymmetric Chern-Simons-matter lagrangian $\eL_{CS}^{N{=}2} + \eL_{M}^{N{=}2}$, one can add a so-called F-term superpotential of the form
\begin{equation}\label{eq:ftermsuperpotential}
  \sW_F = \int d^2 \theta_{\CC} \; \fW_F ( \Xi_{\CC} ) + \int d^2 \theta_{\CC}^* \; \fW_F ( \Xi_{\CC} )^* ,
\end{equation}
which is also $N{=}2$ supersymmetric provided that $\fW_F$ is a $\fg$-invariant holomorphic function of the matter fields.  Scale-invariance again requires $\fW_F$ to be a quartic function.  Notice that \eqref{eq:ftermsuperpotential} does not require an inner product on $V$ but demands it is complex.  Notice also that the chiral superspace measure appearing in \eqref{eq:ftermsuperpotential} guarantees that $\sW_F$ is invariant under the $\fu (1)$ R-symmetry of the $N{=}2$ superalgebra.

One can generally obtain the $N{=}2$ F-term superpotential in \eqref{eq:ftermsuperpotential} off-shell from a particular $N{=}1$ superpotential of the form $\int d^2 \theta \; 2\, \Re \, \fW_F (\Xi )$ when $\fM$ is of complex type, where $\Xi = X + {\bar \theta} \Psi + \half {\bar \theta} \theta F$ is the $N{=}1$ superfield on which the the chiral $N{=}2$ superfield $\Xi_{\CC}$ is constructed.  The extra data here being precisely the quartic $\fg$-invariant holomorphic function $\fW_F$.  It is worth pointing out that one recovers precisely the same F-term superpotential from the aforementioned $N{=}1$ superpotential based on the $N{=}1$ superfield $\Xi = X + {\bar \theta} \Psi + \half {\bar \theta} \theta C$ we had been using before.  This follows from the fact that $F - C = i\sigma \cdot X$ and so the potential discrepancy between the resulting superpotentials is proportional to $\left< i\sigma \cdot X , \partial \fW_F (X) \right>$ which vanishes identically as a consequence of $\fW_F$ being $\fg$-invariant.

In summary, we have seen that, when $\fM \in \Dar(\fg,\CC)$, one can obtain the general $N{=}2$ Chern-Simons-matter lagrangian $\eL_{CS}^{N{=}2} + \eL_{M}^{N{=}2} + \sW_F$ from the choice of $N{=}1$ superpotential
\begin{equation}\label{eq:superpotential2}
  \sW_{\CC} + \sW_F = \int d^2 \theta \; \tfrac{1}{16} (\TT (\Xi ,\Xi ),\TT (\Xi ,\Xi ) ) + 2\, \Re \, \fW_F (\Xi ).
\end{equation}
Whenever $\fM \in \Dar(\fg,\CC)$, one can decompose a generic quartic $N{=}1$ superpotential into its $(4,0)+(3,1)+(2,2)+(1,3)+(0,4)$ components, with respect to the complex structure on $\fM$.  Thus we have found that enhancement to $N{=}2$ supersymmetry for the $N{=}1$ Chern-Simons-matter lagrangian is guaranteed provided the $(3,1)+(1,3)$ component is absent and the $(2,2)$ component is $\sW_\CC$.  Furthermore, as proven in Appendix~\ref{sec:der-nequal2superpotential}, the expression in \eqref{eq:superpotential2} is the unique choice of $N{=}1$ superpotential giving rise to an on-shell lagrangian which is invariant under the $\fu(1)$ R-symmetry that is necessary for $N{=}2$ supersymmetry.

\subsubsection{On-shell N{=}2 supersymmetric lagrangians}
\label{sec:on-shell-3dsusy-complex}

Before going on to look at further types of supersymmetry enhancing superpotentials which exist when $\fM$ is quaternionic, let us conclude this subsection by noting the on-shell form of the generic $N{=}2$ supersymmetric Chern-Simons-matter lagrangian $\eL_{CS}^{N{=}2} + \eL_{M}^{N{=}2} + \sW_F$, after integrating out the auxiliary fields $\chi_{\CC}$, $D$ and $F$.  Their equations of motion are respectively $\chi_{\CC}^* = \half \, \TT (X, \Psi )$, $\sigma = - \tfrac{i}{4} \, \TT (X,X)$ and $\left< F,- \right> = - \partial \fW_F (X)$.  Substituting these expressions into the lagrangian gives
\begin{equation}
  \label{eq:susy-lag-on-shell-complex}
  \begin{aligned}[m]
    \eL_{CS}^{N{=}2} + \eL_{M}^{N{=}2} + \sW_F =&  -\varepsilon^{\mu\nu\rho} \left( A_\mu , \partial_\nu A_\rho + \tfrac{1}{3} [ A_\nu , A_\rho ] \right) -\half \left< D_\mu X , D^\mu X \right> - \fV_D (X) - \fV_F (X) \\
    &+ \half \left< {\bar \Psi} , \gamma^\mu D_\mu \Psi \right> - \half {\bar \Psi}^a \Psi^b \partial_a \partial_b \fW_F (X) - \half {\bar \Psi}^{\bar a} \Psi^{\bar b} \partial_{\bar a} \partial_{\bar b} \fW_F (X)^* \\
    &- \tfrac{1}{4} ( \TT (X, {\bar \Psi} ) , \TT (\Psi , X )) -\tfrac{1}{8} ( \TT (X,X) , \TT ( {\bar \Psi} , \Psi )).
  \end{aligned}
\end{equation}
where we have introduced the positive-definite D-term and F-term sextic scalar potentials $\fV_D (X) = \tfrac{1}{32} \left< \TT (X,X) \cdot X , \TT (X,X) \cdot X \right>$ and $\fV_F (X) = \half \left< \partial \fW_F (X) , \partial \fW_F (X) \right>$ and the indices are with respect to a complex basis $\{ \be_a \}$ on $V$.

Invariance of \eqref{eq:susy-lag-on-shell-complex} under the $N{=}2$ supersymmetry transformations \eqref{eq:susy2g} and \eqref{eq:susy2m} can be established after substituting the equations of motion for $\chi_{\CC}$, $D$ and $F$.  These supersymmetry transformations are
\begin{equation}
  \label{eq:susy2-onshell}
  \begin{aligned}[m]
    \delta X &= {\bar \epsilon_{\CC}}^* \Psi \\
    \delta \Psi &= - ( D_\mu X ) \gamma^\mu \epsilon_{\CC} - \partial \fW_F (X) \, \epsilon_{\CC}^* - \tfrac{1}{4} \TT (X,X) \cdot X \, \epsilon_{\CC} \\
    \delta A_\mu &= \half T \left( X , {\bar \epsilon_{\CC}}^* \gamma_\mu \Psi \right) ,
  \end{aligned}
\end{equation}
and close up to a translation on $\RR^{1,2}$ plus a gauge transformation, using the equations of motion from \eqref{eq:susy-lag-on-shell-complex}.

\subsection{N{=}3 supersymmetry for $\fM$ quaternionic}
\label{sec:nequal3}

Let $\fM = W \in \Dar(\fg,\HH)$.  Recall that we view quaternionic representations as complex representations with a quaternionic structure map $J$.  In the language of Appendix \ref{sec:relat-betw-real}, we work not with $W$ but with $\rh{W}$; although we shall not enforce this notational distinction and talk of fields taking values in $W$ when in fact they take values in $\rh{W}$.  In the case of a quaternionic representation $W$, we have in addition to $\TT (-,-)$ based on just the hermitian structure on $W$, now also the map $\TT (-,J-)$ in terms of which we will write down a superpotential which provides a further $N{=}3$ supersymmetry enhancement.

The map $\TT (-,J-)$ is symmetric and complex bilinear in its arguments.  Thus one can define the following quartic, $\fg$-invariant, holomorphic $N{=}2$ F-term superpotential
\begin{equation}\label{eq:nequal3ftermsuperpotential}
   \fW_F ( \Xi_{\CC} ) = \tfrac{1}{32} ( \TT ( \Xi_\CC , J \Xi_\CC ) , \TT ( \Xi_\CC , J \Xi_\CC ) ),
\end{equation}
where $\Xi_\CC$ is a chiral $N{=}2$ matter superfield, just as in the previous section but here taking values in $W$.  The value of the coefficient is again fixed by the requirement that the F-term superpotential $\sW_F$ for \eqref{eq:nequal3ftermsuperpotential} gives rise to an enhanced $N{=}3$ supersymmetry, when added to $\eL_{CS}^{N{=}2} + \eL_{M}^{N{=}2}$.  Equivalently, following the results of the previous section, this F-term superpotential also arises from the choice of $N{=}1$ superpotential
\begin{equation}
\label{eq:superpotential3}
\begin{aligned}[m]
  \sW_{\HH} &= \tfrac{1}{16} \int d^2 \theta \; \left[ ( \TT (\Xi ,\Xi ), \TT (\Xi ,\Xi ) ) + \Re\, ( \TT (\Xi , J \Xi ), \TT (\Xi , J \Xi ) )  \right] \\
  &= \tfrac{1}{16} \int d^2 \theta \; \left[ - ( T (\Xi , I \Xi ), T (\Xi , I \Xi ) ) + ( T (\Xi , J \Xi ) , T (\Xi , J \Xi ) ) - ( T (\Xi , I J \Xi ) , T (\Xi , I J \Xi ) )  \right] ,
\end{aligned}
\end{equation}
where $\Xi$ is an $N{=}1$ matter superfield, again taking values in $W$.  The superpotential above $\sW_{\HH} = \sW_{\CC} + \sW_F$ just describes a special case of the generic expression in \eqref{eq:superpotential2} for the choice of F-term superpotential \eqref{eq:nequal3ftermsuperpotential} based on the quaternionic structure here. 

The enhanced $N{=}3$ supersymmetry can be seen using a similar method to that which was employed in the previous subsection for understanding enhancement from $N{=}1$ to $N{=}2$ supersymmetry.  In this case however we note that one can obtain the $N{=}2$ F-term superpotential based on \eqref{eq:nequal3ftermsuperpotential} via integrating out an auxiliary chiral $N{=}2$ matter superfield $\Pi_\CC = \tau_\CC + {\bar \theta_{\CC}^*} \zeta_{\CC}^* + \half {\bar \theta_{\CC}^*} \theta_{\CC}^* E_\CC$ that is valued in $\fg_\CC$ rather than $W$ (the $N{=}2$ supersymmetry transformations for these fields just follow from \eqref{eq:susy2m} by taking the action $\cdot$ of $\fg$ to be the adjoint action of $\fg_\CC$ on itself).  The F-term superpotential resulting from
\begin{equation}\label{eq:superpotentialquaternionaux}
  -\half ( \Pi_\CC , \Pi_\CC ) - \tfrac{1}{4} ( \Pi_\CC , \TT (\Xi_\CC , J \Xi_\CC ) ),
\end{equation}
then gives precisely \eqref{eq:nequal3ftermsuperpotential} after integrating out $\Pi_\CC$.  Classical scale invariance here follows from the auxiliary components $(\tau_\CC , \zeta_\CC , E_\CC )$ being assigned weights $( 1, \tfrac{3}{2} ,2)$ while their $\fu(1)$ R-charges are $(1,0,-1)$.

Before integrating out $\Pi_\CC$ though, notice that adding the F-term superpotential associated with the first term in \eqref{eq:superpotentialquaternionaux} to the $N{=}2$ supersymmetric Chern-Simons term \eqref{eq:susy-cs2} gives
\begin{equation}\label{eq:susy-cs3}
  \begin{aligned}[m]
  \eL_{CS}^{N{=}2} &-\half  \int d^2 \theta_\CC \; ( \Pi_\CC , \Pi_\CC ) -\half  \int d^2 \theta_\CC^* \; ( \Pi_\CC^* , \Pi_\CC^* )\\
  =& -\varepsilon^{\mu\nu\rho} \left( A_\mu , \partial_\nu A_\rho + \tfrac{1}{3} [ A_\nu , A_\rho ] \right) +2\, (\sigma ,D) - ( \tau_\CC , E_\CC ) - ( \tau_\CC^* , E_\CC^* ) \\
  & - ( {\bar \chi_{\CC}}^* , \chi_{\CC} ) +\half ( {\bar \zeta_{\CC}} , \zeta_{\CC} ) +\half ( {\bar \zeta_{\CC}}^* , \zeta_{\CC}^* ),
  \end{aligned}
\end{equation}
which is in fact precisely the lagrangian $\eL_{CS}^{N{=}3}$ for $N{=}3$ supersymmetric Chern--Simons theory.  This can be seen by recalling that the off-shell $N{=}4$ supersymmetry transformations for the $N{=}2$ superfield components $(A_\mu , \chi_\CC , \sigma , D)$ and $( \tau_\CC , \zeta_\CC , E_\CC )$ above, which collectively comprise an $N{=}4$ vector supermultiplet in three dimensions, are
\begin{equation}
  \label{eq:susy4g}
  \begin{aligned}[m]
    \delta A_\mu &= {\mathrm{Re}} \left( {\bar \epsilon_{\CC}}^* \gamma_\mu \chi_\CC + {\bar \eta_{\CC}}^* \gamma_\mu \zeta_\CC \right) \\
    \delta \chi_{\CC} &= \half F_{\mu\nu} \gamma^{\mu\nu} \epsilon_{\CC} + i ( D_\mu \sigma ) \gamma^\mu \epsilon_{\CC} - iD\, \epsilon_{\CC} \\
    &\quad + ( D_\mu \tau_\CC^* ) \gamma^\mu \eta_\CC^* - i [ \sigma , \tau_\CC^* ] \eta_\CC^* - E_\CC \, \eta_\CC \\
    \delta \zeta_{\CC} &= \half F_{\mu\nu} \gamma^{\mu\nu} \eta_{\CC} + i ( D_\mu \sigma ) \gamma^\mu \eta_{\CC} + \left( iD - [ \tau_\CC , \tau_\CC^* ] \right) \eta_{\CC} \\
    &\quad - ( D_\mu \tau_\CC^* ) \gamma^\mu \epsilon_\CC^* + i [ \sigma , \tau_\CC^* ] \epsilon_\CC^* + E_\CC^* \, \epsilon_\CC \\
    \delta \sigma &= - {\mathrm{Im}} \left( {\bar \epsilon_{\CC}}^* \chi_{\CC} + {\bar \eta_{\CC}}^* \zeta_{\CC} \right) \\
    \delta \tau_\CC^* &= {\bar \epsilon_{\CC}} \zeta_\CC - {\bar \eta_{\CC}} \chi_\CC \\
    \delta D &= {\mathrm{Im}} \left( {\bar \epsilon_{\CC}}^* \left( \gamma^\mu D_\mu \chi_{\CC} +i [ \sigma , \chi_{\CC} ] \right) - {\bar \eta_{\CC}}^* \left( \gamma^\mu D_\mu \zeta_{\CC} +i [ \sigma , \zeta_{\CC} ] \right) -2\, [ {\bar \eta_{\CC}}^* \chi_\CC^* , \tau_\CC^* ] \right) \\
    \delta E_\CC &= - {\bar \epsilon_{\CC}} \left( \gamma^\mu D_\mu \zeta_\CC^* -i [\sigma , \zeta_\CC^* ] + [ \chi_\CC , \tau_\CC ] \right) + {\bar \eta_{\CC}}^* \left( \gamma^\mu D_\mu \chi_\CC +i [\sigma , \chi_\CC ] - [ \zeta_\CC^* , \tau_\CC^* ] \right) ,
  \end{aligned}
\end{equation}
where the parameters $\epsilon_\CC$ and $\eta_\CC$ are complex spinors on $\RR^{1,2}$, with $\fu(1)$ R-charges $-1$ and $0$ respectively.  Upon setting $\eta_\CC =0$ in \eqref{eq:susy4g} one recovers the $N{=}2$ supersymmetry transformations in \eqref{eq:susy2g} for $(A_\mu , \chi_\CC , \sigma , D)$ and those in \eqref{eq:susy2m} for the components of the auxiliary superfield $\Pi_\CC = \tau_\CC + {\bar \theta_{\CC}^*} \zeta_{\CC}^* + \half {\bar \theta_{\CC}^*} \theta_{\CC}^* E_\CC$.  It is not too difficult to check that the supersymmetry transformations in \eqref{eq:susy4g} close off-shell for any $\epsilon_\CC$ and $\eta_\CC$, thus generating an $N{=}4$ superconformal algebra.  However, invariance of the integral of the Chern--Simons lagrangian in \eqref{eq:susy-cs3} under \eqref{eq:susy4g} is only possible provided ${\mathrm{Re}} \, \eta_\CC = 0$.  Thus it is indeed only invariant under the subset of $N{=}3$ supersymmetry transformations generated by the parameters $\epsilon_\CC$ and ${\mathrm{Im}} \, \eta_\CC$ in \eqref{eq:susy4g}.

Let us now consider the remaining contribution coming from adding the F-term superpotential associated with the second term in \eqref{eq:superpotentialquaternionaux} to the $N{=}2$ supersymmetric matter lagrangian \eqref{eq:susy-m2a}.  This gives
\begin{equation}\label{eq:susy-m3}
  \begin{aligned}[m]
  \eL_{M}^{N{=}2} &-\half  \int d^2 \theta_\CC \; ( \Pi_\CC , \TT (\Xi_\CC , J \Xi_\CC ) ) +\half  \int d^2 \theta_\CC^* \; ( \Pi_\CC^* , \TT ( J \Xi_\CC ,\Xi_\CC ) ) \\
  =& -\half \left< D_\mu X , D^\mu X \right> + \half \left< X , i D \cdot X \right> -\half \left< \sigma \cdot X , \sigma \cdot X \right> + \half \left< F , F \right> \\
  &+ \half \left< {\bar \Psi} , \gamma^\mu D_\mu \Psi \right>  - \half \left< {\bar \Psi} , i \sigma \cdot \Psi \right> + \left< {\bar \Psi} , \chi_{\CC} \cdot X \right> \\
  &- \half \left< {\bar \Psi} , \tau_\CC^* \cdot J \Psi \right> - \left< {\bar \Psi} , \zeta_\CC \cdot J X \right>  - \half \left< E_\CC \cdot X , JX \right> + \left< F , \tau_\CC^* \cdot JX \right>,
  \end{aligned}
\end{equation}
and upon integrating out the auxiliary matter field $F$, thus fixing $F = - \tau_\CC^* \cdot JX$, one obtains the $N{=}4$ supersymmetric matter lagrangian
\begin{equation}\label{eq:susy-m3a}
  \begin{aligned}[m]
  \eL_{M}^{N{=}4} =& -\half \left< D_\mu X , D^\mu X \right> + \half \left< X , i D \cdot X \right> -\half \left< \sigma \cdot X , \sigma \cdot X \right> - \half \left< E_\CC \cdot X , JX \right> -\half \left< \tau_\CC \cdot X , \tau_\CC \cdot X \right> \\
  &+ \half \left< {\bar \Psi} , \gamma^\mu D_\mu \Psi \right>  - \half \left< {\bar \Psi} , i \sigma \cdot \Psi \right> + \left< {\bar \Psi} , \chi_{\CC} \cdot X \right> - \half \left< {\bar \Psi} , \tau_\CC^* \cdot J \Psi \right> - \left< {\bar \Psi} , \zeta_\CC \cdot J X \right>.
  \end{aligned}
\end{equation}
The integral of $\eL_{M}^{N{=}4}$ is indeed invariant under the on-shell $N{=}4$ supersymmetry transformations
\begin{equation}
  \label{eq:susy4m}
  \begin{aligned}[m]
    \delta X &= ( {\bar \epsilon_{\CC}}^* + {\bar \eta_{\CC}} J ) \Psi \\
    \delta \Psi &= - \gamma^\mu ( \epsilon_\CC - \eta_\CC J ) D_\mu X - ( \tau_\CC^* \cdot JX ) \epsilon_\CC^* - ( \tau_\CC^* \cdot X ) \eta_\CC^* - i( \epsilon_\CC - \eta_\CC J ) \sigma \cdot X,
  \end{aligned}
\end{equation}
for the matter fields $X$ and $\Psi$, which comprise an on-shell $N{=}4$ hypermultiplet in three dimensions.  The transformations in \eqref{eq:susy4m} are gauged with respect to the off-shell $N{=}4$ vector supermultiplet described above.  It is straightforward to check that the $N{=}4$ supersymmetry transformations \eqref{eq:susy4m} combined with \eqref{eq:susy4g} close precisely up to the equation of motion
\begin{equation}\label{eq:susy4mpsieom}
   \gamma^\mu D_\mu \Psi  -i\sigma \cdot \Psi + \chi_\CC \cdot X - \tau_\CC^* \cdot J\Psi - \zeta_\CC \cdot JX=0,
\end{equation}
for the fermionic field $\Psi$ resulting from \eqref{eq:susy-m3a}. (We are forced to work on-shell here for the matter fields to realise more than $N{=}2$ supersymmetry since in order to do so off-shell would require the use of rather elaborate harmonic or projective superspace techniques that will be unnecessary for our present analysis.) Notice that one recovers precisely the $N{=}2$ supersymmetry transformations in \eqref{eq:susy2m} for the matter fields upon setting $\eta_\CC = 0$ and imposing the equation of motion $F = - \tau_\CC^* \cdot JX$ for the auxiliary matter field $F$.  Similarly one can obtain the full $N{=}4$ transformations in \eqref{eq:susy4m} from two sets of $N{=}2$ transformations in \eqref{eq:susy2m}, one with matter fields $(X, \Psi )$ and parameter $\epsilon_\CC$ and the other with matter fields $(-JX, \Psi )$ and parameter $\eta_\CC$.

In summary, we have shown that for $\fM = W \in \Dar(\fg,\HH)$, the $N{=}1$ superpotential $\sW_{\HH}$ in \eqref{eq:superpotential3} added to the $N{=}1$ Chern--Simons-matter lagrangian $\eL_{CS} + \eL_M$ (or equivalently the F-term superpotential in \eqref{eq:nequal3ftermsuperpotential} added to the $N{=}2$ Chern--Simons-matter lagrangian $\eL_{CS}^{N{=}2} + \eL_{M}^{N{=}2}$) gives rise to precisely the $N{=}3$ Chern--Simons-matter lagrangian $\eL_{CS}^{N{=}3} + \eL_{M}^{N{=}4}$.  A proof is given in Appendix~\ref{sec:der-nequal3superpotential} that $\sW_\HH$ in \eqref{eq:superpotential3} is the \emph{unique} choice of $N{=}1$ superpotential giving rise to an on-shell lagrangian which is invariant under the $\fusp(2)$ R-symmetry that is necessary for $N{=}3$ supersymmetry.  Thus one establishes that the class of $N{=}3$ superconformal Chern--Simons-matter theories is rigid in the sense that the superpotential function $\sW_\HH$ is fixed uniquely by the requirement of $N{=}3$ supersymmetry.

It is worth stressing that, were it not for the Chern-Simons term, the quaternionic structure of $W$ would have allowed an even greater enhancement to $N{=}4$ supersymmetry here.  The obvious question is therefore whether there are special kinds of quaternionic unitary representations for which the realisation of (at least) $N{=}4$ superconformal symmetry is possible? This is indeed the case and the resulting theories will be detailed in the next section.  For $N{=}4$ supersymmetry, $W$ is found to necessarily involve a special class of quaternionic unitary representations related to anti-Lie triple systems and defined in Appendix~\ref{sec:spec-quat-unit}.

Before moving on to this however, let us first describe the on-shell form of the $N{=}3$ Chern--Simons-matter lagrangian and supersymmetry transformations.  The $N{=}3$ discussion will then be concluded with a description of how the R-symmetry is realised in the Chern--Simons-matter theory above.  This will be useful later when we come to investigate how further supersymmetry enhancement can occur via embedding this in a larger R-symmetry algebra.

\subsubsection{On-shell N{=}3 supersymmetric lagrangian}
\label{sec:on-shell-3dsusy-quaternionic}

Having already integrated out $F$ in order to obtain the matter lagrangian \eqref{eq:susy-m3a}, it remains to impose the equations of motion $\chi_\CC^* = \half \TT ( X , \Psi )$, $\zeta_\CC^* = -\half \TT ( X , J \Psi )$, $\sigma = -\tfrac{i}{4} \TT (X,X)$ and $\tau_\CC = -\tfrac{1}{4} \TT (X, J X)$ for the respective auxiliary fields $\chi_{\CC}$, $\zeta_\CC$, $D$ and $E_\CC$ in the $N{=}4$ vector supermultiplet.  Substituting these expressions into the lagrangian gives
\begin{equation}
  \label{eq:susy-lag-on-shell-quaternionic}
  \begin{aligned}[m]
    \eL_{CS}^{N{=}3} + \eL_{M}^{N{=}4} =&  -\varepsilon^{\mu\nu\rho} \left( A_\mu , \partial_\nu A_\rho + \tfrac{1}{3} [ A_\nu , A_\rho ] \right) -\half \left< D_\mu X , D^\mu X \right> - \fV_D (X) - \fV_F (X) \\
    &+ \half \left< {\bar \Psi} , \gamma^\mu D_\mu \Psi \right> - \tfrac{1}{4} ( \TT (X, {\bar \Psi} ) , \TT (\Psi , X )) -\tfrac{1}{8} ( \TT (X,X) , \TT ( {\bar \Psi} , \Psi )) \\
    &- \tfrac{1}{8} ( \TT (X, J {\bar \Psi} ) , \TT ( X , J \Psi )) -\tfrac{1}{16} ( \TT (X,JX) , \TT ( {\bar \Psi} , J \Psi )) \\
    &- \tfrac{1}{8} ( \TT (JX, {\bar \Psi} ) , \TT ( J X , \Psi )) -\tfrac{1}{16} ( \TT (JX,X) , \TT ( J {\bar \Psi} , \Psi )),
  \end{aligned}
\end{equation}
where the F-term scalar potential here is $\fV_F (X) = \tfrac{1}{32} \left< \TT (X,JX) \cdot X , \TT (X,JX) \cdot X \right>$ while $\fV_D (X)$ is just as in \eqref{eq:susy-lag-on-shell-complex}.

Invariance of \eqref{eq:susy-lag-on-shell-quaternionic} under the $N{=}3$ supersymmetry transformations \eqref{eq:susy4g} and \eqref{eq:susy4m} can be established after substituting the aforementioned equations of motion for $\chi_{\CC}$, $\zeta_\CC$, $D$, $E_\CC$ and $F$.  These supersymmetry transformations are
\begin{equation}
  \label{eq:susy3-onshell}
  \begin{aligned}[m]
    \delta X &= ( {\bar \epsilon_{\CC}}^* + {\bar \eta_{\CC}} J ) \Psi \\
    \delta \Psi &= - \gamma^\mu ( \epsilon_\CC - \eta_\CC J ) D_\mu X - \tfrac{1}{4} \TT (JX,X) \cdot ( \epsilon_\CC^* J + \eta_\CC^* ) X - \tfrac{1}{4} ( \epsilon_\CC + \eta_\CC J ) \TT(X,X) \cdot X \\
    \delta A_\mu &= \half T \left( X , ( {\bar \epsilon_{\CC}}^* - {\bar \eta_{\CC}}^* J ) \gamma_\mu \Psi \right) ,
  \end{aligned}
\end{equation}
and close up to a translation on $\RR^{1,2}$ plus a gauge transformation, using the equations of motion from \eqref{eq:susy-lag-on-shell-quaternionic}.

\subsubsection{R-symmetry}
\label{sec:nequal3rsymmetry}

Relative to a basis $\{ \be_\alpha \}$ on $\CC^2$, let us denote by $v^\alpha$ the components of a complex vector $v$ which transforms in the defining representation of $\fu(2)$ acting on $\CC^2$.  Identifying the complex conjugate with the dual of this vector, the components of the complex conjugate vector $u^*$ are written $u_\alpha$ with the index downstairs (whereby $u_\alpha v^\alpha$ is $\fu(2)$-invariant with repeated indices summed).  With respect to this basis, $\varepsilon_{\alpha\beta} = - \varepsilon_{\beta\alpha}$ denotes the component of the ${\mathfrak{sp}}(2,\CC )$-invariant holomorphic 2-form $\varepsilon = \be^1 \wedge \be^2$ on $\CC^2$.  A tensor $w$ in the adjoint representation ${\bf 3}$ of $\fusp(2) = \fu(2) \cap {\mathfrak{sp}}(2,\CC )$ can be taken to have complex components $w^{\alpha\beta} = w^{\beta\alpha}$ obeying the reality condition $w_{\alpha\beta} = \varepsilon_{\alpha\gamma} \varepsilon_{\beta\delta} w^{\gamma\delta}$.  Consequently, the tensor $w^\alpha{}_\beta := \varepsilon_{\beta\gamma} w^{\alpha\gamma}$ can be thought of as a skew-hermitian $2\times2$ matrix (in the sense that $w^\alpha{}_\beta = - w_\beta{}^\alpha$) as befits the adjoint representation of $\fusp(2) = \fsu(2)$.  A vector $v \in W$ in the defining representation of $\fu(2)$ is in the fundamental representation ${\bf 2}$ of $\fusp(2)$ if it obeys the pseudo-reality condition $J v^\alpha = \varepsilon_{\alpha\beta} v^\beta$.  Consequently $h( u^\alpha , v^\alpha ) = - \varepsilon_{\alpha\beta} \, \omega ( u^\alpha , v^\beta )$ for any $u$ and $v$ in the fundamental representation.

Let us now collect the $N{=}4$ vector supermultiplet auxiliary fields $( \chi_\CC , \zeta_\CC , \sigma , \tau_\CC , D , E_\CC )$ into the $\fusp(2)$ tensors
\begin{equation}\label{eq:nequal3quaterniondata}
\chi_{\alpha\beta} = \begin{pmatrix} \chi_\CC & \zeta_\CC^* \\ - \zeta_\CC & \chi_\CC^* \end{pmatrix}, \quad \Sigma_\alpha{}^\beta = \begin{pmatrix} i \sigma & \tau_\CC^* \\ - \tau_\CC & -i\sigma \end{pmatrix}, \quad \Delta^\alpha{}_\beta = \begin{pmatrix} iD - \half [ \tau_\CC , \tau_\CC^* ] & E_\CC^* \\ - E_\CC & -iD + \half [ \tau_\CC , \tau_\CC^* ] \end{pmatrix},
\end{equation}
and the $N{=}4$ supersymmetry parameter $\epsilon_{\alpha\beta} = \begin{pmatrix} \epsilon_\CC & \eta_\CC^* \\ - \eta_\CC & \epsilon_\CC^* \end{pmatrix}$ which obeys $\epsilon_{\alpha\beta} = \varepsilon_{\alpha\gamma} \varepsilon_{\beta\delta} \, \epsilon^{\gamma \delta}$ with $\epsilon^{\alpha\beta}$ defined as the complex conjugate of $\epsilon_{\alpha\beta}$.  Since they are skew-hermitian, $\Sigma$ and $\Delta$ are taken to transform in the ${\bf 3}$ of $\fusp(2)$ while the fermionic matrices $\chi$ and $\epsilon$ transform in the reducible ${\bf 2} \otimes {\bf 2} = {\bf 3} \oplus {\bf 1}$ representation.  Of course, being an $N{=}4$ supermultiplet, one can more naturally denote the matrices above as $\chi_{\alpha{\dot \beta}}$, $\Sigma_{\dot \alpha}{}^{\dot \beta}$, $\Delta^{\alpha}{}_{\beta}$ and $\epsilon_{\alpha{\dot \beta}}$ in the appropriate representations $({\bf 2},{\bf 2})$, $({\bf 1},{\bf 3})$, $({\bf 3},{\bf 1})$ and $({\bf 2},{\bf 2})$ of the $N{=}4$ R-symmetry algebra $\fso(4) = \fsu(2) \oplus \fsu(2)$ (with indices $\alpha$ and ${\dot \alpha}$ denoting the fundamental representations of the left and right $\fsu(2) = \fusp(2)$ factors in $\fso(4)$).  The $N{=}3$ structure is then recovered by embedding the R-symmetry $\fusp(2)$ as the diagonal subalgebra of $\fso(4)$.  Notice that this selects precisely the one supersymmetry parameter ${\mathrm{Re}} \, \eta_\CC$ which could not preserve the supersymmetric Chern--Simons term in \eqref{eq:susy-cs3} to be the singlet in the decomposition ${\bf 2} \otimes {\bf 2} = {\bf 3} \oplus {\bf 1}$ for the $N{=}4$ parameter $\epsilon_{\alpha {\dot \beta}}$.

In this notation, the $N{=}4$ supersymmetry transformations in \eqref{eq:susy4g} can be written more succinctly as
\begin{equation}
  \label{eq:susy4g1}
  \begin{aligned}[m]
    \delta A_\mu &= \half {\mathrm{Re}} \left( {\bar \epsilon^{\alpha {\dot \beta}}} \gamma_\mu \chi_{\alpha {\dot \beta}} \right) \\
    \delta \chi_{\alpha {\dot \beta}} &= \half F_{\mu\nu} \gamma^{\mu\nu} \epsilon_{\alpha {\dot \beta}} - ( \gamma^\mu \epsilon_{\alpha {\dot \gamma}} ) D_\mu \Sigma^{\dot \gamma}{}_{\dot \beta} + \half \epsilon_{\alpha {\dot \gamma}} [ \Sigma^{\dot \gamma}{}_{\dot \delta} , \Sigma^{\dot \delta}{}_{\dot \beta} ] + \Delta_\alpha{}^\gamma \epsilon_{\gamma {\dot \beta}} \\
    \delta \Sigma^{\dot \alpha}{}_{\dot \beta} &= {\bar \epsilon^{\gamma {\dot \alpha}}} \chi_{\gamma {\dot \beta}} - \half \delta^{\dot \alpha}_{\dot \beta} \, {\bar \epsilon^{\gamma {\dot \delta}}} \chi_{\gamma {\dot \delta}} \\
    \delta \Delta_\alpha{}^\beta &= {\bar \epsilon_{\alpha {\dot \gamma}}} \left( \delta^{\dot \gamma}_{\dot \delta} \, \gamma^\mu D_\mu + \Sigma^{\dot \gamma}{}_{\dot \delta} \right) \chi^{\beta {\dot \delta}} - \half \delta_\alpha^\beta \; {\bar \epsilon_{\gamma {\dot \gamma}}} \left( \delta^{\dot \gamma}_{\dot \delta} \, \gamma^\mu D_\mu + \Sigma^{\dot \gamma}{}_{\dot \delta} \right) \chi^{\gamma {\dot \delta}}.
  \end{aligned}
\end{equation}

The $N{=}3$ supersymmetric Chern--Simons term \eqref{eq:susy-cs3} is
\begin{equation}\label{eq:susy-cs3a}
   \eL_{CS}^{N{=}3} = -\varepsilon^{\mu\nu\rho} \left( A_\mu , \partial_\nu A_\rho + \tfrac{1}{3} [ A_\nu , A_\rho ] \right) +  (\Sigma_\alpha{}^\beta , \Delta_\beta{}^\alpha ) - \tfrac{1}{6} ( \Sigma_\alpha{}^\beta , [ \Sigma_\beta{}^\gamma , \Sigma_\gamma{}^\alpha ] ) - \half ( {\bar \chi^{\beta\alpha}} , \chi_{\alpha\beta} ).
\end{equation}
It is worth emphasising that the $- \tfrac{1}{6} \Sigma^3$ term is purely to account for the fact the diagonal elements of $\Delta$ in \eqref{eq:nequal3quaterniondata} involve the shifted auxiliary field $D + \tfrac{i}{2} [ \tau_\CC , \tau_\CC^* ]$ rather than just $D$.  The reason that this shifted definition is useful is that it allows one to obtain from \eqref{eq:susy4g1} precisely the $N{=}2$ supersymmetry structure with parameter $\epsilon_\CC$ we have found already by simply setting the other parameter $\eta_\CC =0$.  Thus, just as in \eqref{eq:susy-cs3}, there are no cubic terms involving any of the auxiliary fields in \eqref{eq:susy-cs3a}.  Notice that \eqref{eq:susy-cs3a} cannot be expressed invariantly in terms of the vector supermultiplet fields in the representations of $\fso(4)$ described above.  This is another signal of only $N{=}3$ supersymmetry for the Chern--Simons term.

The $N{=}4$ hypermultiplet matter fields $X$ and $\Psi$ can be assembled into the vectors
\begin{equation}\label{eq:nequal3quaterniondata2}
 X^\alpha = \begin{pmatrix} X \\ JX \end{pmatrix}, \quad \Psi_{\dot \alpha} = \begin{pmatrix} \Psi \\ J\Psi \end{pmatrix},
\end{equation}
corresponding respectively to the representations $({\bf 2},{\bf 1})$ and $({\bf 1},{\bf 2})$ of $\fso(4)$.  Thus we satisfy the pseudo-reality conditions $J X^\alpha = \varepsilon_{\alpha\beta} X^\beta$ and $J \Psi_{\dot \alpha} = \varepsilon^{{\dot \alpha}{\dot \beta}} \Psi_{\dot \beta}$ identically.  They can both be thought of as inhabiting the graph of $J$ in $W \oplus W$.

The on-shell $N{=}4$ supersymmetry transformations \eqref{eq:susy4m} for \eqref{eq:nequal3quaterniondata2} can now be more compactly expressed as
\begin{equation}
  \label{eq:susy4mH}
  \begin{aligned}[m]
    \delta X^\alpha &= {\bar \epsilon^{\alpha{\dot \beta}}} \Psi_{\dot \beta} \\
    \delta \Psi_{\dot \alpha} &= - ( \gamma^\mu \epsilon_{\beta {\dot \alpha}} ) D_\mu X^\beta - \left( \Sigma_{\dot \alpha}{}^{\dot \beta} \cdot X^\gamma \right) \epsilon_{\gamma {\dot \beta}},
  \end{aligned}
\end{equation}
and the $N{=}4$ supersymmetric matter lagrangian \eqref{eq:susy-m3a} becomes
\begin{equation}\label{eq:susy-m3b}
  \begin{aligned}[m]
  \eL_{M}^{N{=}4} =& -\tfrac{1}{4} \left< D_\mu X^\alpha , D^\mu X^\alpha \right> + \tfrac{1}{4} \left< X^\alpha , \Delta^\alpha{}_\beta \cdot X^\beta \right> -\tfrac{1}{8} \left< \Sigma_{\dot \alpha}{}^{\dot \beta} \cdot X^\gamma , \Sigma_{\dot \alpha}{}^{\dot \beta} \cdot X^\gamma \right>  \\
  &+ \tfrac{1}{4} \left< {\bar \Psi_{\dot \alpha}} , \gamma^\mu D_\mu \Psi_{\dot \alpha} \right>  - \tfrac{1}{4} \left< {\bar \Psi_{\dot \alpha}} , \Sigma_{\dot \alpha}{}^{\dot \beta} \cdot \Psi_{\dot \beta} \right> + \half \left< {\bar \Psi_{\dot \alpha}} , \chi_{\beta{\dot \alpha}} \cdot X^\beta \right>.
  \end{aligned}
\end{equation}
Notice that, for example, $\left< u^\alpha , v^\alpha \right> = 2 \left< u , v \right> = - \varepsilon_{\alpha\beta} \, {\mathrm{Re}} \, \omega ( u^\alpha , v^\beta )$ for any $u,v \in W$, with $u^\alpha = ( u , Ju)$ and $v^\alpha = ( v , Jv)$, and the contraction of $\alpha$ indices here is $\fusp(2)$-invariant as a consequence of $h(-,-)$ being complex sesquilinear.  The same applies for contracted ${\dot \alpha}$ indices with respect to the other $\fusp(2)$ factor in $\fso(4)$.  Thus the matter lagrangian \eqref{eq:susy-m3b} is manifestly $\fso(4)$-invariant as befits the fact that it is $N{=}4$ supersymmetric.

The lack of off-shell $N{=}4$ supersymmetry and $\fso(4)$-invariance for the Chern--Simons term propagates into the form of the equations of motion for some of the auxiliary fields.  In particular, the equations of motion for $\chi_{\CC}$, $\zeta_\CC$, $D$ and $E_\CC$ collect into the following $\fusp(2)$ representations
\begin{equation}\label{eq:nequal3quaterniondata3}
\chi_{\alpha\beta} = -\half \, \TT ( \Psi_\alpha , X^\beta ), \quad \Sigma_\alpha{}^\beta = \tfrac{1}{4} \, \TT ( X^\beta , X^\alpha ),
\end{equation}
with the indices matching as a consequence of $\TT$ being a complex sesquilinear map.  These $\fusp(2)$-invariant equations are clearly not $\fso(4)$-invariant since they would not make sense after sprinkling dots commensurate with the $\fso(4)$ representations that the fields were declared to inhabit above.  Notice though that the first equation in \eqref{eq:nequal3quaterniondata3} would have been $\fso(4)$-invariant, with $\chi_{\alpha {\dot \beta}} = -\half \, \TT ( \Psi_{\dot \beta} , X^\alpha )$, if it had been the transposed vector appearing on the right hand side.  This seemingly innocuous statement will turn out to be a key feature of realising $N{=}4$ supersymmetry to be described in the next section.

Let us now close by noting the manifestly $\fusp(2)$-invariant form of the on-shell $N{=}3$ lagrangian \eqref{eq:susy-lag-on-shell-quaternionic} given by
\begin{equation}
  \label{eq:susy-lag-on-shell-quaternionic2}
  \begin{aligned}[m]
    \eL^{N{=}3} =&  -\varepsilon^{\mu\nu\rho} \left( A_\mu , \partial_\nu A_\rho + \tfrac{1}{3} [ A_\nu , A_\rho ] \right) -\tfrac{1}{4} \left< D_\mu X^\alpha , D^\mu X^\alpha \right> + \tfrac{1}{4} \left< {\bar \Psi}_\alpha , \gamma^\mu D_\mu \Psi_\alpha \right>  \\
    &- \tfrac{1}{8} ( \TT ( X^\alpha , {\bar \Psi}_\beta ) , \TT ( \Psi_\alpha , X^\beta )) -\tfrac{1}{16} ( \TT ( X^\alpha , X^\beta ) , \TT ( {\bar \Psi}_\alpha , \Psi_\beta )) \\
    &+\tfrac{1}{384} \left( [ \TT ( X^\alpha , X^\beta ) , \TT ( X^\beta , X^\gamma ) ] , \TT ( X^\gamma , X^\alpha ) \right) - \tfrac{1}{128} \left< \TT ( X^\alpha , X^\beta ) \cdot X^\gamma , \TT ( X^\alpha , X^\beta ) \cdot X^\gamma \right>,
  \end{aligned}
\end{equation}
and the $N{=}3$ supersymmetry transformations in \eqref{eq:susy3-onshell} which become 
\begin{equation}
  \label{eq:susy3-onshell2}
  \begin{aligned}[m]
    \delta X^\alpha &= {\bar \epsilon}^{\alpha\beta} \Psi_\beta \\
    \delta \Psi_{\alpha} &= - ( \gamma^\mu \epsilon_{\alpha\beta} ) D_\mu X^\beta - \tfrac{1}{4} \TT ( X^\beta , X^\alpha ) \cdot X^\gamma \, \epsilon_{\beta\gamma} \\
    \delta A_\mu &= \tfrac{1}{4} T \left( X^\alpha , {\bar \epsilon}^{\alpha\beta} \gamma_\mu \Psi_\beta \right) .
  \end{aligned}
\end{equation}

\section{N$>$3 superconformal Chern--Simons-matter theories}
\label{sec:scft-ternary-3plus}

Having described the structure of $N \leq 3$ superconformal Chern--Simons-matter theories with matter in unitary representations of generic type, we are now in a position to examine in more detail the special kinds of matter representations which guarantee further supersymmetry enhancement.

\subsection{N{=}4 supersymmetry}
\label{sec:nequal4}

We shall begin by considering matter representations $\fM = W \in \Dar(\fg,\HH)_{\text{aLTS}}$ in the notation of Appendix \ref{sec:spec-quat-unit}.  This class of representation was shown by Gaiotto and Witten in \cite{GaiottoWitten} to give rise to an enhanced $N{=}4$ superconformal symmetry.  By substituting this data in the $N{=}3$ theory based on the superpotential $\sW_\HH$ in \eqref{eq:superpotential3} described in the previous section, we recover the known examples of $N{=}4$ superconformal Chern--Simons-matter theories.  When $W \in \Irr(\fg,\HH)_{\text{aLTS}}$, the resulting class of indecomposable $N{=}4$ theories will be shown in Section~\ref{sec:antiLTS-GW} to coincide with those found by Gaiotto and Witten in \cite{GaiottoWitten}.  When $W = W_1 \oplus W_2$ with $W_1,W_2 \in\Dar(\fg,\HH)_{\text{aLTS}}$ (but $W$ not necessarily of aLTS class), it will be shown in Section~\ref{sec:twisthyp} that the resulting class of $N{=}4$ theories coincide with those found by Hosomichi, Lee, Lee, Lee and Park in \cite{pre3Lee} by coupling to so-called twisted hypermultiplets.  We will see how this special type of coupling, which exists only for $N=4$ theories, is really what distinguishes the theories of \cite{GaiottoWitten} and \cite{pre3Lee} and, contrary to what the notation above might suggest, it is not merely a question of whether the representations are irreducible or not.  For instance, in terms of the notation in Section~\ref{sec:matter-reps}, we will find that the bosonic matter in the $N=4$ theories of \cite{GaiottoWitten} and \cite{pre3Lee} transform under (the appropriate real form of) the representations $\Delta^{(4)}_+ \otimes W$ and $\Delta^{(4)}_+ \otimes W_1 \oplus \Delta^{(4)}_- \otimes W_2$ respectively of $\fso(4) \oplus \fg$ (i.e. only for the theories of \cite{pre3Lee} do both $\fso(4)$ chiralities $\Delta^{(4)}_\pm$ appear).

\subsubsection{$W \in \Dar(\fg,\HH)_{\text{aLTS}}$}
\label{sec:antiLTS-GW}

Such representations $W$ are characterised by the existence of a Lie superalgebra structure on $\fg_\CC \oplus W$ --- a fact which was first appreciated and utilised in the context of $N{=}4$ superconformal Chern--Simons-matter theories in \cite{GaiottoWitten}.  Recall that a similar (but distinct) embedding Lie superalgebra structure on $V \oplus \fg_\CC \oplus \Vbar$ characterises a complex unitary $\fg$-module $V$ being in the aJTS class.  However, this extreme case is too severe to have merited consideration in Section~\ref{sec:nequal2} simply because it would imply that the minimal superpotential $\sW_{\CC}$ in \eqref{eq:superpotentialcomplex} (that is required to produce the $N{=}2$ gauge-matter couplings) must vanish identically.  In the quaternionic case, there is still not too much room for manoeuvre.  In particular, notice that the aLTS cyclicity condition \eqref{eq:quat-aLTS} means that $\TT (X,JX) \cdot X = 0$ identically for any $W$-valued field $X$ in this special case.  Consequently the F-term superpotential in \eqref{eq:nequal3ftermsuperpotential} vanishes identically.  Thus the superpotential $\sW_\HH$ in \eqref{eq:superpotential3} is equal to the minimal superpotential $\sW_{\CC}$ in \eqref{eq:superpotentialcomplex} for $W\in \Dar(\fg,\HH)_{\text{aLTS}}$.

Some crucial identities implied by the aLTS cyclicity condition \eqref{eq:quat-aLTS} are   
\begin{equation}
  \label{eq:antiLTSid1}
  ( T (X,IX) , T (X,IX) ) = ( T (X,JX) , T (X,JX) ) = ( T (X,IJX) , T (X,IJX) ),
\end{equation}
and
\begin{equation}
  \label{eq:antiLTSid2}
  ( T (X,IX) , T (X,JX) ) = ( T (X,IX) , T (X,IJX) ) = ( T (X,JX) , T (X,IJX) ) =0,
\end{equation}
for any $W$-valued field $X$.  Remarkably these algebraic conditions imply that the minimal superpotential $\sW_\HH = \sW_\CC$ for $W$ is precisely that which was used by Gaiotto and Witten \cite{GaiottoWitten}, thus allowing the realisation of $N{=}4$ superconformal symmetry!  Initially it will be more convenient to express this superpotential as
\begin{equation}\label{eq:superpotential4}
  \sW_{\HH} = -\tfrac{1}{16} \int d^2 \theta \; ( T (\Xi ,J \Xi ), T (\Xi ,J \Xi ) ).
\end{equation}
Writing the $W$-valued $N{=}1$ superfield $\Xi$ appearing above as $\Xi^\alpha = ( \Xi , J \Xi )$ in terms of a $\fusp(2)$-doublet then the superpotential \eqref{eq:superpotential4} can be reexpressed as
\begin{equation}\label{eq:superpotential4a}
  \sW_{\HH} = -\tfrac{1}{48} \int d^2 \theta \; \half \, \varepsilon_{\alpha\beta} \, \varepsilon_{\gamma\delta} \, ( \TT (\Xi^\alpha ,J \Xi^\gamma ), \TT (\Xi^\beta ,J \Xi^\delta ) ).
\end{equation}
Of course, on its own, this is not $\fso(4)$-invariant because the constituent matter fields $X^\alpha$ and $\Psi_{\dot \alpha}$ naturally transform in the fundamental representations of the two different $\fusp(2)$ factors of $\fso(4)$.

Since we are now beyond the realms of a wieldy off-shell supersymmetric framework, we shall proceed to describe the on-shell $N{=}4$ supersymmetric structure of the theory associated with the superpotential in \eqref{eq:superpotential4}.  The on-shell form \eqref{eq:susy-lag-on-shell} of the lagrangian $\eL_{GW}^{N{=}4} = \eL_{CS} + \eL_M + \sW_{\HH}$ is given by
\begin{equation}
\label{eq:susy-lag-csm4}
  \begin{aligned}[m]
   \eL_{GW}^{N{=}4} =& -\varepsilon^{\mu\nu\rho} \left( A_\mu , \partial_\nu A_\rho + \tfrac{1}{3} [ A_\nu , A_\rho ] \right) -\tfrac{1}{2} \left< D_\mu X , D^\mu X \right> + \tfrac{1}{2} \left< {\bar \Psi} , \gamma^\mu D_\mu \Psi \right>  \\
   &+ \tfrac{1}{8} ( T ( {\bar \Psi} , J \Psi ) , T (X,JX) ) + \tfrac{1}{4} ( T ( X ,J {\bar \Psi} ) , T (X,J \Psi ) ) + \tfrac{1}{4} ( T ( X , {\bar \Psi} ) , T ( X , \Psi ) ) \\
   &-\tfrac{1}{32} \left< T (X,JX) \cdot X , T (X,JX) \cdot X \right> .
  \end{aligned}
\end{equation}
This lagrangian can be given a manifestly $\fso(4)$-invariant expression as
\begin{equation}
\label{eq:susy-lag-csm4a}
  \begin{aligned}[m]
   \eL_{GW}^{N{=}4} =& -\varepsilon^{\mu\nu\rho} \left( A_\mu , \partial_\nu A_\rho + \tfrac{1}{3} [ A_\nu , A_\rho ] \right) -\tfrac{1}{4} \left< D_\mu X^\alpha , D^\mu X^\alpha \right> + \tfrac{1}{4} \left< {\bar \Psi}_{\dot \alpha} , \gamma^\mu D_\mu \Psi_{\dot \alpha} \right>  \\
   &+ \tfrac{1}{16} \, \varepsilon_{\alpha\beta} \, \varepsilon^{{\dot \gamma}{\dot \delta}} \, ( \TT ( X^\alpha , J {\bar \Psi}_{\dot \gamma} ) , \TT ( X^\beta , J \Psi_{\dot \delta} ) ) \\
   &-\tfrac{1}{768} \, ( [ \TT( X^\alpha , X^\beta ) , \TT( X^\beta , X^\gamma ) ] , \TT( X^\gamma , X^\alpha ) ),
  \end{aligned}
\end{equation}
which, mutatis mutandis, is indeed the $N{=}4$ Gaiotto--Witten lagrangian in \cite{GaiottoWitten}.  It is worth stressing that the precise value of the coefficient in the superpotential \eqref{eq:superpotential4} is what has allowed the various Yukawa couplings to assemble themselves in an $\fso(4)$-invariant manner in the second line above and has fixed the overall coefficient for this term.

The on-shell $N{=}4$ supersymmetry transformations, under which the integral of the lagrangian $\eL_{GW}^{N{=}4}$ is invariant, are given by
\begin{equation}
  \label{eq:susy4-onshell}
  \begin{aligned}[m]
    \delta X &= ( {\bar \epsilon_{\CC}}^* + {\bar \eta_{\CC}} J ) \Psi \\
    \delta \Psi &= - \gamma^\mu ( \epsilon_\CC - \eta_\CC J ) D_\mu X - \tfrac{1}{4} ( \epsilon_\CC - \eta_\CC J ) \TT(X,X) \cdot X \\
    \delta A_\mu &= \half T \left( X , ( {\bar \epsilon_{\CC}}^* + {\bar \eta_{\CC}} J ) \gamma_\mu \Psi \right) ,
  \end{aligned}
\end{equation}
and close up to a translation on $\RR^{1,2}$ plus a gauge transformation, using the equations of motion from \eqref{eq:susy-lag-csm4}.  Their $\fso(4)$-covariant expressions are
\begin{equation}
  \label{eq:susy4-onshella}
  \begin{aligned}[m]
    \delta X^\alpha &= {\bar \epsilon}^{\alpha {\dot \beta}} \Psi_{\dot \beta} \\
    \delta \Psi_{\dot \alpha} &= - \left[ \gamma^\mu D_\mu X^\beta + \tfrac{1}{12} \, \TT( X^\beta , X^\gamma ) \cdot X^\gamma \right] \epsilon_{\beta {\dot \alpha}} \\
    \delta A_\mu &= \tfrac{1}{4} \TT \left( X^\alpha , {\bar \epsilon^{\alpha {\dot \beta}}} \gamma_\mu \Psi_{\dot \beta} \right) .
  \end{aligned}
\end{equation}

Comparing first the on-shell $N{=}4$ transformations in \eqref{eq:susy4-onshell} with the $N{=}3$ ones in \eqref{eq:susy3-onshell}, we note the following differences.  First, the second term in the transformation of $\Psi$ in \eqref{eq:susy3-onshell} (that arose from imposing the equation of motion $\tau_\CC = - \tfrac{1}{4} \TT (X,JX)$) is absent in \eqref{eq:susy4-onshell} which is consistent with the fact that we have no F-term superpotential here.  Second, the sign of the parameter $\eta_\CC$ has changed in the second term in the variation of $\Psi$ in \eqref{eq:susy4-onshell} relative to \eqref{eq:susy3-onshell}.  Finally, in the on-shell transformation of $A_\mu$ above we have effectively replaced the parameter $\eta_\CC^*$ in \eqref{eq:susy3-onshell} with $- \eta_\CC$ in \eqref{eq:susy4-onshell}.  These changes are all necessary for the realisation of $N{=}4$ supersymmetry on-shell and the final change has a natural interpretation based on the remark at the end of the penultimate paragraph in Section~\ref{sec:nequal3rsymmetry}.  Namely, it is precisely the on-shell $N{=}4$ transformation for $A_\mu$ in \eqref{eq:susy4-onshella} that would have resulted from substituting the $\fso(4)$-covariant equation $\chi_{\alpha {\dot \beta}} = -\half \, \TT ( \Psi_{\dot \beta} , X^\alpha )$ for the auxiliary field into the first line of \eqref{eq:susy4g1}.  Moreover, this equation of motion could be obtained by replacing the $\fusp(2)$-invariant expression $- \half ( {\bar \chi^{\beta \alpha}} , \chi_{\alpha \beta} )$ for the auxiliary fermions in the off-shell $N{=}3$ Chern--Simons lagrangian \eqref{eq:susy-cs3a} with the $\fso(4)$-invariant term $- \half ( {\bar \chi^{\alpha {\dot \beta}}} , \chi_{\alpha {\dot \beta}} )$.  One way to think of this change is from Wick rotating one of the four auxiliary fermions changing the ${\mathfrak{so}}(1,3)$-invariant inner product in the $N{=}3$ case to an $\fso(4)$-invariant one in the $N{=}4$ case.  We should stress though that is just a formal conceit in that there still exists no off-shell formulation of the $N{=}4$ theory above that does not involve more elaborate harmonic or projective superspace methods.

Before moving on, it will be convenient to introduce some notation that will allow us to encapsulate the structure above more compactly.  We will define the tensor $\mu^\alpha{}_\beta := \tfrac{1}{4} \, \TT ( X^\alpha , X^\beta )$ which transforms in the adjoint of one of the $\fusp(2)$ factors of $\fso(4)$.  This tensor is precisely the moment map associated with the action of $\fg$ on the flat hyperkähler manifold $W$.  It will also be useful define a superpartner of sorts for the hyperkähler moment map to be $\nu^{\alpha {\dot \beta}} := \tfrac{1}{4} \TT ( X^\alpha , \Psi_{\dot \beta} )$ (whose complex conjugate is $\nu_{\alpha {\dot \beta}} = - \tfrac{1}{4} \TT ( \Psi_{\dot \beta} , X^\alpha )$).  These two expressions can be collected into a useful $N{=}1$ superfield moment map ${\mathscr{M}}^\alpha{}_\beta := \tfrac{1}{4} \TT ( \Xi^\alpha , \Xi^\beta )$, albeit only $\fusp(2)$-covariant.  In terms of these tensors, the $\fusp(2)$-invariant form of the superpotential in \eqref{eq:superpotential4a} is simply \begin{equation}\label{eq:superpotential4b}
  \sW_{\HH} = \tfrac{1}{6} \int d^2 \theta \; ( {\mathscr{M}}^\alpha{}_\beta , {\mathscr{M}}^\beta{}_\alpha ),
\end{equation}
leading to the Gaiotto--Witten lagrangian \eqref{eq:susy-lag-csm4a} given by
\begin{equation}
\label{eq:susy-lag-csm4b}
  \begin{aligned}[m]
   \eL_{GW}^{N{=}4} =& -\varepsilon^{\mu\nu\rho} \left( A_\mu , \partial_\nu A_\rho + \tfrac{1}{3} [ A_\nu , A_\rho ] \right) -\tfrac{1}{4} \left< D_\mu X^\alpha , D^\mu X^\alpha \right> + \tfrac{1}{4} \left< {\bar \Psi}_{\dot \alpha} , \gamma^\mu D_\mu \Psi_{\dot \alpha} \right>  \\
   &+ ( {\bar \nu^{\alpha {\dot \beta}}} , \nu_{\alpha {\dot \beta}} ) -\tfrac{1}{12} \, ( [ \mu^\alpha{}_\beta , \mu^\beta{}_\gamma ] , \mu^\gamma{}_\alpha ),
  \end{aligned}
\end{equation}
and the $N{=}4$ supersymmetry transformations \eqref{eq:susy4-onshella} are
\begin{equation}
  \label{eq:susy4-onshellb}
  \begin{aligned}[m]
    \delta X^\alpha &= {\bar \epsilon}^{\alpha {\dot \beta}} \Psi_{\dot \beta} \\
    \delta \Psi_{\dot \alpha} &= - \left[ \gamma^\mu D_\mu X^\beta + \tfrac{1}{3} \, \mu^\beta{}_\gamma \cdot X^\gamma \right] \epsilon_{\beta {\dot \alpha}} \\
    \delta A_\mu &= {\bar \epsilon_{\alpha {\dot \beta}}} \gamma_\mu \nu^{\alpha {\dot \beta}}.
  \end{aligned}
\end{equation}

\subsubsection{$W = W_1 \oplus W_2$ with $W_1,W_2 \in \Dar(\fg,\HH)_{\text{aLTS}}$}
\label{sec:twisthyp}

The quaternionic hermitian structure on $W$ is defined in the obvious way such that $(W,h ,J) = ( W_1 \oplus W_2 , h_1 \oplus h_2 , J_1 \oplus J_2 )$ in terms of the corresponding structures on $W_1$ and $W_2$.  The action of $\fg$ on $W$ is defined by $X \cdot ( v_1 , v_2 ) = ( X \cdot v_1 , X \cdot v_2 )$ for any $X \in \fg$.  Consequently the map $\TT = \TT_1 \oplus \TT_2$ decomposes orthogonally in terms of its restrictions to $W_1$ and $W_2$; in other words, $\TT(w_1,w_2) = 0$ for all $w_1\in W_1$ and $w_2 \in W_2$.

Demanding that $W\in\Dar(\fg,\HH)_{\text{aLTS}}$ (and faithful) is too strong and the resulting theory decouples into a Gaiotto--Witten theory for $W_1$ and another for $W_2$.  Indeed, consider the aLTS cyclicity condition \eqref{eq:quat-aLTS}
\begin{equation}
  \TT (u,Jv) \cdot w + \TT (v,Jw) \cdot u + \TT (w,Ju) \cdot v =0
\end{equation}
for all $u,v,w \in W$.  Decomposing this equation on $W_1 \oplus W_2$ shows that it is identically satisfied on the individual components $W_1$ and $W_2$, since they are both aLTS.  However, the contributions from the mixed components imply that $\TT_1 ( u_1 , v_1 ) \cdot w_2 = 0$ and $\TT_2 ( u_2 , v_2 ) \cdot w_1 = 0$, for all $u_1 , v_1 , w_1 \in W_1$ and $u_2 , v_2 , w_2 \in W_2$.  Let $\fg_1 = \TT(W_1,W_1)$ and $\fg_2=\TT(W_2,W_2)$.  Then $\fg_1$ acts trivially on $W_2$ and $\fg_2$ acts trivially on $W_1$.  Since $W$ is faithful, $\fg_1 \cap \fg_2 = 0$ and since $\TT(W_1,W_2)=0$, we see that $\fg_\CC = \fg_1 \oplus \fg_2$.  Furthermore the direct sum is orthogonal with respect to the inner product $(-,-)$.  Consequently the superconformal Chern--Simons-matter theories based on such data would effectively decouple in terms of distinct Gaiotto--Witten $N{=}4$ theories on $W_1$ and $W_2$.  We will therefore exclude the possibility that $W$ itself be aLTS.

Let us now examine the superpotential $\sW_\HH$ in \eqref{eq:superpotential3} for an $N{=}1$ matter superfield $\Xi$ valued in $W = W_1 \oplus W_2$.  Of course, this is no longer identical to the expression in \eqref{eq:superpotential4} when $W$ is not aLTS.  That said, if one decomposes $\sW_\HH$ on $W$ into its component parts on $W_1$ and $W_2$ one obtains a sum of three distinct contributions.  Two of these are simply the decoupled superpotentials on the individual components $W_1$ and $W_2$.  Since we have assumed that both these components are aLTS then clearly these contributions do each agree with the expression \eqref{eq:superpotential4} for the Gaiotto--Witten superpotentials associated with $W_1$ and $W_2$ individually.  In addition one has a contribution from the mixture of components on $W_1$ and $W_2$.  It is this mixed term which can be thought of as providing a non-trivial F-term superpotential contribution in the resulting theory (which is of course absent from the individual $W_1$ and $W_2$ contributions).

The next question is whether this theory can realise $N{=}4$ supersymmetry.  Recall that the superpotential $\sW_\HH$ in \eqref{eq:superpotential3} gives rise to an on-shell $N{=}3$ superconformal theory with manifest $\fusp(2)$ R-symmetry.  However, on its own, $\sW_\HH$ is generically only invariant under the $\fu(1)$ R-symmetry subalgebra arising from the the $N{=}2$ supersymmetric framework from whence it came.  A crucial indicator of the enhanced $N{=}4$ supersymmetry in the Gaiotto--Witten theory is that its superpotential $\sW_\HH$ enjoys the larger global symmetry $\fusp(2) > \fu(1)$ (as was made explicit in \eqref{eq:superpotential4a} and \eqref{eq:superpotential4b}).  The reason for this is simple.  To see why, assume more generally that there exists an on-shell $N{=}n$ superconformal Chern--Simons-matter theory with $\fso(n)$ R-symmetry that one is trying to obtain from an off-shell $N{=}1$ superspace formalism.  Relative to the resulting $N{=}n$ theory, clearly ones choice of $N{=}1$ superspace breaks the $\fso(n)$ R-symmetry down to the $\fso(n-1)$ subalgebra preserving the chosen $N{=}1$ superspace parameter.  Hence, assuming that there exists an off-shell $N{=}1$ superpotential that gives rise to this $N{=}n$ superconformal theory, it must be invariant under precisely this $\fso(n-1)$ subalgebra.  This is exactly what has been described above in the Gaiotto--Witten theory when $n=4$.  The realisation of this isotropy subalgebra generating a symmetry of the $N{=}1$ superpotential will be a useful guiding principle for us in our search here and in later sections for theories with enhanced superconformal symmetry.

With this in mind we note that if one expresses the $N{=}1$ matter superfield $\Xi = ( \Xi_1 , \Xi_2 )$ valued in $W = W_1 \oplus W_2$ as a $\fusp(2)$-doublet in the obvious way, such that $\Xi_1^\alpha = ( \Xi_1 , J_1 \Xi_1 )$ and $\Xi_2^\alpha = ( \Xi_2 , J_2 \Xi_2 )$, then the superpotential $\sW_\HH$ is not $\fusp(2)$-invariant.  The trick is to instead write the $N{=}1$ matter superfield as $\Xi = ( \Xi_1 , J_2 \Xi_2 )$ when evaluating $\sW_\HH$ on $W = W_1 \oplus W_2$.  Of course, this may seem rather trivial since it is nothing but a field redefinition of the $W_2$ superfield component.  However, this seemingly innocuous modification allows the superpotential to be expressed in a manifestly $\fusp(2)$-invariant way as
\begin{equation}\label{eq:superpotential4t}
  \sW_{\HH} \mid_{W_1 \oplus W_2} = \sW_{\HH} \mid_{W_1} + \sW_{\HH} \mid_{W_2} + \sW_{\text{\tiny{mixed}}},
\end{equation}
where
\begin{equation}\label{eq:superpotential4ta}
  \sW_{\HH} \mid_{W_1} = -\tfrac{1}{16} \int d^2 \theta \; ( T_1 (\Xi_1 , J_1 \Xi_1 ), T_1 (\Xi_1 , J_1 \Xi_1 ) ) = \tfrac{1}{6} \int d^2 \theta \; ( ({\mathscr{M}}_1)^\alpha{}_\beta , ({\mathscr{M}}_1)^\beta{}_\alpha ),
\end{equation}
\begin{equation}\label{eq:superpotential4tb}
  \sW_{\HH} \mid_{W_2} = -\tfrac{1}{16} \int d^2 \theta \; ( T_2 (\Xi_2 , J_2 \Xi_2 ), T_2 (\Xi_2 , J_2 \Xi_2 ) ) = \tfrac{1}{6} \int d^2 \theta \; ( ({\mathscr{M}}_2)^\alpha{}_\beta , ({\mathscr{M}}_2)^\beta{}_\alpha ),
\end{equation}
and
\begin{equation}
\label{eq:superpotential4tc}
\begin{aligned}[m]
  \sW_{\text{\tiny{mixed}}} &= - \tfrac{1}{16} \int d^2 \theta \left[ 2\, ( \TT_1 (\Xi_1 ,\Xi_1 ), \TT_2 (\Xi_2 ,\Xi_2 ) ) \right. \\
  &\hspace*{1in} \left. + ( \TT_1 (\Xi_1 , J_1 \Xi_1 ), \TT_2 (J_2 \Xi_2 ,\Xi_2 ) ) + ( \TT_1 (J_1 \Xi_1 ,\Xi_1 ) , \TT_2 (\Xi_2 , J_2 \Xi_2 ) )  \right] \\
  &= - \int d^2 \theta \; ( ({\mathscr{M}}_1)^\alpha{}_\beta , ({\mathscr{M}}_2)^\beta{}_\alpha ),
\end{aligned}
\end{equation}
with the $\fusp(2)$-doublet superfields $\Xi_1^\alpha = ( \Xi_1 , J_1 \Xi_1 )$, $\Xi_2^\alpha = ( \Xi_2 , J_2 \Xi_2 )$ as before and $({\mathscr{M}}_1)^\alpha{}_\beta = \tfrac{1}{4} \TT_1 ( \Xi_1^\alpha , \Xi_1^\beta )$, $({\mathscr{M}}_2)^\alpha{}_\beta = \tfrac{1}{4} \TT_2 ( \Xi_2^\alpha , \Xi_2^\beta )$.  The unmixed term $\sW_{\HH} \mid_{W_1}$ is insensitive to the distinction between choosing the superfield to be $\Xi_1$ or $J_1 \Xi_1$ on $W_1$ (since $T_1 ( J_1 \Xi_1 , J_1^2 \Xi_1 ) = - T_1 ( J_1 \Xi_1 , \Xi_1 ) =  T_1 ( \Xi_1 , J_1 \Xi_1 )$) and likewise for the unmixed $W_2$ contribution.  By the same reasoning, the mixed term $\sW_{\text{\tiny{mixed}}}$ would be the same were we to choose either $( \Xi_1 , J_2 \Xi_2 )$ or $J( \Xi_1 , J_2 \Xi_2 ) = ( J_1 \Xi_1 , - \Xi_2 )$ as the matter superfield.  The significant point here is that there is a relative factor of $J$ between the $W_1$ and $W_2$ superfield contributions.  This is necessary so that the terms in the second line of \eqref{eq:superpotential4tc} take the form $z_1 z_2^* + z_1^* z_2$, rather than the generic form $z_1 z_2 + z_1^* z_2^*$ predicated on \eqref{eq:superpotential3}, thereby combining with the first term in a way that is invariant under $\fusp(2) = \fso(3)$, under which the superfield moment maps ${\mathscr{M}}_1$ and ${\mathscr{M}}_2$ both transform in the adjoint.

The on-shell form \eqref{eq:susy-lag-on-shell} of the lagrangian $\eL_{H3LP}^{N{=}4} = \eL_{CS} + \eL_M + \sW_{\HH} \mid_{W_1 \oplus W_2}$ based on the superpotential \eqref{eq:superpotential4t} can, after some manipulations, be expressed as
\begin{equation}
\label{eq:susy-lag-csm4-hlllp}
  \begin{aligned}[m]
    &-\varepsilon^{\mu\nu\rho} \left( A_\mu , \partial_\nu A_\rho + \tfrac{1}{3} [ A_\nu , A_\rho ] \right) -\tfrac{1}{2} \left< D_\mu X_1 , D^\mu X_1 \right>_1 -\tfrac{1}{2} \left< D_\mu X_2 , D^\mu X_2 \right>_2 \\
   &+ \tfrac{1}{2} \left< {\bar \Psi}_1 , \gamma^\mu D_\mu \Psi_1 \right>_1 + \tfrac{1}{2} \left< {\bar \Psi}_2 , \gamma^\mu D_\mu \Psi_2 \right>_2 \\
   &+ \tfrac{1}{8} ( T_1 ( {\bar \Psi}_1 , J_1 \Psi_1 ) , T_1 ( X_1 , J_1 X_1 ) ) + \tfrac{1}{4} ( T_1 ( X_1 , J_1 {\bar \Psi}_1 ) , T_1 ( X_1 , J_1 \Psi_1 ) ) + \tfrac{1}{4} ( T_1 ( X_1 , {\bar \Psi}_1 ) , T_1 ( X_1 , \Psi_1 ) ) \\
   &+ \tfrac{1}{8} ( T_2 ( {\bar \Psi}_2 , J_2 \Psi_2 ) , T_2 ( X_2 , J_2 X_2 ) ) + \tfrac{1}{4} ( T_2 ( X_2 , J_2 {\bar \Psi}_2 ) , T_2 ( X_2 , J_2 \Psi_2 ) ) + \tfrac{1}{4} ( T_2 ( X_2 , {\bar \Psi}_2 ) , T_2 ( X_2 , \Psi_2 ) )  \\
   &+ \tfrac{1}{8} \left( ( \TT_1 ( {\bar \Psi}_1 , \Psi_1 ) , \TT_2 ( X_2 , X_2 ) ) + \Re \, ( \TT_1 ( {\bar \Psi}_1 , J_1 \Psi_1 ) , \TT_2 ( J_2 X_2 , X_2 ) ) \right) \\
   &+ \tfrac{1}{8} \left( ( \TT_2 ( {\bar \Psi}_2 , \Psi_2 ) , \TT_1 ( X_1 , X_1 ) ) + \Re \, ( \TT_2 ( {\bar \Psi}_2 , J_2 \Psi_2 ) , \TT_1 ( J_1 X_1 , X_1 ) ) \right) \\
   &+ \tfrac{1}{2} \, \Re \left( ( \TT_1 ( X_1 , {\bar \Psi}_1 ) , \TT_2 ( X_2 , \Psi_2 ) ) + ( \TT_1 ( X_1 , J_1 {\bar \Psi}_1 ) , \TT_2 ( J_2 X_2 , \Psi_2 ) ) \right) \\
   &-\tfrac{1}{32} \left< T_1 ( X_1 , J_1 X_1 ) \cdot X_1 , T_1 ( X_1 , J_1 X_1 ) \cdot X_1 \right>_1 -\tfrac{1}{32} \left< T_2 ( X_2 , J_2 X_2 ) \cdot X_2 , T_2 ( X_2 , J_2 X_2 ) \cdot X_2 \right>_2 \\
   &-\tfrac{1}{32} \left< \TT_1 ( X_1 , X_1 ) \cdot X_2 , \TT_1 ( X_1 , X_1 ) \cdot X_2 \right>_2 -\tfrac{1}{32} \left< \TT_2 ( X_2 , X_2 ) \cdot X_1 , \TT_2 ( X_2 , X_2 ) \cdot X_1 \right>_1 \\
   &-\tfrac{1}{64} \left< \TT_1 ( X_1 , J_1 X_1 ) \cdot X_2 , \TT_1 ( X_1 , J_1 X_1 ) \cdot X_2 \right>_2 -\tfrac{1}{64} \left< \TT_2 ( X_2 , J_2 X_2 ) \cdot X_1 , \TT_2 ( X_2 , J_2 X_2 ) \cdot X_1 \right>_1 \\
   &-\tfrac{1}{64} \left< \TT_1 ( J_1 X_1 , X_1 ) \cdot X_2 , \TT_1 ( J_1 X_1 , X_1 ) \cdot X_2 \right>_2 -\tfrac{1}{64} \left< \TT_2 ( J_2 X_2 , X_2 ) \cdot X_1 , \TT_2 ( J_2 X_2 , X_2 ) \cdot X_1 \right>_1 .
  \end{aligned}
\end{equation}
Despite appearances, this lagrangian has some neat and familiar structure.  Notice first that the first four lines together with the eighth line in \eqref{eq:susy-lag-csm4-hlllp} gives just the decoupled matter contributions from the individual components $W_1$ and $W_2$ to the on-shell Gaiotto--Witten lagrangian \eqref{eq:susy-lag-csm4}.  Thus, as we saw in the previous section, these terms can certainly be given a manifestly $\fso(4)$-invariant expression on their own.

In order to write the remaining mixed terms in an $\fso(4)$-invariant manner, let us declare the matter fields $X_1$, $X_2$, $\Psi_1$ and $\Psi_2$ to transform in the following representations of $\fso(4)$:
\begin{equation}\label{eq:Vredtwistedreps}
 X^\alpha = \begin{pmatrix} X_1 \\ J_1 X_1 \end{pmatrix}, \quad \Psi_{\dot \alpha} = \begin{pmatrix} \Psi_1 \\ J_1 \Psi_1 \end{pmatrix}, \quad {\tilde X}_{\dot \alpha} = \begin{pmatrix} X_2 \\ J_2 X_2 \end{pmatrix}, \quad {\tilde \Psi}^{\alpha} = \begin{pmatrix} \Psi_2 \\ J_2 \Psi_2 \end{pmatrix},
\end{equation}
whereby $J_1 X^\alpha =  \varepsilon_{\alpha\beta} X^\beta$, $J_1 \Psi_{\dot \alpha} =  \varepsilon^{{\dot \alpha}{\dot \beta}} \Psi_{\dot \beta}$, $J_2 {\tilde X}_{\dot \alpha} =  \varepsilon^{{\dot \alpha}{\dot \beta}} {\tilde X}_{\dot \beta}$ and $J_2 {\tilde \Psi}^\alpha =  \varepsilon_{\alpha\beta} {\tilde \Psi}^{\beta}$ identically.  The representations $({\bf 2},{\bf 1})$ and $({\bf 1},{\bf 2})$ for the bosonic and fermionic matter fields in $W_1$ is just as we would expect from \eqref{eq:nequal3quaterniondata2} if they are to comprise an $N{=}4$ hypermultiplet.  The fact that we have defined the matter fields in $W_2$ to transform in the opposite representations of $\fso(4)$ stems from the fact that we required the matter superfield in $W_2$ to be $J_2 \Xi_2$ rather than $\Xi_2$ in order to obtain the $\fusp(2)$-invariant superpotential \eqref{eq:superpotential4t}.  Of course, this is still isomorphic to an $N{=}4$ hypermultiplet representation on $W_2$ though it is clearly useful to distinguish between these two types of $\fso(4)$ representations and the latter is often referred to as a {\emph{twisted}} $N{=}4$ hypermultiplet in the literature.  The $N{=}4$ supersymmetry transformations for a twisted hypermultiplet follow by acting with the quaternionic structure $J$ on the untwisted $N{=}4$ hypermultiplet transformations in \eqref{eq:susy4-onshell} and then absorbing the factor of $J$ into the definition of the twisted hypermultiplet matter fields ${\tilde X}$ and ${\tilde \Psi}$.  In terms of the subsequent $\fso(4)$-covariant forms of the $N{=}4$ supersymmetry transformations in \eqref{eq:susy4-onshella} and \eqref{eq:susy4-onshellb}, the corresponding prescription for going from an untwisted to a twisted hypermultiplet consists of switching all upstairs/downstairs and dotted/undotted indices followed by relabelling all the fields with tildes.

Given the $\fso(4)$ representations in \eqref{eq:Vredtwistedreps}, the lagrangian \eqref{eq:susy-lag-csm4-hlllp} can indeed be given the more succinct and manifestly $\fso(4)$-invariant expression
\begin{equation}
\label{eq:susy-lag-csm4-hlllpa}
  \begin{aligned}[m]
   \eL_{H3LP}^{N{=}4} =& -\varepsilon^{\mu\nu\rho} \left( A_\mu , \partial_\nu A_\rho + \tfrac{1}{3} [ A_\nu , A_\rho ] \right) -\tfrac{1}{4} \left< D_\mu X^\alpha , D^\mu X^\alpha \right> -\tfrac{1}{4} \left< D_\mu {\tilde X}_{\dot \alpha} , D^\mu {\tilde X}_{\dot \alpha} \right> \\
   &+ \tfrac{1}{4} \left< {\bar \Psi}_{\dot \alpha} , \gamma^\mu D_\mu \Psi_{\dot \alpha} \right> + \tfrac{1}{4} \left< {\bar {\tilde \Psi}^\alpha} , \gamma^\mu D_\mu {\tilde \Psi}^{\alpha} \right> \\
   &+ ( {\bar \nu^{\alpha {\dot \beta}}} , \nu_{\alpha {\dot \beta}} ) + ( {\bar {\tilde \nu}_{{\dot \alpha} \beta}} , {\tilde \nu}^{{\dot \alpha} \beta} ) + 4\, ( {\bar \nu^{\alpha {\dot \beta}}} , {\tilde \nu}_{{\dot \beta} \alpha} ) + \tfrac{1}{4} ( \mu^\alpha{}_\beta , \TT ( {\bar {\tilde \Psi}^\beta} , {\tilde \Psi}^\alpha ) ) + \tfrac{1}{4} ( {\tilde \mu}_{\dot \alpha}{}^{\dot \beta} , \TT ( {\bar {\Psi}}_{\dot \beta} , \Psi_{\dot \alpha} ) ) \\
   &-\tfrac{1}{12} \, ( [ \mu^\alpha{}_\beta , \mu^\beta{}_\gamma ] , \mu^\gamma{}_\alpha ) -\tfrac{1}{12} \, ( [ {\tilde \mu}_{\dot \alpha}{}^{\dot \beta} , {\tilde \mu}_{\dot \beta}{}^{\dot \gamma} ] , {\tilde \mu}_{\dot \gamma}{}^{\dot \alpha} ) \\
   &- \tfrac{1}{8} \left< \mu^\alpha{}_\beta \cdot {\tilde X}_{\dot \gamma} , \mu^\alpha{}_\beta \cdot {\tilde X}_{\dot \gamma} \right> - \tfrac{1}{8} \left< {\tilde \mu}_{\dot \alpha}{}^{\dot \beta} \cdot X^{\gamma} , {\tilde \mu}_{\dot \alpha}{}^{\dot \beta} \cdot X^{\gamma} \right>,
  \end{aligned}
\end{equation}
where we have defined ${\tilde \mu}_{\dot \alpha}{}^{\dot \beta} := \tfrac{1}{4} \, \TT ( {\tilde X}_{\dot \alpha} , {\tilde X}_{\dot \beta} )$ and ${\tilde \nu}_{{\dot \alpha} \beta} := \tfrac{1}{4} \TT ( {\tilde X}_{\dot \alpha} , {\tilde \Psi}^{\beta} )$ in keeping with the notation introduced at the end of the last section. (To avoid cluttering the expression unnecessarily, we have omitted the subscripts denoting operations involving $W_1$ and $W_2$ since this should be obvious from whether the term involves tildes or not.) Mutatis mutandis, the lagrangian \eqref{eq:susy-lag-csm4-hlllpa} is precisely that found already in \cite{pre3Lee} via the coupling of a twisted and untwisted hypermultiplet in a way that is compatible with $N{=}4$ superconformal symmetry.

The integral of $\eL_{H3LP}^{N{=}4}$ is indeed invariant under the $N{=}4$ supersymmetry transformations
\begin{equation}
  \label{eq:susy4-hlllp}
  \begin{aligned}[m]
    \delta X^\alpha &={\bar \epsilon}^{\alpha {\dot \beta}} \Psi_{\dot \beta} \\
    \delta {\tilde X}_{\dot \alpha} &= {\bar \epsilon}_{\beta {\dot \alpha}} {\tilde \Psi}^{\beta} \\
    \delta \Psi_{\dot \alpha} &= - \left[ \gamma^\mu D_\mu X^\beta + \tfrac{1}{3} \, \mu^\beta{}_\gamma \cdot X^\gamma \right] \epsilon_{\beta {\dot \alpha}} + {\tilde \mu}_{\dot \alpha}{}^{\dot \beta} \cdot X^\alpha \, \epsilon_{\alpha {\dot \beta}} \\
    \delta {\tilde \Psi}^{\alpha} &= -\left[ \gamma^\mu D_\mu {\tilde X}_{\dot \beta} + \tfrac{1}{3} \, {\tilde \mu}_{\dot \beta}{}^{\dot \gamma} \cdot {\tilde X}_{\dot \gamma} \right] \epsilon^{\alpha {\dot \beta}} + \mu^\alpha{}_\beta \cdot {\tilde X}_{\dot \alpha} \, \epsilon^{\beta {\dot \alpha}} \\
    \delta A_\mu &= {\bar \epsilon_{\alpha {\dot \beta}}} \gamma_\mu \left( \nu^{\alpha {\dot \beta}} + {\tilde \nu}^{{\dot \beta} \alpha} \right),
  \end{aligned}
\end{equation}
which close up to a translation on $\RR^{1,2}$ plus a gauge transformation, using the equations of motion from \eqref{eq:susy-lag-csm4-hlllpa}.

In summary, we have established that the superpotential ${\sf W}_\HH$ in \eqref{eq:superpotential3}, which generically guarantees $N{=}3$ supersymmetry when $\fM = W$ is quaternionic unitary, also describes the on-shell theories found in \cite{GaiottoWitten,pre3Lee} with enhanced $N{=}4$ superconformal symmetry for the specific types of $W$ involving the aLTS class assumed above.  Of course, this is to be expected from the point of view that a theory with $N{=}4$ superconformal symmetry is a special kind of $N{=}3$ theory, whose generic superpotential ${\sf W}_\HH$ is rigid.  On the other hand, for example, in the Gaiotto--Witten theory we have seen that the associated $N{=}4$ supersymmetry transformations in \eqref{eq:susy4-onshell} do not simply follow from the those in \eqref{eq:susy3-onshell} for the generic $N{=}3$ theory by restricting $W$ to be aLTS.  Hence the aforementioned deduction is perhaps not so trivial as it may appear.  Moreover, it is worth stressing that the guiding principle that has led us to the $N{=}4$ theories in \cite{GaiottoWitten,pre3Lee} has simply been to look for special cases of quaternionic unitary $\fg$-modules for which the superpotential ${\sf W}_\HH$ can be written in a $\fusp(2)$-invariant way.

Let us conclude by detailing the possible gauge-theoretic structures on which the $N{=}4$ theories of \cite{GaiottoWitten,pre3Lee} described above can be based.  In the case of the $N{=}4$ Gaiotto--Witten theory described in Section~\ref{sec:antiLTS-GW}, the fundamental ingredients describing indecomposable $N{=}4$ Gaiotto--Witten theories are $W \in \Irr(\fg;\HH)_{\text{aLTS}}$.  By Theorem~\ref{thm:aLTS-simplicity}, $W \in \Irr(\fg;\HH)_{\text{aLTS}}$ are in one-to-one correspondence with metric complex simple Lie superalgebras $\fg_\CC \oplus W$.  These have been classified and thus one has an indecomposable $N{=}4$ Gaiotto--Witten theory for each of the classical complex simple Lie superalgebras whose odd component admits a quaternionic structure: namely, $A(m,n)$, $B(m,n)$, $C(n+1)$, $D(m,n)$, $F(4)$, $G(3)$ and $D(2,1;\alpha)$.

For the more general $N{=}4$ superconformal Chern--Simons-matter theories of \cite{pre3Lee} described in Section~\ref{sec:twisthyp}, let us begin by considering the simplest nontrivial setup wherein $W = W_1 \oplus W_2$ with $W_1,W_2 \in \Irr(\fg,\HH)_{\text{aLTS}}$.  Theorem~\ref{thm:aLTS-simplicity} allows us to attach to $W_1$ and $W_2$ complex metric simple Lie superalgebras $G_1$ and $G_2$, respectively.  Despite $W = W_1 \oplus W_2$ being a faithful representation of $\fg$, the irreducible aLTS constituents $W_1$ and $W_2$ need not be.  Thus the real forms $\fg_1$ and $\fg_2$ of the even parts of $G_1$ and $G_2$ need not be isomorphic to $\fg$ itself (even though $\fg_1$ and $\fg_2$ do collectively span $\fg$).  The special case of $W_1 \cong W_2$ where $G_1 \cong G_2$ and thus $\fg_1 \cong \fg_2 \cong \fg$ will be the topic of the next section where it will be shown to give rise to an enhanced $N{=}5$ superconformal symmetry.  In order that the coupling terms between the untwisted and twisted hypermultiplets in the $N{=}4$ lagrangian \eqref{eq:susy-lag-csm4-hlllpa} are non-vanishing, such that we obtain an indecomposable theory, it is necessary for the semisimple Lie algebras $\fg_1$ and $\fg_2$ to have at least one common simple factor.  Since both $G_1$ and $G_2$ are simple, they must be one of the classical Lie superalgebras listed at the end of the last paragraph.  Thus, from the regular simple Lie superalgebras, one may choose any of the pairs $( G_1 , G_2 ) = ( A(m,p) , A(n,p) ) , ( B(m,p) , B(n,p) ) , ( B(p,m) , B(p,n) ) , ( B(m,n) , C(n+1) ) , ( B(m,p) , D(n,p) ) , ( C(m+1) , C(m+1) ) , ( D(m,n) , C(n+1) ) , ( D(m,p) , D(n,p) ) , ( D(p,m) , D(p,n) )$ and in each case identify the simple Lie algebra factor they have in common.  Clearly this technique can be continued to incorporate all the exceptional Lie superalgebras too as well as the additional possibilities which follow from utilising the various low-dimensional Lie algebra isomorphisms.  However, it will not be useful for us to elaborate further on these other possibilities.

The generalisation of this construction when either of the aLTS constituents $W_1$ or $W_2$ is reducible is straightforward.  For example, assume that the untwisted hypermultiplet matter is in an irreducible aLTS $W_1$ but the twisted hypermultiplet matter is taken to be in a reducible aLTS of the form $W_2 \oplus W_3$ where both $W_2$ and $W_3$ are irreducible aLTS representations.  Associated with $W_1$, $W_2$ and $W_3$ we have three simple Lie superalgebras $G_1$, $G_2$ and $G_3$ and the construction above can be employed for the two pairs $( G_1 , G_2 )$ and $( G_1 , G_3 )$ such that the ordered triple $( G_3 , G_1 , G_2 )$ is constrained only by the requirement that adjacent simple Lie superalgebras must have identified at least one simple Lie algebra factor in their even components (e.g. $( G_3 , G_1 , G_2 )$ could be $( A(m,p) , A(p,q) , A(q,n) )$ or $( B(m,p) , D(q,p) , D(q,n) )$ to name but two of many possibilities).  The most general situation can be described such that the faithful reducible representation takes the form $W = \bigoplus_{i=1}^n W_i$ in terms $n$ irreducible aLTS representations $\{ W_i \mid i=1,...,n \}$ (with $n$ associated simple Lie superalgebras $\{ G_i \mid i=1,...,n \}$) where, for convenience, one can assume there is a relative twist between the hypermultiplet matter in adjacent $W_i$ and $W_{i+1}$ for $i=1,...,n-1$.  Thus one has an indecomposable $N{=}4$ superconformal Chern--Simons-matter theory for each ordered $n$-tuple of simple Lie superalgebras $( G_1 ,..., G_n )$ such that each adjacent pair have identified at least one simple Lie algebra in their even parts and where there is a relative twist between the hypermultiplet matter in adjacent pairs. (Thus, when $n$ is even, there is also the possibility of identifying a simple Lie algebra factor in the even parts of $G_1$ and $G_n$ at the extremities.) Of course, there are many possibilities but most of the generic ones, described first in \cite{pre3Lee}, involve chains of simple Lie superalgebras of the same classical type.  To the best of our knowledge, these possibilities comprise all the known examples of $N{=}4$ superconformal Chern--Simons-matter theories.

\subsection{N{=}5 supersymmetry for $W_1 \cong W_2 \in \Dar(\fg,\HH)_{\text{aLTS}}$}
\label{sec:nequal5}

As discussed at the end of Section~\ref{sec:constr-from-supersym} and particularly in the diagram \eqref{eq:4to5}, to obtain a theory with enhanced $N{=}5$ superconformal symmetry, we reconsider the $N{=}4$ theory described in Section~\ref{sec:twisthyp} but now we will assume that $W_1, W_2 \in \Dar(\fg,\HH)_{\text{aLTS}}$ are isomorphic or, in other words, that the matter content is now taken to transform under two copies of $W\in\Dar(\fg,\HH)_{\text{aLTS}}$.  This prescription follows \cite{3Lee} wherein a new class of $N{=}5$ superconformal Chern--Simons-matter theories is constructed in precisely this way, as a special case of the class of $N{=}4$ theories they had found previously in \cite{pre3Lee} by coupling to a twisted hypermultiplet.

We shall therefore employ the notation of the previous section and identify $W_1 \cong W_2 \cong W$ as quaternionic unitary $\fg$-modules.  The untwisted and twisted hypermultiplet matter fields are $X = X_1$, $\Psi = \Psi_1$, ${\tilde X} = X_2$ and ${\tilde \Psi} = \Psi_2$, relative to their counterparts in the previous section, and each field here takes values in $W$.  They can be assembled into a pair of $W$-valued $N{=}1$ matter superfields $\Xi$ and ${\tilde \Xi}$ in the usual way.  It is important to stress that, in identifying $W_1 \cong W_2 \cong W$, one is no longer obliged to take $h ( W_1 , W_2 ) =0$ which followed from defining $W_1 \oplus W_2$ as an orthogonal direct sum with respect to the hermitian inner product.  Indeed, what is needed to describe the $N{=}5$ theory when identifying $W_1 \cong W_2 \cong W$ is to simply evaluate all the inner products involving the matter fields appearing in Section~\ref{sec:twisthyp} using the single hermitian inner product $h$ on $W$.

With respect to this structure, the expression in \eqref{eq:superpotential4t} for the superpotential $\sW_\HH$ reads
\begin{equation}
\label{eq:superpotential5}
   \begin{aligned}[m]
   \sW_{\HH} &= \tfrac{1}{16} \int d^2 \theta \; \left[ ( \TT ( \Xi  ,\Xi ) - \TT ( {\tilde \Xi}  , {\tilde \Xi} ) , \TT ( \Xi  ,\Xi ) - \TT ( {\tilde \Xi}  , {\tilde \Xi} ) ) \right. \\
   &\hspace*{1in}\left. -2\, ( \TT ( \Xi  , {\tilde \Xi} ) , \TT ( \Xi  , {\tilde \Xi} ) ) -2\, ( \TT ( {\tilde \Xi}  , \Xi ) , \TT ( {\tilde \Xi}  , \Xi ) ) \right] ,
   \end{aligned}
\end{equation}
where $\TT$ is the map associated with $W$.  The second line contains the contribution from the F-term superpotential and has been simplified using the identity $( \TT (u,Ju) , \TT (Jv,v) ) = 2\, ( \TT (u,v) , \TT (u,v) )$, for any $u,v \in W$, which follows using the aLTS cyclicity condition \eqref{eq:quat-aLTS} for $W$.

Whereas the superpotential $\sW_\HH$ in \eqref{eq:superpotential4t} for $W_1 \oplus W_2$ could only be given a $\fusp(2)$-invariant expression, we will now show that the superpotential in \eqref{eq:superpotential5} is actually $\fso(4)$-invariant.  This is to be expected of course if the theory is to have an $N{=}5$ superconformal symmetry since by writing the on-shell theory in terms of $N{=}1$ superfields one necessarily breaks the $\fso(5)$ R-symmetry down to an $\fso(4)$ isotropy subalgebra preserving the choice of $N{=}1$ superspace parameter.  However, since the enhanced $\fso(4)$-invariance of \eqref{eq:superpotential5} (relative to the manifestly $\fusp(2)$-invariant expression in \eqref{eq:superpotential4t}) is not immediately apparent, it will be enlightening to see explicitly how this works.  The trick is to not immediately try to write the hypermultiplet matter fields in terms of the representations of $\fso(4)$ that appeared in \eqref{eq:Vredtwistedreps} for the $N{=}4$ theory.  Instead, one must first define the linear combinations $\Xi_\pm := \tfrac{1}{\sqrt{2}} \left( \Xi \pm {\tilde \Xi} \right)$ of the $W$-valued $N{=}1$ superfields above (such combinations do not exist on $W_1 \oplus W_2$).  One then defines $\Xi_+^\alpha = ( \Xi_+ , J \Xi_+ )$ to transform in the $({\bf 2},{\bf 1})$ representation of $\fso(4)$ while $\Xi_{-\, {\dot \alpha}} = ( \Xi_-, J \Xi_- )$ is defined to transform in the opposite $({\bf 1},{\bf 2})$ representation.  In terms of these combinations of $N{=}1$ matter superfields, the superpotential \eqref{eq:superpotential5} takes a manifestly $\fso(4)$-invariant form given by
\begin{equation}
\label{eq:superpotential5a}
   \begin{aligned}[m]
   \sW_{\HH} &= \tfrac{1}{16} \int d^2 \theta \; \left[ ( \TT ( \Xi_+^\alpha  , \Xi_{-\, {\dot \beta}} ) , \TT ( \Xi_{-\, {\dot \beta}}  , \Xi_+^\alpha ) ) \right. \\
   &\hspace*{.9in}\left. -\tfrac{1}{6} \, ( \TT ( \Xi_+^\alpha  , \Xi_+^\beta ) , \TT ( \Xi_+^\beta  ,\Xi_+^\alpha ) ) -\tfrac{1}{6} \, ( \TT ( \Xi_{-\, {\dot \alpha}}  , \Xi_{-\, {\dot \beta}} ) , \TT ( \Xi_{-\, {\dot \beta}}  , \Xi_{-\, {\dot \alpha}} ) ) \right] .
   \end{aligned}
\end{equation}

Let us now consider the on-shell form \eqref{eq:susy-lag-on-shell} of the lagrangian $\eL^{N{=}5} = \eL_{CS} + \eL_M + \sW_{\HH}$ based on the superpotential in \eqref{eq:superpotential5}.  Not surprisingly, this gives precisely the lagrangian in \eqref{eq:susy-lag-csm4-hlllp} after identifying $W_1 \cong W_2 \cong W$.  Thus it can subsequently also be written just as in \eqref{eq:susy-lag-csm4-hlllpa} in terms of the $\fso(4)$ representations defined in \eqref{eq:Vredtwistedreps}.  As befits the expected $N{=}5$ superconformal symmetry of this on-shell lagrangian, one can assemble the constituent matter fields into representations of the expected $\fso(5) = \fusp(4)$ R-symmetry algebra.  We shall define representations of $\fusp(4)$ via straightforward extension of the way we defined representations of $\fusp(2)$.  That is, relative to a basis $\{ \be_A \}$ on $\CC^4$, we denote by $v^A$ the components of a complex vector $v$ transforming in the defining representation of $\fu(4)$ (while components of the complex conjugate vector $v^*$ have a downstairs index).  With respect to this basis, we take the $\fsp(4,\CC )$-invariant complex symplectic form to be $\Omega = \be^1 \wedge \be^2 + \be^3 \wedge \be^4$.  A vector $v \in W$ in the defining representation of $\fu(4)$ is in the fundamental representation ${\bf 4}$ of $\fusp(4) = \fu(4) \cap \fsp(4,\CC )$ if it obeys $J v^A = \Omega_{AB} v^B$.  The embedding of $\fso(4)$ in $\fso(5)$ we shall require corresponds to fixing a subalgebra $\fusp(2) \oplus \fusp(2) < \fusp(4)$ which defines a decomposition of $\CC^4 = \CC^2 \oplus \CC^2$ with the two $\fusp(2)$ factors in $\fso(4)$ acting on the respective $\CC^2$ components (for convenience we will identify $\{ \be_\alpha \}$ with $\be_1$ and $\be_2$ and $\{ \be^{\dot \alpha} \}$ with $\be_3$ and $\be_4$).  In terms of this embedding, we assemble the matter fields into
\begin{equation}\label{eq:Nequals5reps}
 X^A = \begin{pmatrix} X^\alpha \\ {\tilde X}_{\dot \alpha} \end{pmatrix}, \quad \Psi^A = \begin{pmatrix} {\tilde \Psi^\alpha} \\ \Psi_{\dot \alpha} \end{pmatrix},
\end{equation}
with both bosons and fermions transforming in the fundamental representation of $\fusp(4)$ since the pseudo-reality conditions $J X^A = \Omega_{AB} X^B$ and $J \Psi^A = \Omega_{AB} \Psi^B$ are identically satisfied.

In terms of the representations in \eqref{eq:Nequals5reps}, the on-shell lagrangian takes the manifestly $\fusp(4)$-invariant expression
\begin{multline}
\label{eq:susy-lag-csm5}
   \eL^{N{=}5} = -\varepsilon^{\mu\nu\rho} \left( A_\mu , \partial_\nu A_\rho + \tfrac{1}{3} [ A_\nu , A_\rho ] \right) -\tfrac{1}{4} \left< D_\mu X^A , D^\mu X^A \right> + \tfrac{1}{4} \left< {\bar \Psi}^A , \gamma^\mu D_\mu \Psi^A \right> \\
  \qquad + ( {\bar \nu}^A{}_B , \nu_A{}^B ) + 2\, ( {\bar \nu}^A{}_B , \nu^B{}_A ) +\tfrac{1}{15} \, ( [ \mu^A{}_B , \mu^B{}_C ] , \mu^C{}_A ) - \tfrac{3}{40} \left<  \mu^A{}_B \cdot X^C , \mu^A{}_B \cdot X^C \right>,
\end{multline}
where we have defined $\mu^A{}_B := \tfrac{1}{4} \, \TT ( X^A , X^B )$ and $\nu^A{}_B := \tfrac{1}{4} \TT ( X^A , \Psi^B )$.  Mutatis mutandis, the lagrangian \eqref{eq:susy-lag-csm5} is indeed precisely the one for the $N{=}5$ superconformal Chern--Simons-matter theory found in \cite{3Lee} and its integral is invariant under the $N{=}5$ supersymmetry transformations
\begin{equation}
  \label{eq:susy5}
  \begin{aligned}[m]
    \delta X^A &={\bar \epsilon}^A{}_B \Psi^B \\
    \delta \Psi^A &= - \left[ \gamma^\mu D_\mu X^B + \tfrac{1}{3} \, \mu^B{}_C \cdot X^C \right] \epsilon_B{}^A + \tfrac{2}{3} \, \mu^A{}_B \cdot X^C \, \epsilon_C{}^B \\
    \delta A_\mu &= {\bar \epsilon}_A{}^B \gamma_\mu \nu^A{}_B,
  \end{aligned}
\end{equation}
where the complex $N{=}5$ supersymmetry parameter $\epsilon_{AB} := \Omega_{BC} \, \epsilon_A{}^C$ is skewsymmetric $\epsilon_{AB} = -\epsilon_{BA}$, symplectic traceless $\Omega^{AB} \epsilon_{AB} =0$ and obeys the reality condition $\epsilon_{AB} = \Omega_{AC} \Omega_{BD} \, \epsilon^{CD}$, thus describing five linearly independent Majorana spinors on $\RR^{1,2}$.

We conclude this section by summarising the consequences of the (ir)reducibility of $W$ for the $N{=}5$ theory.  If $W \in \Dar(\fg,\HH)_{\text{aLTS}}$ is reducible, so that $W = W_1 \oplus W_2$, then as discussed in the second paragraph of Section~\ref{sec:twisthyp} the $N{=}5$ lagrangian in \eqref{eq:susy-lag-csm5} decouples into the sum of $N{=}5$ lagrangians on the individual irreducible components.  (Recall that $W_1,W_2 \in \Dar(\fg,\HH)_{\text{aLTS}}$ since so is $W$.)  The potential mixed terms in the lagrangian vanish identically as a consequence of the mixed components for the different irreducible factors in the aLTS cyclicity condition for $W$.  It is therefore enough to consider $W \in \Irr(\fg,\HH)_{\text{aLTS}}$.  As stated in Proposition~\ref{prop:irreducible}, if $W \in \Irr(\fg,\HH)_{\text{aLTS}}$, the underlying complex representation $\rh{W} \in \Dar(\fg,\CC)$ is irreducible unless $W = V_\HH$, so that $\rh{W} \cong V \oplus \Vbar$ for some $V \in \Irr(\fg,\CC)$.  This case will be examined in the next section where it will be found to give rise to an enhanced $N{=}6$ superconformal symmetry.  Thus we deduce that the indecomposable Chern--Simons-matter theories with \emph{precisely} $N{=}5$ superconformal symmetry are in one-to-one correspondence with $W\in\Irr(\fg,\HH)_{\text{aLTS}}$ for which $\rh{W} \in \Irr(\fg,\CC)$, which according to Theorem~\ref{thm:aLTS-simplicity}, are in turn in one-to-one correspondence with complex simple Lie superalgebras $\fg_\CC \oplus W$.  Such Lie superalgebras have been classified and are given by the classical simple Lie superalgebras $A(m,n)$, $B(m,n)$, $C(n+1)$, $D(m,n)$, $F(4)$, $G(3)$ and $D(2,1;\alpha )$.  This list exhausts the examples of $N{=}5$ superconformal Chern--Simons-matter theories that have been obtained already in \cite{3Lee,BHRSS}.

\subsection{N{=}6 supersymmetry for $W = V \oplus \Vbar$ with $V \in \Irr(\fg,\CC)_{\text{aJTS}}$}
\label{sec:nequal6}

Following again the prescription in \cite{3Lee}, and as discussed in Section \ref{sec:constr-from-supersym} and particularly diagram \eqref{eq:5to6}, one can enhance the supersymmetry of an $N{=}5$ theory to $N{=}6$ by considering matter representations $W \in \Irr(\fg,\HH)_{\text{aLTS}}$ which are quaternionifications $W = V_\HH$ of some $V \in \Irr(\fg,\CC)$, so that $\rh{W} \cong V \oplus \Vbar$.  As shown in Proposition~\ref{prop:relations}(i), this means that in fact $V \in \Irr(\fg,\CC)_{\text{aJTS}}$. 

Let us now investigate the structure of the superpotential $\sW_\HH$ in \eqref{eq:superpotential5} for $W = V \oplus \Vbar$. (The quaternionic unitary structure associated with this representation is defined in the proof of Proposition~\ref{prop:relations}(i) in Appendix~\ref{sec:some-relat-betw}.) The constituent $N{=}1$ matter superfields will be written $\Xi = ( \Xi^1 , {\bar \Xi^2} )$ and ${\tilde \Xi} = ( \Xi^3 , {\bar \Xi^4} )$ in terms of four $N{=}1$ superfields $\Xi^1 , \Xi^2 , \Xi^3 , \Xi^4 \in V$.  In terms of these superfields, the superpotential \eqref{eq:superpotential5} becomes
\begin{equation}
\label{eq:superpotential6}
   \begin{aligned}[m]
   \sW_{\HH} &= - \tfrac{1}{8} \int d^2 \theta \; \left[ ( \TT ( \Xi^1 , \Xi^1 ) ,  \TT ( \Xi^2 , \Xi^2 ) ) + ( \TT ( \Xi^1 , \Xi^1 ) ,  \TT ( \Xi^3 , \Xi^3 ) ) - ( \TT ( \Xi^1 , \Xi^1 ) ,  \TT ( \Xi^4 , \Xi^4 ) ) \right. \\
   &\hspace*{.9in}\left. -( \TT ( \Xi^2 , \Xi^2 ) ,  \TT ( \Xi^3 , \Xi^3 ) ) + ( \TT ( \Xi^2 , \Xi^2 ) ,  \TT ( \Xi^4 , \Xi^4 ) ) + ( \TT ( \Xi^3 , \Xi^3 ) ,  \TT ( \Xi^4 , \Xi^4 ) ) \right. \\
   &\hspace*{.9in}\left. +2\, ( \TT ( \Xi^1  , \Xi^2 ) , \TT ( \Xi^4  , \Xi^3 ) ) +2\, ( \TT ( \Xi^2  , \Xi^1 ) , \TT ( \Xi^3  , \Xi^4 ) ) \right] ,
   \end{aligned}
\end{equation}
where $\TT$ is the map associated with $V$ and the aJTS skewsymmetry condition $\TT ( x , y ) \cdot z =- \TT ( z , y ) \cdot x$ has been used.

The third line in \eqref{eq:superpotential6} represents the contribution from the F-term superpotential which was shown to admit a manifestly $\fso(4)$-invariant expression for the $N{=}6$ theory in \cite{MaldacenaBL}.  To demonstrate how this works, and mimicking the nomenclature in \cite{MaldacenaBL}, let us collect the four $V$-valued $N{=}1$ superfields into the two pairs $A = ( \Xi^1 , \Xi^4 )$ and $B = ( \Xi^2 , -\Xi^3 )$.  It will be convenient to take the pairs $A$ and $B$ to transform separately in the fundamental representation of two different copies of $\fsp(2, \CC )$.  With respect to the orthonormal bases $\{ \be_\alpha \}$ and $\{ \be_{\dot \alpha} \}$ associated with these two fundamental representations, we take the respective $\fsp(2,\CC)$-invariant symplectic forms to be $\varepsilon = \be^1 \wedge \be^2$ and ${\tilde \varepsilon} = \be^{\dot 1} \wedge \be^{\dot 2}$.  In terms of this structure, the third line in \eqref{eq:superpotential6} can be written as
\begin{equation}
\label{eq:superpotential6fterm}
\tfrac{1}{8} \int d^2 \theta \; \varepsilon_{\alpha\beta} \, {\tilde \varepsilon}^{{\dot \alpha}{\dot \beta}} \, \Re \, ( \TT ( A^\alpha  , B^{\dot \alpha} ) , \TT ( A^\beta  , B^{\dot \beta} ) ).
\end{equation}
which is manifestly invariant under $\fsp(2,\CC ) \oplus \fsp(2,\CC )$.  The addition of the kinetic terms for the matter fields to this F-term superpotential however breaks each $\fsp(2, \CC )$ down to $\fusp(2)$ (since both $h( A^\alpha , A^\alpha )$ and $h( B^{\dot \alpha} , B^{\dot \alpha})$ must also be invariant).  Hence the resulting symmetry realised by \eqref{eq:superpotential6fterm} is indeed $\fusp(2) \oplus \fusp(2) \cong \fso(4)$.

Of course, for an $N{=}6$ superconformal Chern--Simons-matter theory written in terms of $N{=}1$ superfields we expect that the full superpotential should be invariant under the $\fso(5)$ isotropy subalgebra of the $\fso(6)$ R-symmetry preserving our choice of $N{=}1$ superspace parameter.  This is indeed the case and follows by assembling the superfields into the array $( \Xi^1 , \Xi^2 , \Xi^3 , \Xi^4 )$ which is to be thought of as a $V$-valued element of $\CC^4$ whose components we denote by $\Xi^A$ with respect to a basis $\{ \be_A \}$ on $\CC^4$.  This $\CC^4$ is to be equipped with an action of $\fsp(4,\CC)$ such that $\Xi^A$ transforms in the fundamental representation.  We will define the $\fsp(4,\CC)$ subalgebra as those complex linear transformations which preserve the complex symplectic form $\Omega = \be^1 \wedge \be^3 + \be^2 \wedge \be^4$.  The full superpotential in \eqref{eq:superpotential6} then takes the explicitly $\fsp(4,\CC)$-invariant form
\begin{equation}
\label{eq:superpotential6a}
\sW_{\HH} =  -\tfrac{1}{16} \int d^2 \theta \; \left( \delta_A^C \delta_B^D -  \Omega_{AB} \Omega^{CD} \right) ( \TT ( \Xi^A , \Xi^C ) ,  \TT ( \Xi^B , \Xi^D ) ).
\end{equation}
Again, the addition of the kinetic terms for the matter fields breaks this $\fsp(4, \CC )$ symmetry down to the expected $\fusp(4) \cong \fso(5)$ (since $h ( \Xi^A , \Xi^A )$ must also be invariant).

Let us now consider the on-shell form \eqref{eq:susy-lag-on-shell} of the lagrangian $\eL^{N{=}6} = \eL_{CS} + \eL_M + \sW_{\HH}$ based on the superpotential in \eqref{eq:superpotential6}.  This simply amounts to rewriting \eqref{eq:susy-lag-csm5} for the special case of $W = V \oplus \Vbar$ in a form which is explicitly invariant under the $\fso(6) \cong \fsu(4)$ R-symmetry of the $N{=}6$ superconformal algebra.  To this end, let us begin by writing the untwisted and twisted hypermultiplet matter fields from the original $N{=}4$ theory as $X = ( {\bf X}^1 , {\overline{{\bf X}^2}} )$, ${\tilde X} = ( {\bf X}^3 , {\overline{{\bf X}^4}} )$, $\Psi = ( {\bf \Psi}_4 , - {\overline{{\bf \Psi}_3}} )$ and ${\tilde \Psi} = ( {\bf \Psi}_2 , - {\overline{{\bf \Psi}_1}} )$ on $W = V \oplus \Vbar$ in terms of the four $V$-valued bosons $( {\bf X}^1 , {\bf X}^2 , {\bf X}^3 , {\bf X}^4 )$ and fermions $( {\bf \Psi}_1 , {\bf \Psi}_2 , {\bf \Psi}_3 , {\bf \Psi}_4 )$ whose components we will denote by ${\bf X}^A$ and ${\bf \Psi}_A$ respectively.  This is a convenient parametrisation inasmuch as it allows one to assemble the components into the $\fusp(4)$-covariant expressions
\begin{equation}\label{eq:Nequals6reps}
 X^A = ( {\bf X}^A , {\overline{\Omega_{AB} {\bf X}^B}} ), \quad \Psi^A = ( \Omega^{AB} \, {\bf \Psi}_B , - {\overline{{\bf \Psi}_A}} ),
\end{equation}
with the components on the left hand sides being defined just as in \eqref{eq:Nequals5reps}.  The pseudo-reality conditions $J X^A = \Omega_{AB} X^B$ and $J \Psi^A = \Omega_{AB} \Psi^B$ are thus identically satisfied, with no constraint on ${\bf X}^A$ and ${\bf \Psi}_A$, from the definition of $J$ acting on $V \oplus \Vbar$.  The expressions in \eqref{eq:Nequals6reps} can be understood as describing the canonical embedding of the fundamental representation of $\fusp(4)$ into the real form of the fundamental representation of $\fsu(4)$, wherein ${\bf X}^A$ and ${\bf \Psi}_A$ respectively transform in the complex representations corresponding to the fundamental ${\bf 4}$ and antifundamental ${\bf {\bar 4}}$ of $\fsu(4)$.

In terms of these representations, the on-shell lagrangian \eqref{eq:susy-lag-csm5} can be given the following $\fsu(4)$-invariant expression
\begin{equation}
\label{eq:susy-lag-csm6}
  \begin{aligned}[m]
   \eL^{N{=}6} =& -\varepsilon^{\mu\nu\rho} \left( A_\mu , \partial_\nu A_\rho + \tfrac{1}{3} [ A_\nu , A_\rho ] \right) -\tfrac{1}{2} \left< D_\mu {\bf X}^A , D^\mu {\bf X}^A \right> + \tfrac{1}{2} \left< {\bar {\bf \Psi}}_A , \gamma^\mu D_\mu {\bf \Psi}_A \right> \\
   &+ 2\, ( {\bar {\boldsymbol{\nu}}}^{AB} , {\boldsymbol{\nu}}_{AB} - 2\, {\boldsymbol{\nu}}_{BA} ) + \varepsilon_{ABCD} ( {\bar {\boldsymbol{\nu}}}^{AC} , {\boldsymbol{\nu}}^{BD} ) + \varepsilon^{ABCD} ( {\bar {\boldsymbol{\nu}}}_{AC} , {\boldsymbol{\nu}}_{BD} ) \\
   &+\tfrac{2}{3}\, ( [ {\boldsymbol{\mu}}^A{}_B , {\boldsymbol{\mu}}^B{}_C ] , {\boldsymbol{\mu}}^C{}_A ) - \tfrac{1}{2} \left< {\boldsymbol{\mu}}^A{}_B \cdot {\bf X}^C , {\boldsymbol{\mu}}^A{}_B \cdot {\bf X}^C \right>,
  \end{aligned}
\end{equation}
where we have defined ${\boldsymbol{\mu}}^A{}_B := \tfrac{1}{4} \, \TT ( {\bf X}^A , {\bf X}^B )$ and ${\boldsymbol{\nu}}^{AB} := \tfrac{1}{4} \TT ( {\bf X}^A , {\bf \Psi}_B )$ in terms of the map $\TT$ associated with $V$ and $\varepsilon = \be^1 \wedge \be^2 \wedge \be^3 \wedge \be^4$ is the $\fsu(4)$-invariant 4-form with respect an orthonormal basis $\{ \be^A \}$ on the dual of $\CC^4$.  Under the embedding of $\fusp(4)$ in $\fsu(4)$ described above, the moment maps here are related to their $N{=}5$ counterparts defined below \eqref{eq:susy-lag-csm5} such that $\mu^A{}_B = {\boldsymbol{\mu}}^A{}_B - \Omega^{AC} \Omega_{BD} {\boldsymbol{\mu}}^D{}_C$ and $\nu^A{}_B = \Omega_{BC} {\boldsymbol{\nu}}^{AC} - \Omega^{AC} {\boldsymbol{\nu}}_{CB}$ (where ${\boldsymbol{\nu}}_{AB} = - \tfrac{1}{4} \TT ( {\bf \Psi}_B , {\bf X}^A )$ is the complex conjugate of ${\boldsymbol{\nu}}^{AB}$).  The components of the $\fsu(4)$-invariant 4-form can be expressed as $\varepsilon_{ABCD} = \Omega_{AB} \Omega_{CD} + \Omega_{AC} \Omega_{DB} + \Omega_{AD} \Omega_{BC}$ in terms of the $\fusp(4)$-invariant symplectic form under the aforementioned embedding.

Notice that the $N{=}6$ lagrangian above has a global $\fu(1)$ symmetry under which the gauge field is uncharged while the bosonic and fermionic matter fields ${\bf X}^A$ and ${\bf \Psi}_A$ both have the same charge.  It is to be distinguished from the $\fu(1) < \fsu(4)$ R-symmetry subalgebra that is realised in the description of this theory in terms of $N{=}2$ superfields under which the bosons and fermions have opposite charges $\half$ and $-\half$.  Indeed, this global $\fu(1)$ is a flavour symmetry since it commutes with the $N{=}6$ superconformal algebra. 

The lagrangian \eqref{eq:susy-lag-csm6} describes precisely the $N{=}6$ theory in \cite{MaldacenaBL,BL4,3Lee} and its integral is invariant under the $N{=}6$ supersymmetry transformations
\begin{equation}
  \label{eq:susy6}
  \begin{aligned}[m]
    \delta {\bf X}^A &={\bar \epsilon}^{AB} {\bf \Psi}_B \\
    \delta {\bf \Psi}_A &= - \left[ \gamma^\mu D_\mu {\bf X}^B + {\boldsymbol{\mu}}^B{}_C \cdot {\bf X}^C \right] \epsilon_{AB} - {\boldsymbol{\mu}}^B{}_A \cdot {\bf X}^C \, \epsilon_{BC} \\
    \delta A_\mu &= - {\bar \epsilon}_{AB} \gamma_\mu {\boldsymbol{\nu}}^{AB} - {\bar \epsilon}^{AB} \gamma_\mu {\boldsymbol{\nu}}_{AB},
  \end{aligned}
\end{equation}
where the complex $N{=}6$ supersymmetry parameter $\epsilon_{AB}$ here is skewsymmetric $\epsilon_{AB} = -\epsilon_{BA}$ and obeys the reality condition $\epsilon_{AB} = \half \varepsilon_{ABCD} \, \epsilon^{CD}$, thus describing six linearly independent Majorana spinors on $\RR^{1,2}$.  The $N{=}5$ supersymmetry transformations in \eqref{eq:susy5} are recovered following the embedding of $\fusp(4)$ in $\fsu(4)$ described above and then imposing the symplectic tracelessness condition on $\epsilon_{AB}$.

Given the construction of the $N{=}6$ theory as an enhanced $N{=}5$ theory, indecomposability of the $N{=}6$ theory follows from that of the $N{=}5$ theory, which required $W \in \Irr(\fg,\HH)_{\text{aLTS}}$.  Enhancement further requires $W = V_\HH$ and hence $V \in \Irr(\fg,\CC)_{\text{aJTS}}$.  As stated in Proposition~\ref{prop:irreducible}, the underlying real representation $\rf{V} \in \Dar(\fg,\RR)$ is still irreducible unless $V = U_\CC$ is the complexification of $U \in \Irr(\fg,\RR)$.  As we will see in the next section, in this case  supersymmetry will be enhanced to $N{=}8$.  Thus we deduce that the Chern--Simons-matter theories with \emph{precisely} $N{=}6$ superconformal symmetry are in one-to-one correspondence with $V \in \Irr(\fg,\CC)_{\text{aJTS}}$ which are not the complexification of a real representation.  According to Theorem \ref{thm:BL4-simplicity}, such $V\in\Irr(\fg,\CC)_{\text{aJTS}}$ are in turn in one-to-one correspondence with complex simple 3-graded Lie superalgebras $V \oplus \fg_\CC \oplus \Vbar$.  These have been classified \cite[Theorem~4]{KacSuperSketch} and are given by the two classical simple Lie superalgebras $A(m,n)$ and $C(n+1)$.  Again this conclusion is in accordance with the earlier classification of $N{=}6$ superconformal Chern--Simons-matter theories in \cite{SchnablTachikawa} and remarks in \cite{GaiottoWitten,3Lee}.

\subsection{N{=}8 supersymmetry for $V=U_\CC$ with $U \in \Irr(\fg,\RR)_{\text{3LA}}$}
\label{sec:nequal8}

We will now describe how to obtain the theory that was first discovered by Bagger and Lambert \cite{BL1,BL2} and Gustavsson \cite{GustavssonAlgM2} with maximal $N{=}8$ superconformal symmetry as a special case of the $N{=}6$ theory encountered in the previous section by taking the representation $V \in \Irr(\fg,\CC)_{\text{aJTS}}$ on which that theory is based to be the complexification $V = U_\CC$ of $U \in \Irr(\fg,\RR)_{\text{3LA}}$.  That is $V$ here will be taken to be irreducible as a complex representation but reducible as a real representation.  The canonical complex unitary structure on $U_\CC$ is defined in Appendix~\ref{sec:real-as-complex}.  Proposition~\ref{prop:relations}(f) in Appendix~\ref{sec:some-relat-betw} establishes the crucial relation that $V = U_\CC \in \Dar(\fg,\CC)_{\text{aJTS}}$ if and only if $U\in \Dar(\fg,\RR)_{\text{3LA}}$.  

Given the incremental amounts of supersymmetry enhancement we have been observing, based on assuming increasingly specialised types of unitary representations, one might wonder why we have omitted the possibility of realising $N{=}7$ supersymmetry.  Indeed, the on-shell theory we shall initially find has a manifest global $\fso(7)$ symmetry and is invariant under an $N{=}7$ superconformal algebra.  However, it will be shown that, without further representation-theoretic assumptions, the lagrangian for this theory can be rewritten in a manifestly $\fso(8)$-invariant way that is precisely the maximal $N{=}8$ superconformal Chern--Simons-matter theory in \cite{BL1,BL2,GustavssonAlgM2}.

Let us now investigate the structure of the superpotential $\sW_\HH$ in \eqref{eq:superpotential6} for the special case of $V = U_\CC$.  The four $V$-valued $N{=}1$ matter superfields will be written $\Xi^1 = \xi^1 + i {\hat \xi}^1$, $\Xi^2 = \xi^2 + i {\hat \xi}^2$, $\Xi^3 = \xi^3 + i {\hat \xi}^3$ and $\Xi^4 = \xi^4 + i {\hat \xi}^4$ in terms of eight $U$-valued $N{=}1$ superfields $\xi^1 , \xi^2 , \xi^3 , \xi^4 , {\hat \xi}^1 , {\hat \xi}^2 , {\hat \xi}^3 , {\hat \xi}^4$.  In terms of these superfields, the superpotential \eqref{eq:superpotential6} becomes
\begin{equation}
\label{eq:superpotential8}
   \begin{aligned}[m]
   \sW_{\HH} &= \tfrac{1}{2} \int d^2 \theta \; \left[ ( T( \xi^1 , {\hat \xi}^1 ) ,  T( \xi^2 , {\hat \xi}^2 ) ) + ( T( \xi^1 , {\hat \xi}^1 ) ,  T( \xi^3 , {\hat \xi}^3 ) ) - ( T( \xi^2 , {\hat \xi}^2 ) ,  T( \xi^3 , {\hat \xi}^3 ) ) \right. \\
   &\hspace*{.8in}\left. + ( T( \xi^2 , {\hat \xi}^2 ) ,  T( \xi^4 , {\hat \xi}^4 ) ) + ( T( \xi^3 , {\hat \xi}^3 ) ,  T( \xi^4 , {\hat \xi}^4 ) ) - ( T( \xi^1 , {\hat \xi}^1 ) ,  T( \xi^4 , {\hat \xi}^4 ) ) \right. \\
   &\hspace*{.8in}\left. + ( T( \xi^1 , \xi^2 ) ,  T( {\hat \xi}^3 , {\hat \xi}^4 ) ) + ( T( \xi^3 , \xi^4 ) ,  T( {\hat \xi}^1 , {\hat \xi}^2 ) ) - ( T( \xi^1 , \xi^4 ) ,  T( {\hat \xi}^2 , {\hat \xi}^3 ) ) \right. \\
   &\hspace*{.8in}\left. - ( T( \xi^1 , \xi^3 ) ,  T( {\hat \xi}^2 , {\hat \xi}^4 ) ) - ( T( \xi^2 , \xi^4 ) ,  T( {\hat \xi}^1 , {\hat \xi}^3 ) ) - ( T( \xi^2 , \xi^3 ) ,  T( {\hat \xi}^1 , {\hat \xi}^4 ) ) \right. \\
   &\hspace*{.8in}\left. +( T( \xi^1 , \xi^2 ) ,  T( \xi^3 , \xi^4 ) ) +( T( {\hat \xi}^1 , {\hat \xi}^2 ) ,  T( {\hat \xi}^3 , {\hat \xi}^4 )) \right] ,
   \end{aligned}
\end{equation}
where $T$ is the map associated with $U$ and the total skewsymmetry of $T(x,y)\cdot z$ has been used.  For an $N{=}8$ superconformal Chern--Simons-matter theory written in terms of $N{=}1$ superfields we expect that the superpotential should be invariant under the $\fso(7)$ subalgebra of the $\fso(8)$ R-symmetry which preserves the $N{=}1$ superspace parameter.  This is indeed the case and follows by assembling the eight real constituent $N{=}1$ superfields into the array $( \xi^3 , \xi^4 , \xi^1 , \xi^2 , {\hat \xi}^1 , {\hat \xi}^2 , {\hat \xi}^3 , {\hat \xi}^4 )$ which is to be thought of as a $U$-valued element of $\RR^8$.  Let us denote its components by $\Xi^I$ with respect to a basis $\{ e_I \}$ on $\RR^8$.  This $\RR^8$ is to be equipped with the action of $\fso(7)$ that defines the real spinor representation in seven dimensions.  A natural quartic tensor on $\RR^8$ that is preserved by this action of $\fso(7)$ is the Cayley 4-form $\Omega$.  If we take $\{ e^I \}$ to define an orthonormal basis on the dual of $\RR^8$ then the Cayley form can be taken to be
\begin{equation}
\label{eq:cayleyform}
\Omega = e^{1234} +  ( e^{12} - e^{34} ) \wedge ( e^{56} - e^{78} ) +  ( e^{13} + e^{24} ) \wedge ( e^{57} + e^{68} ) +  ( e^{14} - e^{23} ) \wedge ( e^{58} - e^{67} ) + e^{5678},
\end{equation}
where multiple indices denote wedge products of the corresponding basis elements.  In terms of the Cayley form, the superpotential in \eqref{eq:superpotential8} can be written in the manifestly $\fso(7)$-invariant form
\begin{equation}
\label{eq:superpotential8a}
\sW_{\HH} =  \tfrac{1}{48} \int d^2 \theta \; \Omega_{IJKL} (T( \Xi^I , \Xi^J ),T( \Xi^K , \Xi^L ) ).
\end{equation}
This $\fso(7)$-invariant expression for the superpotential that gives rise to the $N{=}8$ Bagger--Lambert lagrangian has appeared already in \cite{MauriPetkouBL}.  What we have shown above is that it is precisely this superpotential which follows from evaluating $\sW_\HH$ in \eqref{eq:superpotential6} for the special case of $V = U_\CC$.

We will now examine the on-shell form \eqref{eq:susy-lag-on-shell} of the lagrangian $\eL^{N{=}8} = \eL_{CS} + \eL_M + \sW_{\HH}$ based on the superpotential \eqref{eq:superpotential8}.  There are two ways to evaluate this.  The simplest is to integrate out the auxiliary fields in the generic $N{=}1$ Chern--Simons-matter lagrangian with the form of the superpotential in \eqref{eq:superpotential8a} based on an $N{=}1$ matter superfield $\Xi^I$ valued in $\RR^8 \otimes U$ (the terms in the matter part of the lagrangian being evaluated with respect to the obvious tensor product inner product involving the unit inner product on $\RR^8$, with components $\delta_{IJ}$, and the inner product $\left< -,- \right>$ on $U$).  The other way is to just evaluate the on-shell $N{=}6$ lagrangian \eqref{eq:susy-lag-csm6} for $V = U_\CC$.  Of course, as consistency dictates, both methods give the same answer
\begin{equation}
\label{eq:susy-lag-csm8}
  \begin{aligned}[m]
   \eL^{N{=}8} =& -\varepsilon^{\mu\nu\rho} \left( A_\mu , \partial_\nu A_\rho + \tfrac{1}{3} [ A_\nu , A_\rho ] \right) -\tfrac{1}{2} \left< D_\mu X^I , D^\mu X^I \right> + \tfrac{1}{2} \left< {\bar \Psi}^I , \gamma^\mu D_\mu \Psi^I \right> \\
   &+ 2 \left( 2\, \delta_{IK} \delta_{JL} + \Omega_{IJKL} \right) ( {\bar \nu}^{IK} , \nu^{JL} ) -\tfrac{1}{3} \left< \mu^{IJ} \cdot X^K , \mu^{IJ} \cdot X^K \right>,
  \end{aligned}
\end{equation}
where $\mu^{IJ} := \tfrac{1}{4} \, T ( X^I , X^J )$ and $\nu^{IJ} := \tfrac{1}{4} \, T ( X^I , \Psi^J )$ in terms of the map $T$ associated with $U$.  Of course, this form of the lagrangian is only manifestly $\fso(7)$-invariant and we shall detail its $\fso(8)$-invariant expression in a moment.  First though it will be useful to elaborate a little on how the matter fields appearing in \eqref{eq:susy-lag-csm8} are realised in the two different derivations noted above.  In the first and simpler derivation, $X^I$ and $\Psi^I$ are just the obvious bosonic and fermionic components of the $N{=}1$ superfield $\Xi^I$.  To obtain the scalar potential in \eqref{eq:susy-lag-csm8} from this approach requires the identity
\begin{equation}
\label{eq:cayleyid}
\Omega_{IJKL} \Omega^{MNPL} = 6\, \delta_{[I}^M \delta_J^N \delta_{K]}^P - 9\, \delta_{[I}^{[M} \Omega_{JK]}{}^{NP]},
\end{equation}
for the components of the Cayley form to be employed and the contribution of the second term on the right hand side above to the scalar potential can be shown to vanish identically as a consequence of the fundamental identity \eqref{eq:FI-R} for the 3-Lie bracket associated with $T$ on $U$.  In the second derivation, after taking $V = U_\CC$, in order that the kinetic terms for the matter fields in the lagrangians \eqref{eq:susy-lag-csm6} and \eqref{eq:susy-lag-csm8} agree, the identification of $\CC^4$ with $\RR^8$ must be isometric, corresponding to the canonical embedding of $\fsu(4)$ in $\fso(8)$.  A convenient way to achieve this for the bosonic fields is to identify the first and last four of the eight real scalars $X^I$ respectively with the real and imaginary parts of the four complex scalars ${\bf X}^A$ in the $N{=}6$ theory.  For the fermionic fields though one identifies the first and last four fermions $\Psi^I$ respectively with the imaginary and and real parts of the four complex fermions ${\bf \Psi}_A$ in the $N{=}6$ theory. (This distinction between the way the bosons and fermions are identified is necessary in order to then rewrite the Yukawa couplings for the $N{=}6$ theory in an $\fso(7)$-invariant way.) In deriving the expression for the Yukawa couplings in the second line of \eqref{eq:susy-lag-csm8} from \eqref{eq:susy-lag-csm6}, use has been made of the identity $\Omega = \Re\, \varepsilon - \half k \wedge k$ for the Cayley form \eqref{eq:cayleyform} on $\RR^8$ in terms of (the real part of) the holomorphic 4-form $\varepsilon = \be^1 \wedge \be^2 \wedge \be^3 \wedge \be^4$ and the Kähler form $k = \tfrac{i}{2} \be^A \wedge \be_A$ on $\CC^4$ (the latter being a real 2-form and can be expressed as $e^{15} + e^{26} + e^{37} + e^{48}$ on $\RR^8$ under the aforesaid identification).  Finally, in deriving the scalar potential in \eqref{eq:susy-lag-csm8}, it was useful to first note that the $N{=}6$ scalar potential in the third line in \eqref{eq:susy-lag-csm6} can be rewritten as $-\tfrac{1}{3} \left( \left< {\boldsymbol{\mu}}^A{}_B \cdot {\bf X}^C , {\boldsymbol{\mu}}^A{}_B \cdot {\bf X}^C \right> - \half \left< {\boldsymbol{\mu}}^A{}_B \cdot {\bf X}^B , {\boldsymbol{\mu}}^A{}_C \cdot {\bf X}^C \right> \right)$.

The integral of the lagrangian in \eqref{eq:susy-lag-csm8} is invariant under the $N{=}7$ supersymmetry transformations
\begin{equation}
  \label{eq:susy7}
  \begin{aligned}[m]
    \delta X^I &={\bar \epsilon}^{IJ} \Psi^J \\
    \delta \Psi^I &= - \gamma^\mu D_\mu X^J \, \epsilon^{IJ} + \tfrac{2}{3} \, \mu^{IJ} \cdot X^K \, \epsilon^{JK} + \tfrac{1}{3} \, \Omega^{IJKL} \, \mu^{JK} \cdot X^M \, \epsilon^{LM} \\
    \delta A_\mu &= - 2\, {\bar \epsilon}^{IJ} \gamma_\mu \nu^{IJ},
  \end{aligned}
\end{equation}
where the real $N{=}7$ supersymmetry parameter $\epsilon_{IJ}$ is skewsymmetric $\epsilon_{IJ} = -\epsilon_{JI}$ and obeys $\epsilon_{IJ} = - \tfrac{1}{6} \Omega_{IJKL} \, \epsilon_{KL}$ which defines the projection onto the seven-dimensional $\fso(7)$-invariant subspace of $\Lambda^2 \RR^8$.  Thus it describes seven linearly independent Majorana spinors on $\RR^{1,2}$.  The $N{=}6$ supersymmetry transformations in \eqref{eq:susy6} can be recovered on $V = U_\CC$ under the identification of $\RR^8$ with $\CC^4$ described above after imposing the the extra condition $k^{IJ} \epsilon_{IJ} =0$ using the Kähler form to eliminate the seventh supersymmetry parameter.

\subsubsection{$\fso(8)$ R-symmetry}
\label{sec:so8rsymmetry}

To rewrite the lagrangian \eqref{eq:susy-lag-csm8} in a manifestly $\fso(8)$-invariant way first requires a choice of embedding for $\fso(7)$ in $\fso(8)$.  Recall that there are three real eight-dimensional irreducible representations of $\fso(8)$: the vector representation $\boldsymbol{8}_v$ and the positive and negative chirality spinor representations $\boldsymbol{8}_s$ and $\boldsymbol{8}_c$.  Consequently there are three distinct embeddings of $\fso(7)$ in $\fso(8)$ that can be understood as the subalgebras preserving a fixed nonzero element in either $\boldsymbol{8}_v$, $\boldsymbol{8}_s$ or $\boldsymbol{8}_c$.  Of course, 'distinct' only up to the triality symmetry which relates these three representations.  In each case, whichever of the three representations the fixed element resides in breaks into the vector and singlet representations ${\bf 7} \oplus {\bf 1}$ of $\fso(7)$ while the remaining two representations both reduce to the spinor representation $\boldsymbol{8}$.  Both the bosonic and fermionic matter fields $X^I$ and $\Psi^I$ above transform in the $\boldsymbol{8}$ while the supersymmetry parameter $\epsilon_{IJ}$ transforms in the ${\bf 7}$ of $\fso(7)$.  We are therefore free to choose any of the three embeddings so long as it is the supersymmetry parameter which embeds into whichever of the three representations of $\fso(8)$ that contains a singlet under the $\fso(7)$ subalgebra.  As in \cite{Nahm}, it has been convenient for us so far to assume that the supercharges in an $N$-extended superconformal algebra transform in the vector representation of the $\fso(N)$ R-symmetry.  In the case at hand of $N{=}8$ this would suggest we lift the supercharges into the $\boldsymbol{8}_v$ of $\fso(8)$ with the matter fields lifting to the $\boldsymbol{8}_s$ and $\boldsymbol{8}_c$.  However, it will prove more convenient here to lift the supercharges into the $\boldsymbol{8}_s$ of $\fso(8)$ with the matter fields $X^I$ and $\Psi^I$ lifting respectively to the $\boldsymbol{8}_v$ and $\boldsymbol{8}_c$ representations.  The main reason for this is technical inasmuch as it allows one to rewrite things more neatly in terms of elements of the Clifford algebra in eight dimensions and whence recover the well-known form of the $N{=}8$ lagrangian originally presented in \cite{BL2}.  Of course, one can always apply triality to obtain whichever of the three representations of $\fso(8)$ one wants for the supercharges (for example, this has been done explicitly in \cite{Gustavsson:2009pm} in the context of so-called \lq trial BLG' theories).

We shall adopt the conventions of \cite{AFOS} for the Clifford algebra $\Cl(8)$.  An $\fso(8)$-covariant basis for $\Cl(8)$ can be constructed in terms of products of the $8$ real $16\times 16$ skewsymmetric matrices $\Gamma_I$ obeying $\Gamma_I \Gamma_J + \Gamma_J \Gamma_I = -2 \delta_{IJ} \, 1$. (The index $I$ will be used here to denote the vector representation of $\fso(8)$ and $\delta_{IJ}$ denotes the components of the unit $\fso(8)$-invariant inner product.) The real 16-dimensional vector space acted upon by the matrices $\Gamma_I$ corresponds to the spinor representation of $\fso(8)$.  The chirality matrix is defined by $\Gamma := \Gamma_1 ... \Gamma_8$, which is idempotent and anticommutes with each $\Gamma_I$.  The chiral and antichiral representations $\boldsymbol{8}_s$ and $\boldsymbol{8}_c$ correspond respectively to the positive and negative chirality eigenspaces of $\Gamma$.  The supersymmetry parameter $\epsilon_{IJ}$ is to be lifted to $\boldsymbol{\epsilon} = \Gamma \boldsymbol{\epsilon}$ in $\boldsymbol{8}_s$ while the fermions $\Psi^I$ are to be lifted to $\boldsymbol{\Psi} = - \Gamma \boldsymbol{\Psi}$ in $\boldsymbol{8}_c$ of $\fso(8)$ (of course, both are also Majorana spinors on $\RR^{1,2}$).  The lift of the bosons $X^I$ is rather more trivial requiring only the reinterpretation of the index $I$ from the $\boldsymbol{8}$ of $\fso(7)$ to the $\boldsymbol{8}_v$ of $\fso(8)$.

Let us now assume the existence of a fixed (commuting) chiral spinor $\vartheta \in \boldsymbol{8}_s$ which we take to be unit normalised such that $\vartheta^t \vartheta = 1$ (Majorana conjugation is just transposition here since the charge conjugation matrix for $\Cl(8)$ can be taken to be the identity).  This defines the desired embedding of $\fso(7)$ in $\fso(8)$ as the stabiliser of $\vartheta$.  In terms of this fixed chiral spinor, one can deduce the precise identifications for the supersymmetry parameter and fermions to be
\begin{equation}\label{eq:Nequals8reps}
 \epsilon_{IJ} = \vartheta^t \Gamma_{IJ} \boldsymbol{\epsilon} , \quad \Psi^I = \vartheta^t \Gamma^I \boldsymbol{\Psi} ,
\end{equation}
where $\Gamma_{IJ} :=  \Gamma_{[I} \Gamma_{J]}$ and the Cayley form $\Omega_{IJKL} = \vartheta^t \Gamma_{IJKL} \vartheta$.  Making use of the Fierz identity $\vartheta \vartheta^t = \tfrac{1}{16} \left( 1+ \Gamma + \tfrac{1}{4!} \Omega_{IJKL} \Gamma_{IJKL} \right)$, one can check that the right-hand side of the first equation in \eqref{eq:Nequals8reps} obeys the same projection condition $\epsilon_{IJ} = - \tfrac{1}{6} \Omega_{IJKL} \, \epsilon_{KL}$ satisfied by the left-hand side.  Hence the eighth supersymmetry parameter in $\boldsymbol{\epsilon}$ is automatically projected out on the right-hand side.

Substituting this into the lagrangian \eqref{eq:susy-lag-csm8} then gives the sought after $\fso(8)$-invariant expression
\begin{equation}
\label{eq:susy-lag-csm8-bis}
  \begin{aligned}[m]
   \eL^{N{=}8} =& -\varepsilon^{\mu\nu\rho} \left( A_\mu , \partial_\nu A_\rho + \tfrac{1}{3} [ A_\nu , A_\rho ] \right) -\tfrac{1}{2} \left< D_\mu X^I , D^\mu X^I \right> + \tfrac{1}{2} \left< {\bar {\boldsymbol{\Psi}}}^t , \gamma^\mu D_\mu \boldsymbol{\Psi} \right> \\
   &+\half ( \mu^{IJ} , T( {\bar {\boldsymbol{\Psi}}}^t , \Gamma_{IJ} \boldsymbol{\Psi} )) -\tfrac{1}{3} \left< \mu^{IJ} \cdot X^K , \mu^{IJ} \cdot X^K \right>,
  \end{aligned}
\end{equation}
whose integral is indeed invariant under the $N{=}8$ supersymmetry transformations
\begin{equation}
  \label{eq:susy8}
  \begin{aligned}[m]
    \delta X^I &= {\bar {\boldsymbol{\epsilon}}}^t \Gamma^I \boldsymbol{\Psi} \\
    \delta \boldsymbol{\Psi} &= - \gamma^\mu D_\mu X^I \, \Gamma_I \boldsymbol{\epsilon} - \tfrac{1}{3} \, \mu^{IJ} \cdot X^K \, \Gamma_{IJK} \boldsymbol{\epsilon}  \\
    \delta A_\mu &= - \half \, T ( X^I , {\bar {\boldsymbol{\epsilon}}}^t \gamma_\mu \Gamma_I \boldsymbol{\Psi} ).
  \end{aligned}
\end{equation}
The $N{=}8$ supersymmetry parameter above can be decomposed as $\boldsymbol{\epsilon} = - \tfrac{1}{8} \epsilon_{IJ} \Gamma_{IJ} \vartheta + \eta \vartheta$ with respect to the $\fso(7)$ subalgebra, where $\eta$ is a single fermionic Majorana spinor on $\RR^{1,2}$.  Setting $\eta = 0$ one recovers precisely the $N{=}7$ supersymmetry transformations in \eqref{eq:susy7}.  Mutatis mutandis, the lagrangian \eqref{eq:susy-lag-csm8-bis} and supersymmetry transformations \eqref{eq:susy8} indeed agree with those in \cite{BL2}.  To be precise, equations (45) and (42) in \cite{BL2} match up with \eqref{eq:susy-lag-csm8-bis} and \eqref{eq:susy8} by identifying their 3-bracket $[-,-,-]$ with $-\half T(-,-)\cdot -$ on $U$ and their Lie algebra inner product with $2 (-,-)$ on $\fg$ here.  Of course, the form presented in \cite{BL2} is in terms of projected Majorana spinors of $\Cl(1,10)$ broken to $\Cl(1,2) \otimes \Cl(0,8)$ and one identifies with our expressions above such that the gamma matrices of $\Cl(1,10)$ take the form $\gamma_\mu \otimes \Gamma$ and $1 \otimes i \Gamma_I$.

\section*{Acknowledgments}

We would like to thank Patricia Ritter for collaborating in the early stages of this work.  PdM would like to thank Jan de Boer, Amihay Hanany, Chris Hull and Neil Lambert for useful discussions related to this work.  PdM is supported by a Seggie-Brown Postdoctoral Fellowship of the School of Mathematics of the University of Edinburgh.  The work of JMF was supported by a World Premier International Research Center Initiative (WPI Initiative), MEXT, Japan.  JMF takes pleasure in thanking Hitoshi Murayama for the invitation to visit IPMU and the Leverhulme Trust for the award of a Research Fellowship freeing him from his duties at the University of Edinburgh.

\appendix

\section{Unitary representations of metric Lie algebras}
\label{sec:rep-theory}

The organisational principle advocated in this paper for the classification of superconformal Chern--Simons theory is the representation theory of metric Lie algebras.  This is a refinement of the usual representation theory of Lie algebras, on which it is based.  Therefore in this appendix we collect the basic facts about unitary representations of a metric Lie algebra.  We will fix once and for all a real finite-dimensional metric Lie algebra $\fg$ with ad-invariant inner product $(-,-)$.  After reviewing the useful yoga of real, complex and quaternionic unitary representations, we discuss special types of these representations which can only be defined when the Lie algebra is metric.  They turn out to be related to certain triple systems which embed in Lie (super)algebras.

\subsection{Notation and basic notions}
\label{sec:notat-basic-noti}

Unitary representations of real Lie algebras come in three types, depending on the ground field: real, complex or quaternionic.  It is possible to view all three as special classes of real or complex representations.  Let us review this briefly only to set the notation.  All our representations will be assumed finite-dimensional.

\subsubsection{Real representations}
\label{sec:real-representations}

We shall usually denote real representations by $U$ and denote the (symmetric) inner product by $\left<-,-\right>$.  This defines a Lie algebra $\fso(U)$ consisting of linear transformations of $U$ which are skewsymmetric relative to the inner product.  An orthogonal representation of $\fg$ on $U$ is a Lie algebra homomorphism $\fg \to \fso(U)$.  This means that for each $X \in \fg$ and $u,v \in U$,
\begin{equation}
  \label{eq:orthogonal}
  \left<X\cdot u, v\right> = - \left<u, X \cdot v\right>,
\end{equation}
where the $\cdot$ denotes the action of $\fg$.

\subsubsection{Complex representations}
\label{sec:compl-repr}

We shall usually denote complex representations by $V$ and denote the hermitian inner product by $h$.  This is a sesquilinear form $h: V \times V \to \CC$, which is complex linear in the first entry and complex antilinear in the second.  It defines a Lie algebra $\fu(V)$ consisting of complex linear transformations of $V$ which are skewhermitian relative to $h$.  A unitary representation of $\fg$ on the hermitian vector space $V,h$ is a Lie algebra homomorphism $\fg \to \fu(V)$, which translates to
\begin{equation}
  \label{eq:unitary}
  h(X\cdot u, v) = - h(u, X \cdot v).
\end{equation}
Given a complex representation $V$ we shall denote by $\Vbar$ the complex conjugate representation.  This shares the same underlying set as $V$, but the complex number $z\in\CC$ acts on $\Vbar$ in the way that its complex conjugate $\bar z$ acts on $V$.  It is therefore convenient notationally to parametrise the vectors in $\Vbar$ by the vectors in $V$, but denoting them as $\overline v$, where $v \in V$.  In this way, $z \overline v = \overline{\bar z v}$ and the action of $\fg$ is such that $X \cdot \overline v = \overline{X \cdot v}$.  If $V$ is a complex unitary representation, then so is $\Vbar$, with hermitian structure
\begin{equation}
  \label{eq:hermitianVbar}
  h(\overline u, \overline v) = h(v,u) = \overline{h(u,v)}.
\end{equation}

\subsubsection{Quaternionic representations as complex representations}
\label{sec:quat-repr}

For reasons which will hopefully become clearer below, the nonexistence of quaternionic Lie algebras forbid us from working over the quaternions.  Instead we will consider a quaternionic representation to be given by a complex hermitian vector space $V,h$ as in Section~\ref{sec:compl-repr} above and a quaternionic structure, that is, a complex antilinear map $J: V \to V$ obeying $J^2 = -1$ that is both $\fg$-invariant and compatible with the hermitian inner product in the sense that
\begin{equation} \label{eq:J-compat}
  h(Ju,Jv) = h(v,u).
\end{equation}
In particular, we have a complex symplectic structure
\begin{equation}
  \label{eq:complex-symplectic}
  \omega(u,v) := h(u,Jv),
\end{equation}
whence $V$ must have even complex dimension.  The Lie subalgebra of $\fu(V)$ consisting of endomorphisms which commute with $J$ is called $\fusp(V)$.  The notation stems from the fact that $\fusp(V) = \fu(V) \cap \fsp(V)$, where $\fsp(V)$ is the Lie subalgebra of $\fgl(V)$ which preserves $\omega$.  For us in this paper, a quaternionic representation of $\fg$ on $V$ is then a Lie algebra homomorphism $\fg \to \fusp(V)$.

\subsubsection{Real representations as complex representations}
\label{sec:real-as-complex}

It is possible to view real orthogonal representations as a special kind of complex unitary representations.  If $U$ is a real orthogonal representation of $\fg$, then let $U_\CC = \CC \otimes_\RR U$ denote its complexification.  This becomes a complex representation of $\fg$ by declaring $X \cdot (z \otimes u) = z \otimes (X \cdot u)$ for all $z \in \CC$ and $u \in U$ and letting $z'\in\CC$ act on $U_\CC$ by $z' (z \otimes u) = (z'z)\otimes u$.  It is clear that this action of $\fg$ is complex linear.  We define a hermitian structure on $U_\CC$ by extending the following additively:
\begin{equation}
  \label{eq:h-from-real}
  h(z_1 \otimes u, z_2 \otimes v) = z_1 \overline{z_2} \left<u,v\right>,
\end{equation}
which makes it clear that $\fg$ preserves $h$ as well, whence $U_\CC,h$ is a complex unitary representation of $\fg$.

Conversely, complex unitary representations $V$ which are of this type are characterised by the existence of a real structure, that is, a complex antilinear map $R: V \to V$ satisfying $R^2 = 1$ that both commutes with the action of $\fg$ and is compatible with the hermitian structure in that
\begin{equation}
  \label{eq:c-compat}
  h(Ru,Rv) = h(v,u).
\end{equation}
In the case of $U_\CC$, $R$ is simply complex conjugation: $R(z \otimes u) = \overline{z}\otimes u$.  In the abstract case, if $V,h,R$ is such a representation, then we may define $U$ to be the eigenspace of $R$ with eigenvalue $1$.  Since $R$ is complex antilinear, this is only a real subspace of $V$.  The eigenspace with eigenvalue $-1$ is given by $iU$, whence $V = U \oplus iU \cong U_\CC$.  On $V$ we can define a $\fg$-invariant complex bilinear inner product by
\begin{equation}
  \label{eq:c-bilinear-IP}
  b(u,v) = h(u,Rv),
\end{equation}
whose restriction to $U$, by condition \eqref{eq:c-compat}, is real and hence makes $U$ into a real orthogonal representation.  We will often employ the notation $U = [V]$ for this sort of real representations.

\subsubsection{Complex and quaternionic representations as real representations}
\label{sec:compl-quat-repr}

Finally, it is possible to discuss both complex and quaternionic unitary representations as particular cases of real orthogonal representations.  A complex unitary representation $V,h$ gives rise to a real representation by simply restricting scalars to the real numbers.  Relative to a basis, one is simply taking real and imaginary parts of the vectors in $V$.  This real representation is usually denoted $\rf{V}$ to remind us that its real dimension is twice the complex dimension of $V$.  Multiplication by $i$ then defines a real linear endomorphism $I$ of $\rf{V}$, satisfying $I^2 = -1$; that is, a complex structure.  Since $\fg$ acts complex linearly on $V$, it commutes with the action of $I$.  The real part of $h$ defines a symmetric inner product $\left<-,-\right>$ on $\rf{V}$, relative to which $I$ is orthogonal.

Conversely if $U$ is a real orthogonal representation with a $\fg$-invariant orthogonal complex structure $I$, then we can define on $U$ the structure of a complex vector space by having the complex number $a + i b \in \CC$ act on $u\in U$ by $(a+i b) u = a u + b I u$.  We can also define a hermitian structure $h$ on $U$ by \begin{equation}
  \label{eq:h-from-I}
  h(u,v) = \left<u,v\right> + i \left<u, I v\right>.
\end{equation}

Similarly, a quaternionic unitary representation is a special type of real orthogonal representation $U$ where we have two invariant orthogonal complex structures $I$ and $J$ which anticommute, in that $IJ = -JI$.  This allows us to define a complex symplectic structure by
\begin{equation}
  \label{eq:omega-from-I-and-J}
  \omega(u,v) = \left<u, J v\right> + i \left<u, I J v\right>.
\end{equation}
If so inclined, we could now define on $U$ the structure of a quaternionic vector space with all the trimmings, but we will not do so in this paper.

\subsubsection{Reality conditions}
\label{sec:reality}

If $W_1$ and $W_2$ are two quaternionic unitary representations in the sense of Appendix~\ref{sec:quat-repr}, then their tensor product $W_1 \otimes_\CC W_2$ is a complex representation with a real structure.  Indeed, if $J_1,J_2$ are the quaternionic structure maps of $W_1,W_2$, respectively, then $R = J_1 \otimes J_2$ is a real structure map.  The underlying real representation $[W_1\otimes_\CC W_2]$ is the real subspace given by the eigenspace of $R$ with eigenvalue $1$.

A common device to obtain a real representation out of a quaternionic representation $W$ is to tensor with the quaternions, understood as a trivial quaternionic representation.  Indeed, give $\HH$ the structure of a two-dimensional complex representation by having $\CC$ act on the right and consider $V = \HH \otimes_\CC W$.  Let $J$ denote the quaternionic structure on $W$ and $j$ that on $\HH$ and let $R = j \otimes J$.  As a representation of $\fg$, $V \cong W \oplus W$, with the isomorphism being given explicitly by \begin{equation}
  \label{eq:HW-isom-WW}
  1 \otimes w_1 + j \otimes w_2 \leftrightarrow \begin{pmatrix} w_1 \\ w_2  \end{pmatrix}.
\end{equation}
In this picture, the conjugation $c$ acts by extending the following expression complex antilinearly:
\begin{equation}
  R \begin{pmatrix}  w_1 \\ w_2  \end{pmatrix} = \begin{pmatrix} - Jw_2 \\ Jw_1  \end{pmatrix}.
\end{equation}
The real subspace $U$ is the eigenspace of $c$ with eigenvalue $1$, which means that $w_2 = Jw_1$, so that
\begin{equation}
  U = \left\{ \begin{pmatrix} w \\ Jw  \end{pmatrix} \middle |  w \in W\right\}
\end{equation}
is the graph of $J$ as a subspace of $W \oplus W$.  Since $J$ is antilinear, $U$ is only a real subspace.

\subsubsection{Relations between real, complex and quaternionic representations}
\label{sec:relat-betw-real}

As seen above, there are natural maps between representations obtained by altering the ground field.  We can extend, restrict and, in the case of the complex field, also conjugate.  These maps are summarised succinctly in the following (noncommutative!) diagram, borrowed from \cite{Adams} via \cite{MR781344}:
\begin{equation}
  \label{eq:adams}
  \xymatrix{ & \Dar(\fg,\CC) & \\
    \Dar(\fg,\RR) \ar[ur]^c & & \Dar(\fg,\HH) \ar[ul]_{r'} \\
    & \Dar(\fg,\CC) \ar[uu]^t \ar[ul]^r \ar[ur]_q }
\end{equation}
where $\Dar(\fg,\KK)$ means the category of representations of the Lie algebra $\fg$ over the field $\KK$, and the arrows denote the following functors:
\begin{itemize}
\item[$t$:] if $V \in \Dar(\fg,\CC)$, then $t(V) = \Vbar$ denote the complex conjugate representation;
\item[$q$:] if $V \in \Dar(\fg,\CC)$, then $q(V) = V_\HH = \HH \otimes_\CC V \in \Dar(\fg,\HH)$, where $\HH$ is a right $\CC$-module;
\item[$c$:] if $U\in \Dar(\fg,\RR)$, then $c(U) = U_\CC = \CC \otimes_\RR U$ is its complexification;
\item[$r$:] if $V \in \Dar(\fg,\CC)$, then $r(V) = \rf{V} \in \Dar(\fg,\RR)$ is obtained by restricting scalars; and
\item[$r'$:] if $W \in \Dar(\fg,\HH)$, then $r'(W) = \rh{W} \in \Dar(\fg,\CC)$ is obtained by restricting scalars.
\end{itemize}
The map $t$ does not change the dimension, and neither do $c$ or $q$ in the sense that $\dim_\CC U_\CC = \dim_\RR U$ and $\dim_\HH V_\HH = \dim_\CC V$.  However $r$ and $r'$ double the dimension: $\dim_\RR \rf{V} = 2 \dim_\CC V$ and $\dim_\CC \rh{W} = 2 \dim_\HH W$.  In this paper we are not working with quaternionic representations themselves but with their image under $r'$; although this will not always be reflected in our notation.  In other words, we will often write simply $W$ for $\rh{W}$ if in so doing the possibility of confusion is minimal.

The above maps obey some relations, which are more or less obvious (see \cite[Proposition~3.6]{Adams} or \cite[Proposition~(6.1)]{MR781344}).

\begin{proposition}
\label{prop:diamond}
  The following relations hold:
  \begin{enumerate}
  \item $t^2 = 1$ or $\overline{\overline{V}} \cong V$ for all $V \in \Dar(\fg,\CC)$;
  \item $tc = c$ or $\overline{U_\CC} \cong U_\CC$ for all $U \in \Dar(\fg,\RR)$;
  \item $qt = q$ or ${\overline V}_\HH \cong V_\HH$ for all $V \in \Dar(\fg,\CC)$;
  \item $rc = 2$ or $\rf{U_\CC} \cong U \oplus U$ for all $U \in \Dar(\fg,\RR)$;
  \item $cr = 1 + t$ or $\rf{V}_\CC \cong V \oplus \Vbar$ for all $V \in \Dar(\fg,\CC)$;
  \item $rt = r$ or $\rf{\Vbar} \cong \rf{V}$ for all $V \in \Dar(\fg,\CC)$;
  \item $t r' = r'$ or $\overline{\rh{W}} \cong \rh{W}$ for all $W \in \Dar(\fg,\HH)$;
  \item $q r' = 2$ or $\rh{W}_\HH \cong W \oplus W$ for all $W \in \Dar(\fg,\HH)$; and
  \item $r' q = 1 + t$ or $\rh{V_\HH} \cong V \oplus \Vbar$ for all $V \in \Dar(\fg,\CC)$.
\end{enumerate}
\end{proposition}

Recall that a real (resp. complex, quaternionic) representation is \textbf{irreducible} if it admits no proper real (resp. complex, quaternionic) subrepresentations.  We shall denote by $\Irr(\fg,\KK)$ the irreducible representations of $\fg$ of type $\KK$.  The following proposition states what happens to irreducible representations under the above maps.  It is not always the case that irreducibles go to irreducibles, but their images are under control in any case.

\begin{proposition}
\label{prop:irreducible}
  The following hold:
  \begin{enumerate}
  \item $V \in \Irr(\fg,\CC) \iff \Vbar \in \Irr(\fg,\CC)$;
  \item if $U \in \Irr(\fg,\RR)$ then $U_\CC \in \Irr(\fg,\CC)$, unless $U = \rf{V}$ for some $V\in\Irr(\fg,\CC)$, in which case $U_\CC \cong V \oplus \Vbar$;
  \item if $V\in \Irr(\fg,\CC)$ then $\rf{V} \in \Irr(\fg,\RR)$, unless $V = U_\CC$ for some $U\in\Irr(\fg,\RR)$, in which case $\rf{V} \cong U \oplus U$;
  \item if $W \in \Irr(\fg,\HH)$ then $\rh{W} \in \Irr(\fg,\CC)$, unless $W = V_\HH$ for some $V\in\Irr(\fg,\CC)$, in which case $\rh{W} \cong V \oplus \Vbar$; and
  \item if $V\in \Irr(\fg,\CC)$ then $V_\HH \in \Irr(\fg,\HH)$, unless $V = \rh{W}$ for some $W \in \Irr(\fg,\HH)$, in which case $V_\HH \cong V \oplus V$.
  \end{enumerate}
\end{proposition}

See \cite[Proposition~(6.6)]{MR781344} for a partial proof.

\subsection{Lie-embeddable representations of metric Lie algebras}
\label{sec:special-reps}

We will now use that the Lie algebra $\fg$ admits an ad-invariant inner product $(-,-)$.  This allows us to distinguish certain privileged types of unitary representations, which are summarised in Table~\ref{tab:special} in the Introduction.  For reasons that will become clear in Appendix~\ref{sec:embedd-lie-super}, we call them \textbf{Lie-embeddable representations} in this paper; although we are not claiming that these are the only representations which could be given this name.

\subsubsection{Real orthogonal representations}
\label{sec:spec-real-orth}

Let $U \in \Dar(\fg,\RR)$ with corresponding homomorphism $\fg \to \fso(U)$.  Using the inner products on $U$ and on $\fg$ we may dualise this map to arrive at a $\fg$-equivariant bilinear map $T: U \times U \to \fg$.  Explicitly, we have that for all $u,v \in U$ and $X \in \fg$,
\begin{equation}
  \label{eq:T-map-R}
  \left(T(u, v), X\right) = \left<X\cdot u, v\right>,
\end{equation}
from where it follows at once that it is alternating:
\begin{equation}
  \label{eq:T-skew}
  T(v,w) = - T(w,v).
\end{equation}
It also follows that
\begin{equation}
  \label{eq:real-symmetry}
  \left(T(u, v), T(w,x)\right) = \left<T(w,x)\cdot u, v\right> = \left<T(u,v)\cdot w, x\right>,
\end{equation}
whence the fourth-rank tensor $\eR(u,v,w,x) := \left<T(u,v)\cdot w, x\right>$ belongs to
\begin{equation}
  \label{eq:real-rank4-rep}
  S^2 \Lambda^2 U \cong \Lambda^4 U \oplus U^{\yng(2,2)}.
\end{equation}
For generic real orthogonal representations $U$, the tensor $\eR \in S^2\Lambda^2 U$ will have components in both representations, but for some special representations one of the components will vanish.  There are thus two classes of special real orthogonal representations to be described below.
In order to explain the names we will give to these special classes, it will be convenient to review the construction in \cite{Lie3Algs}.

The map $T$ above allows us to define a trilinear product on $U$ by
\begin{equation}
  \label{eq:3-bracket-R}
  [u,v,w] := T(u,v) \cdot w
\end{equation}
for all $u,v,w \in U$, defining a triple system on $U$.  The resulting triple system, which appeared originally in \cite{FaulknerIdeals} but more recently in \cite{CherSaem} in the context of superconformal Chern--Simons-matter theories, satisfies the following axioms for all $x,y,z,v,w\in U$:
\begin{enumerate}
\item the \emph{unitarity} condition
  \begin{equation}
    \label{eq:unitarity-R}
    \left<[x,y,z], w\right> =  - \left<z, [x,y,w]\right>;
  \end{equation}
\item the \emph{symmetry} condition
  \begin{equation}
    \label{eq:symmetry-R}
    \left<[x,y,z], w\right> = \left<[z,w,x], y\right>;
  \end{equation}
\item and the \emph{fundamental identity}
  \begin{equation}
    \label{eq:FI-R}
    [x,y,[v,w,z]] - [v,w,[x,y,z]] = [[x,y,v],w,z] + [v,[x,y,w],z].
  \end{equation}
\end{enumerate}
The fundamental identity is simply the $\fg$-equivariance of the map $T$, whereas the symmetry condition is a rewriting of equation \eqref{eq:real-symmetry} and the unitarity condition is just the fact that $\fg$ preserves the inner product.

The first class of special representations consists of those where $\eR \in U^{\yng(2,2)}$, which is equivalent to the Bianchi-like identity
\begin{equation}
  \label{eq:real-LTS}
  T(u,v)\cdot w + T(v,w)\cdot u + T(w,u)\cdot v = 0,
\end{equation}
or, in terms of the 3-bracket,
\begin{equation}
  [u,v,w] + [v,w,u] + [w,u,v] = 0.
\end{equation}
Such a 3-bracket defines on $U$ the structure of a \textbf{Lie triple system} (LTS) and hence we say that the unitary representation $U$ is \textbf{LTS}, written $U \in \Dar(\fg,\RR)_{\text{LTS}}$.  Lie triple systems are linear approximations to riemannian symmetric spaces and indeed the tensor $\eR$ is nothing but the Riemann curvature tensor.

The other special class consists of representations $U$ where $\eR \in \Lambda^4U$ or, equivalently, for which the 3-bracket if totally skewsymmetric.  Such triple systems are known as \textbf{3-Lie algebras} (3LA) \cite{Filippov} and hence such representations are said to be \textbf{3LA}, written $U \in \Dar(\fg,\RR)_{\text{3LA}}$.  As conjectured in \cite{FOPPluecker} and shown in \cite{NagykLie,GP3Lie,GG3Lie}, there is a unique positive-definite representation $U \in \Irr(\fg,\RR)_{\text{3LA}}$, corresponding to $\fg = \fso(4) = \fso(3) \oplus \fso(3)$ with the inner product being equal to the Killing form on one of the $\fso(3)$ and the negative of the Killing form in the other, and $U = \RR^4$ being the vector representation.  Dropping the positive-definite condition, one obtains many such representations, of which there are a number of partial classifications \cite{Lor3Lie,2p3Lie,2pBL}.  They are the relevant representations for $N{=}8$ supersymmetry and the totally skewsymmetric 3-bracket is precisely the same as that which figured in the original description of the $N{=}8$ theory of Bagger and Lambert \cite{BL1,BL2} and Gustavsson \cite{GustavssonAlgM2}.

\subsubsection{Complex unitary representations}
\label{sec:spec-compl-unit}

Let $V \in \Dar(\fg,\CC)$ with corresponding homomorphism $\fg \to \fu(V)$, whose transpose is now a sesquilinear map $\TT: V \times V \to \fg_\CC$ to the complexification $\fg_\CC = \CC \otimes_\RR \fg$ of $\fg$.  We extend the bracket and the inner product on $\fg$ complex bilinearly in such a way as to make $\fg_\CC$ into a complex metric Lie algebra.  In this way, the map $\TT$ is defined explicitly by
\begin{equation}
  \label{eq:T-map-C}
  \left(\TT(u,v),\XX\right) = h(\XX \cdot u, v),
\end{equation}
for all $u,v \in V$ and $\XX \in \fg_\CC$.  In particular,
\begin{equation}
  \label{eq:complex-symmetry}
  \left(\TT(u, v), \TT(w,x)\right) = h(\TT(w,x)\cdot u, v) = h(\TT(u,v)\cdot w, x),
\end{equation}
whence the fourth-rank tensor $\eR(u,v,w,x) := h(\TT(u,v)\cdot w, x)$ belongs to
\begin{equation}
  \label{eq:complex-rank4-rep}
  S^2 (V \otimes \Vbar) \cong \left(S^2V \otimes S^2\Vbar\right) \oplus \left(\Lambda^2V \otimes \Lambda^2\Vbar\right)
\end{equation}
For generic $V\in\Dar(\fg,\CC)$, the tensor $\eR \in S^2(V \otimes \Vbar)$ will have both components, but for special representations one of the components will vanish.  There are two such types of special complex unitary representations which we describe below.

The map $\TT$ above allows us to define a complex sesquibilinear product on $V$ by
\begin{equation}
  \label{eq:3-bracket-C}
  [\![u,v,w]\!] := \TT(u,v) \cdot w
\end{equation}
for all $u,v,w \in V$.  This means that it is complex linear in $u,w$ and complex antilinear in $v$.  In the notation of \cite{Lie3Algs}, which was chosen to  ease the comparison with \cite{BL4}, the bracket $[\![u,v,w]\!] = [w,u;v]$.  The bracket \eqref{eq:3-bracket-C} satisfies a number of identities.  For all $x,y,z,v,w \in V$, we have
\begin{enumerate}
\item the \emph{unitarity} condition
  \begin{equation}
    \label{eq:unitarity-C}
    h([\![v,w,x]\!],y) = h(x,[\![w,v,y]\!]);
  \end{equation}
\item the \emph{symmetry} condition
  \begin{equation}
    \label{eq:symmetry-C}
    h([\![v,w,x]\!],y) = h([\![x,y,v]\!],w);
  \end{equation}
\item and the \emph{fundamental identity}
  \begin{equation}
    \label{eq:FI-C}
    [\![x,y,[\![v,w,z]\!]]\!] - [\![v,w,[\![x,y,z]\!]]\!] = [\![[\![x,y,v]\!],w,z]\!] - [\![v,[\![y,x,w]\!],z]\!].
  \end{equation}
\end{enumerate}
Again the fundamental identity is basically the $\fg$-equivariance of $\TT$, whereas the other two identities follow as before from the fact that $\overline{\TT(u,v)} = - \TT(v,u)$ and that for all $\XX \in \fg_\CC$, we have that
\begin{equation}
  h(\XX\cdot u, v) = - h(u, \overline{\XX} \cdot v),
\end{equation}
and from equation \eqref{eq:complex-symmetry}.

The first special class is when $\eR \in S^2V \otimes S^2\Vbar$, which corresponds to those $V$ where
\begin{equation}
  \label{eq:T-HSS}
  \TT(u,v)\cdot w = \TT(w,v)\cdot u 
\end{equation}
or equivalently where $[\![u,v,w]\!] = [\![w,v,u]\!]$.  Such a bracket defines on $V$ the structure of a \textbf{Jordan triple system} (JTS) \cite{JacobsonLTS} and we will say such a representation $V$ is \textbf{JTS}, written $V \in \Dar(\fg,\CC)_{\text{JTS}}$.  These representations are linear approximations to hermitian symmetric spaces and, as in case of Lie triple systems, the tensor $\eR$ coincides with the Riemann curvature tensor, this time on the complexified tangent bundle of the symmetric space.

The other special class is when $\eR \in \Lambda^2V \otimes \Lambda^2\Vbar$, which corresponds to those $V$ where
\begin{equation}
  \label{eq:T-BL4}
  \TT(u,v)\cdot w = - \TT(w,v)\cdot u
\end{equation}
or, equivalently, where $[\![u,v,w]\!] = - [\![w,v,u]\!]$.  Such a 3-bracket defines on $V$ the structure of an \textbf{anti-Jordan triple system} (aJTS) (see, e.g., \cite[Remark~4.3]{FaulknerFerrarAJP}) and hence we will say that these representations are \textbf{aJTS}, written $V \in \Dar(\fg,\CC)_{\text{aJTS}}$.  They are the relevant representations for $N{=}6$ supersymmetry and the skewsymmetry condition \eqref{eq:T-BL4} identifies this class of triple systems with those used by Bagger and Lambert in \cite{BL4} to recover the $N{=}6$ theories discovered by Aharony, Bergman, Jafferis and Maldacena in \cite{MaldacenaBL}.

\subsubsection{Complex representations in real terms}
\label{sec:spec-complex-as-real}

It is useful to rewrite the complex unitary construction above in terms of real representations.  Let $V\in\Dar(\fg,\CC)$ and let $U=\rf{V} \in \Dar(\fg,\RR)$ with inner product given by the real part of the hermitian inner product on $V$.  Complex multiplication in $V$ is implemented via an orthogonal $\fg$-invariant complex structure $I : U \to U$.  This means that $I^2 = -1$, $\left<Iv,Iw\right>=\left<v,w\right>$ and $X \cdot (Iv) = I(X\cdot v)$ for all $v,w \in U$ and $X \in \fg$.  In particular, $\left<Iv,w\right> = - \left<v,Iw\right>$, whence $I$ is both orthogonal and skewsymmetric.  In terms of $\left<-,-\right>$, the hermitian structure is given by equation \eqref{eq:h-from-I}.

Let $\fg_\CC$ denote the complexification of $\fg$, which we make into a complex metric Lie algebra by extending both the Lie bracket and the ad-invariant inner product $\left(-,-\right)$ complex bilinearly.  There is a natural action of $\fg_\CC$ on $U$ as follows.  If $\XX = X + i Y \in \fg_\CC$, with $X,Y\in \fg$, then for all $v \in U$,
\begin{equation*}
  (X + i Y)\cdot v = X \cdot v + I (Y \cdot v) = X \cdot v + Y \cdot (Iv),
\end{equation*}
the second equation following from the $\fg$-invariance of $I$.  The real bilinear map $\TT: U \times U \to \fg_\CC$ defined in equation \eqref{eq:T-map-C} is written in reals terms as
\begin{equation}
  \label{eq:T-map-C-as-R}
  \left(\TT(u,v),\XX\right) = \left<\XX \cdot u, v\right> + i \left<\XX \cdot u, Iv\right>,
\end{equation}
for all $u,v\in U$, $\XX \in \fg_\CC$.

\begin{lemma}\label{le:T-map-C}
  For all $u,v\in U$, we have $\TT(u,v) = T(u,v) + i T(u, Iv)$.
\end{lemma}

\begin{proof}
  Write $\TT(u,v) = A(u,v) + i B(u,v)$ and expand equation \eqref{eq:T-map-C} for $\XX = X + i Y$ complex bilinearly.  The left-hand side becomes
  \begin{align*}
    \left(\TT(u,v),\XX\right) &= \left(A(u,v) + i B(u,v),X + i Y\right) \\
    &= \left(A(u,v),X \right) + i \left(A(u,v), Y\right)  + i \left(B(u,v),X \right)  - \left(B(u,v), Y\right),
  \end{align*}
  whereas the right-hand side becomes
  \begin{align*}
    h(\XX \cdot u, v) &= \left<(X + i Y) \cdot u,v\right> + i \left<(X + i Y) \cdot u,Iv\right>\\
    &= \left<X \cdot u,v\right> + \left<I (Y \cdot u),v\right> + i \left<X \cdot u,Iv\right> +  i \left<I (Y \cdot u),Iv\right>\\
    &= \left<X \cdot u,v\right> - \left<Y \cdot u, I v\right> + i \left<X \cdot u, Iv \right> +  i \left<Y \cdot u,v\right>.
  \end{align*}
  Comparing real and imaginary parts and the terms depending on $X$ and $Y$, we arrive at the following two equations:
  \begin{equation*}
    \left(A(u,v),X \right)  = \left<X \cdot u,v\right> \qquad\text{and}\qquad \left(B(u,v),X \right)  = \left<X \cdot u, Iv \right>.
  \end{equation*}
  Now using equation \eqref{eq:T-map-R}, we see that the first of the above equations says that $A(u,v) = T(u,v)$, whereas the second equation says that $B(u,v) = T(u, I v)$.
\end{proof}

\begin{proposition}\label{pr:T-map-C}
  The map $\TT: V \times V \to \fg_\CC$ satisfies the following properties
  \begin{equation*}
   \overline{\TT(u,v)} = - \TT(v,u)\qquad  \TT(I u,v) = i \TT(u,v) \qquad \TT(u, I v) = -i \TT(u,v).
  \end{equation*}
\end{proposition}

\begin{proof}
  Again it is sufficient to prove the first two.  Notice first of all that
  \begin{align*}
    \left(T(u,Iv),X\right) &= \left<X \cdot u , I v\right>  && \text{by equation \eqref{eq:T-map-R}}\\
    &= - \left<I (X \cdot u) , v\right> && \text{by skewsymmetry of $I$}\\
    &= - \left<X \cdot I u , v\right> && \text{by $\fg$-invariance of $I$}\\
    &= - \left(T(Iu,v),X\right) && \text{again by equation \eqref{eq:T-map-R},}
  \end{align*}
  whence
  \begin{equation}
    \label{eq:T-I}
    T(u,Iv) = - T(Iu, v).
  \end{equation}
  Together with equation \eqref{eq:T-skew}, we have in addition that
  \begin{equation}
    \label{eq:T-I-too}
    T(u,Iv) = T(v,Iu).
  \end{equation}
  To show the first identity, we calculate
  \begin{align*}
    \overline{\TT(u,v)} &= \overline{T(u,v) + i T(u,Iv)}\\
    &= T(u,v) - i T(u,Iv)\\
    &= - T(v,u) -i T(v,Iu) && \text{using equations \eqref{eq:T-skew} and \eqref{eq:T-I-too}}\\
    &= - \TT(v,u);
  \end{align*}
  and to show the second, we calculate
  \begin{align*}
    \TT(I u,v) &= T(Iu,v) + i T(Iu,Iv)\\
    &= - T(u,Iv) + i T(u,v) && \text{using equation \eqref{eq:T-I}}\\
    &= i ( T(u,v) + i T(u,Iv) ) \\
    &= i \TT(u,v).
  \end{align*}
\end{proof}

We again define a 3-bracket on $V$ by equation \eqref{eq:3-bracket-C}, which, by Lemma \ref{le:T-map-C}, can be rewritten as
\begin{equation}
  \label{eq:3-brackets-R-C}
  [\![u,v,w]\!] = [u,v,w] + [u,Iv,Iw],
\end{equation}
in terms of the 3-bracket $[u,v,w]$ in \eqref{eq:3-bracket-R}.  The bracket \eqref{eq:3-bracket-C} of course satisfies the unitarity \eqref{eq:unitarity-C}, symmetry \eqref{eq:symmetry-C} and fundamental \eqref{eq:FI-C} identities, but in addition, as an easy consequence of Proposition~\ref{pr:T-map-C}, the following:
\begin{equation*}
  [\![I u,v,w]\!] =  I [\![u,v,w]\!] \qquad   [\![u, Iv,w]\!] =  - I [\![u,v,w]\!] \qquad   [\![u,v,I w]\!] =  I [\![u,v,w]\!],
\end{equation*}
which is simply the sesquibilinearity of the 3-bracket.

\subsubsection{Quaternionic unitary representations}
\label{sec:spec-quat-unit}

Let $W,h,J$ be a quaternionic unitary representation.  Again we recall that for us in this paper this means that $W,h$ is a complex hermitian vector space and $J$ a quaternionic structure map.  The reason we work with complex representations with a quaternionic structure instead of a quaternionic representation should now become clear.  In the case of an honest quaternionic representation, the dualising procedure employed here would map to a quaternionification of the Lie algebra, but such an object does not exist; although see \cite{MR1877855} for a possibly related concept.  Let $\TT$ be the sesquilinear map defined in \eqref{eq:T-map-C}.  By suitably inserting the quaternionic structure map, one can turn $\TT$ into a complex bilinear map: $(u,v)\mapsto \TT(u,Jv)$.  This map can be understood as the transpose of the representation homomorphism $\fg \to \fusp(W)$ with respect to the complex symplectic structure $\omega$ defined in \eqref{eq:complex-symplectic}:
\begin{equation}
  \label{eq:T-map-H}
  \left(\TT(u,Jv), \XX\right) = \omega(\XX\cdot u, v).
\end{equation}
Since $\omega$ is symplectic, it follows that
\begin{equation}
  \label{eq:TTJ-symmetric}
  \TT(u,Jv) = \TT(v,Ju),
\end{equation}
whence it defines a symmetric map $V \times V \to \fg_\CC$.  Compatibility with the quaternionic structure says that $J \circ \TT(u,Jv) = \overline{\TT(u,Jv)} \circ J$, where $\overline{\TT(u,Jv)} = -\TT(Jv,u)$.  Again we have that
\begin{equation}
  \label{eq:quat-symmetry}
  \left(\TT(u, Jv), \TT(w,Jx)\right) = \omega(\TT(w,Jx)\cdot u, v) = \omega(\TT(u,Jv)\cdot w, x),
\end{equation}
whence the fourth-rank tensor $\eR(u,v,w,x) := \omega(\TT(u,Jv)\cdot w, x)$ belongs to
\begin{equation}
  \label{eq:quat-rank4-rep}
  S^2 S^2 W \cong W^{\yng(2,2)} \oplus S^4 W.
\end{equation}
For generic quaternionic unitary representations $W$, the tensor $\eR \in S^2S^2 W$ will have components in both representations, but for the special representations one of the components will vanish.  This defines two special classes: one where $\eR \in S^4W$ and one where $\eR \in W^{\yng(2,2)}$.

Again inserting $J$ in the complex sesquilinear 3-bracket \eqref{eq:3-bracket-C}, defines a complex complex trilinear 3-bracket on $W$ by $(u,v,w)\mapsto [\![u,Jv,w]\!]$, for all $u,v,w \in W$.  The resulting triple system satisfies the axioms inherited from \eqref{eq:unitarity-C}, \eqref{eq:symmetry-C} and \eqref{eq:FI-C} and the compatibility with the quaternionic structure --- namely,
\begin{enumerate}
\item the \emph{symplectic} condition
  \begin{equation}
    \label{eq:symplectic-H}
    \omega([\![x,Jy,z]\!],w) = \omega([\![x,Jy,w]\!],z);
  \end{equation}
\item the \emph{symmetry} condition
  \begin{equation}
    \label{eq:symmetry-H}
    \omega([\![x,Jy,z]\!],w) = \omega([\![z,Jw,x]\!],y);
  \end{equation}
\item the \emph{fundamental identity}
  \begin{equation}
    \label{eq:FI-H}
    [\![x,Jy,[\![v,Jw,z]\!]]\!] - [\![v,Jw,[\![x,Jy,z]\!]]\!] = [\![[\![x,Jy,v]\!],Jw,z]\!] + [\![v,J[\![x,Jy,w]\!],z]\!],
  \end{equation}
\item and the \emph{quaternionic} condition
  \begin{equation}
    \label{eq:bracket-J}
    J[\![x,Jy,z]\!] = - [\![Jx, y, Jz]\!].
  \end{equation}
\end{enumerate}
It follows from the symplectic and symmetry conditions that for all $x,y,z\in W$,
\begin{equation}
  \label{eq:symmetry-3bracket}
  [\![x,Jy,z]\!] = [\![y,Jx,z]\!],
\end{equation}
which is precisely the symmetry property \eqref{eq:TTJ-symmetric} noted above.

The first special class is when $\eR \in S^4W$, so that the 3-bracket is totally symmetric.  We will see that they correspond to hyperkähler symmetric spaces and hence this class is trivial in positive-definite signature \cite{MR1913815}.  Nevertheless we will refer to the triple systems that they give rise to as \textbf{quaternionic triple systems} (QTS), for lack of a more appropriate name, and the representations will be said to be \textbf{QTS}, written $W \in \Dar(\fg,\HH)_{\text{QTS}}$.  The tensor $\eR$ is related to the Riemann curvature tensor of the symmetric space.

The other special class corresponds to the case $\eR \in W^{\yng(2,2)}$, or equivalently,
\begin{equation}
  \label{eq:quat-aLTS}
  \TT(u,Jv) \cdot w + \TT(v,Jw) \cdot u + \TT(w,Ju) \cdot v = 0,
\end{equation}
or in terms of the 3-bracket
\begin{equation}
  [\![u,Jv,w]\!] + [\![v,Jw,u]\!] + [\![w,Ju,v]\!] = 0.
\end{equation}
Equivalently, taking the symmetry conditions \eqref{eq:TTJ-symmetric} or \eqref{eq:symmetry-3bracket} into account, we may write the above two conditions as
as
\begin{equation}
  \TT(u,Ju) \cdot u = 0 \qquad\text{and}\qquad [\![u,Ju,u]\!] = 0,
\end{equation}
respectively, for all $u \in W$.  Either of these conditions defines a \textbf{(quaternionic) anti-Lie triple system} (aLTS) and hence we wll say that $W$ is an \textbf{aLTS} representation, written $W \in \Dar(\fg,\HH)_{\text{aLTS}}$.  They are the relevant representations for the $N{=}4,5$ theories.

\subsubsection{Some relations between these representations}
\label{sec:some-relat-betw}

Some of the special cases are related to each other via the maps in Appendix~\ref{sec:relat-betw-real} consistent with the requirements of supersymmetry of the corresponding Chern--Simons theory.  Let us formalise the notation we have been using until now and let $U$, $V$ and $W$ stand for a real, complex or quaternionic representation, respectively.  Remember, however, that for us quaternionic representations are always complex representations in the image of $r'$.  Let $\Dar(\fg,\KK)_{\text{C}}$ denote the unitary representations of $\fg$ of type $\KK$ and class $\text{C}$, where $\KK = \RR, \CC, \HH$ and $\text{C}$ can be either 3LA, LTS, aJTS, JTS, aLTS or QTS.

\begin{proposition}
\label{prop:relations}
  The following relations hold:
  \begin{enumerate}
  \item $V \in \Dar(\fg,\CC)_{\text{aJTS}} \iff \Vbar \in \Dar(\fg,\CC)_{\text{aJTS}}$
  \item $V \in \Dar(\fg,\CC)_{\text{JTS}} \iff \Vbar \in  \Dar(\fg,\CC)_{\text{JTS}}$
  \item $V \in \Dar(\fg,\CC)_{\text{aJTS}} \Leftarrow \rf{V} \in \Dar(\fg,\RR)_{\text{3LA}}$
  \item $V \in \Dar(\fg,\CC)_{\text{JTS}} \iff \rf{V} \in \Dar(\fg,\RR)_{\text{LTS}}$
  \item $U_\CC \in \Dar(\fg,\CC)_{\text{JTS}} \implies U$ is trivial
  \item $U \in \Dar(\fg,\RR)_{\text{3LA}} \iff U_\CC \in \Dar(\fg,\CC)_{\text{aJTS}}$
  \item $W \in \Dar(\fg,\HH)_{\text{QTS}} \iff \rh{W} \in \Dar(\fg,\CC)_{\text{JTS}}$
  \item $\rh{W} \in \Dar(\fg,\CC)_{\text{aJTS}} \implies W$ is trivial
  \item $V \in \Dar(\fg,\CC)_{\text{aJTS}} \iff V_\HH \in \Dar(\fg,\HH)_{\text{aLTS}}$
  \end{enumerate}
\end{proposition}

\begin{proof}
  First of all (a) and (b) are obvious because in both cases the fourth-rank tensor $\eR$ lives in a self-conjugate representation: $\Lambda^2V \otimes \Lambda^2\Vbar$ or $S^2V \otimes S^2\Vbar$, respectively.

  To prove (c), let $V \in \Dar(\fg,\CC)$ and let $U = \rf{V} \in \Dar(\fg,\RR)_{\text{3LA}}$.  The relation between $\TT$ on $V$ and $T$ on $\rf{V}$ is given by Lemma~\ref{le:T-map-C}, whence
  \begin{align*}
    \TT(u,v)\cdot w &= T(u,v) \cdot w + T(u,Iv)\cdot Iw \\
    &= T(u,v) \cdot w + I T(u,Iv)\cdot w &&\text{since $I$ is $\fg$-invariant}\\
    &= - T(w,v) \cdot u - I T(w,Iv)\cdot u && \text{since $r(V) \in \Dar(\fg,\RR)_{\text{3LA}}$}\\
    &= - T(w,v) \cdot u - T(w,Iv)\cdot I u && \text{since $I$ is $\fg$-invariant}\\
    &= - \TT(w,v) \cdot u,
  \end{align*}
  whence $V \in \Dar(\fg,\CC)_{\text{aJTS}}$.

  We prove (d) along similar lines.  In one direction, let $\rf{V} \in \Dar(\fg,\RR)_{\text{LTS}}$ and use Lemma~\ref{le:T-map-C} and the $\fg$-invariance of $I$ to calculate
  \begin{equation*}
    \TT(u,v)\cdot w - \TT(w,v)\cdot u = T(u,v)\cdot w + I T(u,Iv)\cdot w - T(w,v)\cdot u - I T(w,Iv)\cdot u.
  \end{equation*}
  The first and third terms and the second and fourth terms combine, using equation \eqref{eq:real-LTS}, to produce
  \begin{equation*}
    \TT(u,v)\cdot w - \TT(w,v)\cdot u = - T(w,u) \cdot v - I T(w,u) \cdot Iv,
  \end{equation*}
  which vanishes due to the $\fg$-invariance of the complex structure $I$.  In the other direction, from Lemma~\ref{le:T-map-C} we see that $T(u,v)$ is the real part of $\TT(u,v)$:
  \begin{equation*}
    T(u,v) = \half \left(\TT(u,v) - \TT(v,u)\right),
  \end{equation*}
  whence writing down the LTS condition \eqref{eq:real-LTS} in full, we find
  \begin{multline*}
    T(u,v) \cdot w + T(v,w) \cdot u +   T(w,u) \cdot v = \half \left(\TT(u,v) - \TT(v,u) \right) \cdot w + \half \left(\TT(v,w) - \TT(w,v) \right) \cdot u \\
    + \half \left(\TT(w,u) - \TT(u,w) \right) \cdot v,
  \end{multline*}
  which cancels pairwise using the JTS condition \eqref{eq:T-HSS}.
  
  To prove (e), let $V = U_\CC \in \Dar(\fg,\CC)$.  Then $V$ admits a real structure $R$ satisfying equation \eqref{eq:c-compat} from where it follows that the map $\TT: V \times V \to \fg_\CC$ obeys $\TT(Ru,Rv) = \overline{\TT(u,v)}$.  Under the action of $R$, $V$ decomposes as $V = U \oplus i U$, where $U$ and $iU$ are the real eigenspaces of $R$ with eigenvalues $\pm1$, respectively.  It follows that if $u,v\in U$ then $\TT(u,v)$ is real and, since $\overline{\TT(u,v)} = - \TT(v,u)$, it is skewsymmetric.  Hence it defines a real alternating map $U \times U \to \fg$.  This map is seen to be the map $T$ in \eqref{eq:T-map-R}, since equation \eqref{eq:c-compat} says that $h$ on $U$ is real, so that it agrees with the inner product $\left<-,-\right>$ on $U$.  Now let $w \in U$ and consider $\TT(u,v)\cdot w$.  Now, if $V \in \Dar(\fg,\CC)_{\text{JTS}}$, then in particular $\TT(u,v)\cdot w = + \TT(w,v)\cdot u$, but since $\TT(u,v) \cdot w = - \TT(v,u)\cdot w$ we see that $\TT(u,v) \cdot w=0$. 
  
  If, on the contrary, $V = U_\CC \in \Dar(\fg,\CC)_{\text{aJTS}}$, then $\TT(u,v) \cdot w = - \TT(w,v)\cdot u$, whence it is totally skewsymmetric and $U \in \Dar(\fg,\RR)_{\text{3LA}}$, proving the reverse implication in (f).  To finish proving (f), notice that if $U \in \Dar(\fg,\RR)_{\text{3LA}}$, then $\TT(u,v) \cdot w = - \TT(w,v) \cdot u$ for all $u,v,w \in U$.  Now use the sesquibilinearity of $\TT(u,v) \cdot w$ to show that this is satisfied for all $u,v,w\in V$.
  
  To prove (g) simply notice that if $\rh{W} \in \Dar(\fg,\CC)_{\text{JTS}}$, so that the map $\TT$ satisfies the JTS condition \eqref{eq:T-HSS}, then in particular
  \begin{equation*}
    \TT(u,Jv) \cdot w = + \TT(w,Jv) \cdot u,
  \end{equation*}
  which together with the symmetry condition $\TT(u,Jv) = \TT(v,Ju)$ says that $\TT(u,Jv)\cdot w$ is totally symmetric, whence $W \in \Dar(\fg,\HH)_{\text{QTS}}$.  The argument is clearly reversible, so we get both implications.
  
  If instead $\rh{W} \in \Dar(\fg,\CC)_{\text{aJTS}}$, then $\TT(u,Jv)\cdot w$ is symmetric in $u\leftrightarrow v$ but skewsymmetric in $u\leftrightarrow w$, whence it has to vanish, which says that $\rh{W}$ and hence $W$ is a trivial representation.  This proves (h).
  
  Finally, let us prove (i).   Let $W = V_\HH$.  Recall that we do not work with $W$ but with its image $\rh{W}$ under $r'$, which from Proposition~\ref{prop:diamond}(i) is given by $\rh{V_\HH} \cong V \oplus \Vbar$.  As explained in Appendix~\ref{sec:compl-repr}, we will denote vectors in $\Vbar$ by $\overline v$, for $v \in V$.  Then the quaternionic structure $J$ on $V \oplus \Vbar$ is defined by
  \begin{equation*}
    J v = \overline v \qquad\text{and}\qquad J \overline v = - v \qquad \text{for all $v \in V$}.
  \end{equation*}
  The hermitian structure on $V \oplus \Vbar$ is given by the hermitian structures on $V$ and $\Vbar$ and declaring the direct sum to be orthogonal.  The complex symplectic structure on $V \oplus \Vbar$ is such that $V$ and $\Vbar$ are lagrangian submodules and
  \begin{equation*}
    \omega(u, \overline v) = - h(u,v).
  \end{equation*}
  The only nonzero components of the map $\TT$ are
  \begin{equation}
    \label{eq:Propi}
    \TT(u,J\overline v) = - \TT(u,v).
  \end{equation}
  The aLTS condition \eqref{eq:quat-aLTS} is satisfied if and only if
  \begin{equation*}
    \TT(u,J\overline v) \cdot w +   \TT(\overline v, Jw) \cdot u +   \TT(w, Ju) \cdot \overline v = 0.
  \end{equation*}
  The last term vanishes since $V$ is a lagrangian submodule, hence the aLTS condition is equivalent to
  \begin{equation*}
    \TT(u,J\overline v) \cdot w +   \TT(\overline v, Jw) \cdot u = 0,
  \end{equation*}
  which using equation \eqref{eq:Propi} is equivalent to
  \begin{equation*}
    \TT(u,v) \cdot w + \TT(w, v) \cdot u = 0,
  \end{equation*}
  which is equivalent to the aJTS condition \eqref{eq:T-BL4} on $V$.
\end{proof}

It follows form this proposition (and some geometrical results) that if $W \in \Dar(\fg,\HH)_{\text{QTS}}$ then $V = \rh{W} \in \Dar(\fg,\CC)_{\text{JTS}}$ and $U = \rf{V} \in \Dar(\fg,\RR)_{\text{LTS}}$.  This means that the action of $\fg$ on $U$ is the holonomy representation of a symmetric space $M$.  Since $U$ admits an invariant quaternionic structure, the holonomy representation maps $\fg$ to $\fsp(n) \subset \fso(4n)$, with $\dim_\RR U = 4n$.  This means that $M$ is a hyperkähler symmetric space and hence, in particular, Ricci-flat.  However any homogeneous Ricci-flat riemannian manifold is actually flat \cite[Theorem~7.61]{Besse}, a result due originally to Alekseevsky and Kimelfeld \cite{AlekseevskyKimelfeld}.  This allows us to conclude the following.

\begin{corollary}
  \label{cor:noQTS}
  If $W \in \Dar(\fg,\HH)_{\text{QTS}}$ is positive-definite, then $W$ is the trivial representation.
\end{corollary}

There do exist indefinite QTS representations, associated with indefinite hyperkähler manifolds \cite{MR1913815}.

\subsubsection{Embedding Lie (super)algebras}
\label{sec:embedd-lie-super}

The special classes of unitary representations defined above all share one characteristic: namely, the fact that together with $\fg$, they define a Lie (super)algebra.  This is well-known for the case of the representations associated with the symmetric spaces, as we now briefly recall.

Indeed, if $U \in \Dar(\fg,\RR)_{\text{LTS}}$, then on the 2-graded vector space $\fg \oplus U$, with $\fg$ having degree 0 and $U$ having degree 1, one can define the structure of a graded metric Lie algebra in the following way.  The Lie bracket is given by the Lie bracket of $\fg$, the action of $\fg$ on $U$ and the map $T$ defined by equation \eqref{eq:T-map-R} above.  Then identity \eqref{eq:real-LTS} is the one component of the Jacobi identity which is not implicit in the construction.  The inner product consisting of the one on $\fg$ and the one on $U$, with both spaces being mutually perpendicular, is invariant under the adjoint action.  Conversely, given any 2-graded metric Lie algebra, the degree-1 subspace as a representation of the degree-0 Lie subalgebra is an LTS representation.

Now let $V \in \Dar(\fg,\CC)_{\text{JTS}}$.  In this case we can define on the 3-graded vector space $V \oplus \fg_\CC \oplus \Vbar$ --- with degrees $-1,0,1$, respectively --- the structure of a Lie algebra by adding to the Lie bracket on $\fg_\CC$ and the action of $\fg_\CC$ on $V$ and $\Vbar$, the map $\TT$ defined in equation \eqref{eq:T-map-C}, but viewed here as a complex bilinear map $V \times \Vbar \to \fg_\CC$.  Then the defining condition \eqref{eq:T-HSS} for an JTS representation implies the two components of the Jacobi identity which are not already trivially satisfied.  The 3-graded Lie algebra $V \oplus \fg_\CC \oplus \Vbar$ is metric relative to the complex inner product defined by the one on $\fg_\CC$ and by $h$, thought of as a complex bilinear inner product $V \times \Vbar \to \CC$.  Relative to this inner product, the subspaces $V$ and $\Vbar$ are isotropic abelian Lie subalgebras.  These representations are in one-to-one correspondence with hermitian symmetric spaces.  Indeed, the 3-graded Lie algebra $V \oplus \fg_\CC \oplus \Vbar$ is the complexification of a 2-graded real metric Lie algebra $\fg \oplus \rf{V}$, and the inner product is given by the one on $\fg$ together with the real part of the hermitian inner product on $V$.
We remark that if $W$ is a QTS representation of $\fg$, then by Proposition~\ref{prop:relations}(g), $\rh{W}$ gives rise to a JTS representation, hence it admits an embedding Lie algebra with a 3-grading, but with both the subspaces of degree $\pm 1$ isomorphic to $\rh{W}$.

It turns out that similar results hold for the classes of representations relevant to superconformal Chern--Simons theories, except that now the result of the construction will be a metric Lie superalgebra, suggesting a larger rôle in the superconformal theory for these Lie superalgebras.  The embedding Lie superalgebras were constructed in detail in \cite{Lie3Algs} for the case of aJTS and aLTS representations: see Theorem~22 for the aJTS case and the discussion around equation (45) for the aLTS case.  For completeness we review these constructions here.

Let $V \in \Dar(\fg,\CC)_{\text{aJTS}}$.  Then on the 3-graded vector space $V \oplus \fg_\CC \oplus \Vbar$ we define the structure of a complex 3-graded Lie superalgebra using the Lie algebra structure on $\fg_\CC$, the action of $\fg_\CC$ on $V$ and $\Vbar$ and the map $\TT$ defined in equation \eqref{eq:T-map-C}, but again thought of as a complex bilinear map $\TT: V \times \Vbar \to \fg_\CC$.  The identity \eqref{eq:T-BL4} corresponds now to the one component of the Jacobi identity which is not already trivially satisfied by the construction.  The resulting complex 3-graded Lie superalgebra is metric relative to the inner product on $\fg_\CC$ and the symplectic structure on $V \oplus \Vbar$ defined by declaring $V$ and $\Vbar$ to be lagrangian subspaces and $\left(u,\bar v\right) = h(u,v)$.  This complex Lie superalgebra is the complexification of a metric Lie superalgebra with underlying vector space $\fg \oplus \rf{V}$ and with inner product defined by the one on $\fg$ together with the imaginary part of the hermitian inner product on $V$, which is a symplectic structure on $\rf{V}$.  This construction appeared already in \cite{Lie3Algs} and was considered further in \cite{Palmkvist} and \cite{JMFSimplicity}.

Similarly, let $W\in \Dar(\fg,\HH)_{\text{aLTS}}$.  Consider the 2-graded complex vector space $\fg_\CC \oplus W$, with $\fg_\CC$ in degree 0 and $W$ in degree 1.  We define the Lie bracket by extending the one on $\fg_\CC$ and the action of $\fg_\CC$ on $W$ by $[u,v] = \TT(u,Jv)$ for $u,v \in W$.  Then the identity \eqref{eq:quat-aLTS} is the one component of the Jacobi identity for a complex Lie superalgebra which is not automatically satisfied in the case of any $W \in \Dar(\fg,\HH)$.  This Lie superalgebra is metric relative to the inner product on $\fg_\CC$ and to the complex symplectic form $\omega$ on $W$.   This construction appeared in \cite{Lie3Algs} already and was considered further in \cite{JMFSimplicity}.

Finally, we discuss the case of 3LA representations.  It follows from Proposition~\ref{prop:relations}(f) that if $U \in \Dar(\fg,\RR)_{\text{3LA}}$, then its complexification $V = U_\CC \in \Dar(\fg,\CC)_{\text{aJTS}}$.  By Theorem~22 in \cite{Lie3Algs}, recalled briefly above, and using that $\Vbar \cong V$ in this case, we may define a 3-graded metric Lie superalgebra structure on $V \oplus \fg_\CC \oplus V$.  Furthermore, since $V$ here is the complexification of a real representation, this complex Lie superalgebra is the complexification of a metric Lie superalgebra which, unlike in the general case of aJTS representations, is also 3-graded.  We can see this explicitly as follows.  Consider the 3-graded real vector space $U_{-1} \oplus \fg_0 \oplus U_1$, where the subscripts reflect the degree.  For every $u\in U$, we will write $u_1$ and $u_2$, respectively, the corresponding vectors in $U_1$ and $U_{-1}$.  We write $u_a$ generically, where $a=1,2$.  Then we define the following Lie brackets in addition to the ones of $\fg$:
\begin{equation}
  [X, u_a] = (X\cdot u)_a \qquad\text{and}\qquad [u_a, v_b] = \epsilon_{ab} T(u,v),
\end{equation}
with $\epsilon_{ab}$ the Levi-Civita symbol with $\epsilon_{12} = 1$, say.  The inner product is defined to be the one on $\fg$ extended by
\begin{equation}
  \left(u_a, v_b\right) = \epsilon_{ab} \left<u,v\right>.
\end{equation}
It is a simple exercise to verify that the Jacobi identity is satisfied and that the resulting inner product is ad-invariant.

Conversely, given any 3-graded metric Lie superalgebra $U_{-1} \oplus \fg_0 \oplus U_1$ with $U_1$ and $U_{-1}$ both isomorphic to an orthogonal representation $U$ of $\fg$, then the 3-bracket $[u,v,w]$ on $U$ defined by
\begin{equation}
  [[u_a,v_b],w_c] = \epsilon_{ab} [u,v,w]_c,
\end{equation}
defines a metric 3-Lie algebra structure on $U$.

In summary, we have the following characterisation of metric 3-Lie algebras, which gives one answer to an open question stated in \cite{Lie3Algs} and allows us to paraphrase Kantor, as quoted in \cite{MR0466235}, and suggest that there are no 3-Lie algebras, only Lie superalgebras.

\begin{theorem}
  \label{th:3LAchar}
  Metric 3-Lie algebras $\left(U,[u,v,w],\left<u,v\right>\right)$ are in one-to-one correspondence with metric 3-graded Lie superalgebras $U_{-1} \oplus \fg_0 \oplus U_1$, where $U_1$ and $U_{-1}$ are both isomorphic to $U$, a faithful orthogonal representation of $\fg$.
\end{theorem}

\begin{example}
  \label{ex:S4embedding}
  As shown by Ling \cite{LingSimple} there is a unique complex simple 3-Lie algebra.  There is a unique real form of this 3-Lie algebra which is metric relative to a positive-definite inner product.  The corresponding vector space is $\RR^4$ with the standard euclidean inner product and $\fg  = \fso(4)$ the Lie algebra of skewsymmetric endomorphisms, with inner product  given under the isomorphism $\fso(4) = \fsu(2) \oplus \fsu(2)$ by the Killing form on on the first $\fso(3)$ and the negative of the Killing form on the second.  The corresponding 3-graded Lie superalgebra is a ``compact'' real form of $A(1,1)$ in the Kac classification \cite{KacSuperSketch}.  Notice that the Killing form of $A(1,1)$ vanishes identically, but here we see that it does nevertheless have a non-degenerate inner product.
\end{example}

\subsubsection{Simplicity}
\label{sec:simplicity}

We have seen above that to every Lie-embeddable representation of a metric Lie algebra we can attach a triple system and a Lie (super)algebra.  In principle, there are three separate notions of simplicity or irreducibility we can consider: irreducibility of the representation, simplicity of the embedding Lie (super)algebra and simplicity of the triple system --- this latter one being defined as the nonexistence of proper ideals in the triple system, ideals being defined as kernels of homomorphisms.  For the case of \emph{positive-definite} aJTS representations, this has been discussed recently in \cite{Palmkvist} and from the present point of view in \cite{JMFSimplicity}, where LTS and aLTS representations are also treated.  The following theorems are proved in \cite{JMFSimplicity}; although the first is of course classical.  In that paper we had not yet identified anti-Jordan triple systems for what they were and referred to them as $N{=}6$ triple systems instead.

\begin{theorem}
  \label{thm:LTS-simplicity}
  Let $\fg$ be a metric Lie algebra, $U \in \Dar(\fg,\RR)_{\text{LTS}}$ faithful and positive-definite and let $\fk = \fg \oplus U$ denote its embedding 2-graded Lie algebra.  The following are equivalent:
  \begin{enumerate}
  \item $U \in \Irr(\fg,\RR)_{\text{LTS}}$,
  \item $U$ is a simple Lie triple system,
  \item $\fk$ is a simple Lie algebra or else $U \cong \fg$ (as $\fg$-modules) and $\fg$ is a simple Lie algebra.
  \end{enumerate}
\end{theorem}

\begin{theorem}
  \label{thm:BL4-simplicity}
  Let $\fg$ be a metric Lie algebra, $V \in \Dar(\fg,\CC)_{\text{aJTS}}$ faithful and positive-definite and let $\fk = V \oplus \fg_\CC \oplus \Vbar$ denote its embedding 3-graded Lie superalgebra.  The following are equivalent:
  \begin{enumerate}
  \item $V \in \Irr(\fg,\CC)_{\text{aJTS}}$,
  \item $V$ is a simple anti-Jordan triple system,
  \item $\fk$ is a simple Lie superalgebra.
  \end{enumerate}
\end{theorem}

\begin{theorem}
  \label{thm:aLTS-simplicity}
  Let $\fg$ be a metric Lie algebra, $W \in \Dar(\fg,\HH)_{\text{aLTS}}$ positive-definite and let $\fk = \fg_\CC \oplus W$ denote its embedding Lie superalgebra.  The following are equivalent:
  \begin{enumerate}
  \item $W \in \Irr(\fg,\HH)_{\text{aLTS}}$,
  \item $W$ is a simple quaternionic anti-Lie triple system,
  \item $\fk$ is a simple Lie superalgebra.
  \end{enumerate}
\end{theorem}

Similar results can be proved also for the Jordan triple systems, but as they do not play such an important role in the study of superconformal Chern--Simons theories, we will not mention them in this paper.  Also since by Corollary~\ref{cor:noQTS} there are no positive-definite QTS representations, this question does not arise in this case.  There is, however, a classification of hyperkähler symmetric spaces \cite{MR1913815}.  Finally, example \ref{ex:S4embedding} shows that the same result holds for the unique positive-definite $U \in \Irr(\fg,\RR)_{\text{3LA}}$, whose associated triple system is the unique positive-definite nonabelian simple 3-Lie algebra in \cite{Filippov} and which embeds in the simple Lie superalgebra $A(1,1)$.

The above results allow a classification of positive-definite irreducible Lie-embeddable representations which is summarised in Table \ref{tab:irreducible-LE}, where we only the ones of classes 3LA, aJTS and aLTS, as these are the ones relevant for the study of three-dimensional superconformal Chern--Simons-matter theories.  The other positive-definite Lie-embeddable classes are associated with the riemannian and hermitian symmetric spaces and that classification is classical and can be found, for example, in \cite{Helgason}.

\section{Deriving the superpotentials}
\label{sec:der-superpotentials}

This appendix provides a derivation of the $N{=}1$ superpotentials $\sW_\CC + \sW_F$ in \eqref{eq:superpotential2} and $\sW_\HH$ in \eqref{eq:superpotential3} which respectively provide enhanced $N{=}2$ and $N{=}3$ superconformal symmetry when the matter representation $\fM$ is of generic complex and quaternionic type.  The proof follows from the requirement of obtaining the appropriate R-symmetry commensurate with the degree of supersymmetry enhancement in the on-shell lagrangian.

\subsection{$N{=}2$ superpotential}
\label{sec:der-nequal2superpotential}

Let us begin by assuming $\fM =V \in \Dar(\fg,\CC)$.  If we wish to obtain an enhanced $N{=}2$ superconformal symmetry in the on-shell $N{=}1$ lagrangian in \eqref{eq:susy-lag-on-shell} then it is necessary (but not sufficient) for it to be invariant under the associated $\fu(1)$ R-symmetry.  With respect to this desired enhancement we can ascribe the $\fu(1)$ R-charge $\half$ to the bosonic matter field $X$ and $-\half$ to the fermionic matter field $\Psi$ (with their complex conjugates having the opposite R-charges).  Notice that this assignment implies that all the terms in \eqref{eq:susy-lag-on-shell} are automatically $\fu(1)$-invariant (as a consequence of the hermitian inner product $h$ on $V$ being complex-sesquilinear) except for the scalar-fermion Yukawa couplings.

It is convenient to break up the Yukawa couplings into the contributions transforming with different overall $\fu(1)$ charges.  To this end, let us first decompose the quartic superpotential $\fW = \fW_{4,0} + \fW_{3,1} + \fW_{2,2} + \fW_{1,3} + \fW_{0,4}$ with respect to the complex structure on $V$, where $\fW_{p,4-p} = {\overline{\fW_{4-p,p}}}$ for all $p=0,1,2,3,4$ since $\fW$ is real.  Each component $\fW_{p,4-p}$ is a quartic tensor which is taken to be complex-linear in its first $p$ arguments and complex-antilinear in its remaining $4-p$ arguments.  Thus $\fW_{2,2} (X)$ is real and uncharged, $\fW_{3,1} (X)$ has charge $1$ and $\fW_{4,0} (X)$ has charge $2$.  The other contribution to the Yukawa couplings involves $\TT (X, \Psi )$ which has charge $1$ (and its complex conjugate $-\TT (\Psi ,X)$ with charge $-1$).  Assembling these contributions to the Yukawa couplings, we see that there are separate contributions from terms with overall charges $0$, $\pm 1$ and $\pm 2$.  The uncharged contributions are therefore unconstrained and the $\fu(1)$ R-symmetry can only be realised if the complex terms with charges $-1$ and $-2$ (and their complex conjugates with charges $1$ and $2$) vanish identically.

The component $\fW_{4,0}$ (and its complex conjugate) only appear in a term with no overall $\fu(1)$ charge.  This unconstrained component is to be identified with the F-term superpotential in the $N{=}2$ theory.  There is only one contribution to the charge $-1$ term involving the component $\fW_{3,1}$ and it is straightforward to check that this term can vanish only if $\fW_{3,1} =0$.  The remaining contributions to the charge $-2$ terms involve $\fW_{2,2}$ and it is easily checked that they vanish only if $\fW_{2,2} (X) = \tfrac{1}{16} ( \TT (X,X) , \TT (X,X) )$.  Thus we have established that $\sW_\CC + \sW_F$ in \eqref{eq:superpotential2} is the most general $N{=}1$ superpotential which can realise the aforementioned $\fu(1)$ R-symmetry in the on-shell lagrangian.  The fact that we have already established that this theory is invariant under the $N{=}2$ superconformal algebra thus means that the realisation of this $\fu(1)$ R-symmetry is in fact necessary and sufficient in this instance for $N{=}2$ enhancement.

\subsection{$N{=}3$ superpotential}
\label{sec:der-nequal3superpotential}

Let us now take $\fM =W \in \Dar(\fg,\HH)$ but as usual understood as the complex representation $\rh{W}$ with a quaternionic structure.  Since a theory with enhanced $N{=}3$ superconformal symmetry can be thought of as a special kind of $N{=}2$ superconformal Chern--Simons-matter theory, we can utilise the result of the previous appendix to deduce the constraints which an enhanced $\fusp(2) > \fu(1)$ R-symmetry in the on-shell $N{=}2$ lagrangian \eqref{sec:on-shell-3dsusy-complex} puts on the generic holomorphic $N{=}2$ F-term superpotential $\fW_F$.

With respect to this desired enhancement we can collect the $W$-valued bosonic and fermionic matter fields $X$ and $\Psi$ into the fields $X^\alpha = (X,JX)$ and $\Psi_\alpha = ( \Psi , J \Psi )$, obeying identically the pseudo-reality conditions $J X^\alpha = \varepsilon_{\alpha\beta} X^\beta$ and $J \Psi_\alpha = \varepsilon^{\alpha\beta} \Psi_\beta$, corresponding to the fundamental representation of $\fusp(2)$.  Consequently, the kinetic terms $\left< D_\mu X , D^\mu X \right> = \half \left< D_\mu X^\alpha , D^\mu X^\alpha \right>$ and $\left< {\bar \Psi} , \gamma^\mu D_\mu \Psi \right> = \half \left< {\bar \Psi}_\alpha , \gamma^\mu D_\mu \Psi_\alpha \right>$ in \eqref{sec:on-shell-3dsusy-complex} are automatically $\fusp(2)$-invariant, but the same cannot be said of the Yukawa couplings and scalar potential.  The extra conditions for $\fusp(2)$ R-symmetry in these terms can be deduced most easily by focusing initially on the Yukawa couplings.  We will see that the condition that they be $\fusp(2)$-invariant then guarantees that the scalar potential is too.

Consider the Yukawa couplings in the third line of \eqref{sec:on-shell-3dsusy-complex} which do not involve the F-term superpotential.  Since $\TT ( X^\alpha , X^\alpha )$ vanishes identically, it turns out that there are not many options for writing down manifestly $\fusp(2)$-invariant terms which recover these Yukawa couplings.  In particular, one finds that only from $-\tfrac{1}{16} ( \TT ( X^\alpha , X^\beta ) , \TT ( {\bar \Psi}_\alpha , \Psi_\beta ) )$ can one recover the second term while the first term must come from $- \tfrac{1}{8} ( \TT ( X^\alpha , {\bar \Psi}_\beta ) , \TT ( \Psi_\alpha , X^\beta ) )$ (other possible $\fusp(2)$-invariant permutations of indices can be rewritten in terms of these using the pseudo-reality condition for the matter fields).  Of course, these contributions do not give just the third line of \eqref{sec:on-shell-3dsusy-complex} or else it would already be $\fusp(2)$-invariant.  Indeed, the point of doing this is to isolate the terms responsible for obstructing the enhanced $\fusp(2)$ R-symmetry.  Combining these extra terms with the contribution from the F-term superpotential (which one finds also cannot be $\fusp(2)$-invariant on its own) to the Yukawa couplings gives an overall obstruction term which vanishes only if $\fW_F (X) = \tfrac{1}{32} ( \TT (X,JX), \TT (X,JX) )$. 

This is precisely the F-term superpotential in \eqref{eq:nequal3ftermsuperpotential} which gives rise to the $N{=}1$ superpotential $\sW_\HH$ in \eqref{eq:superpotential3}.  The fact that we have already established this theory to be invariant under the $N{=}3$ superconformal algebra means that the realisation of this $\fusp(2)$ R-symmetry in the Yukawa couplings is in fact necessary and sufficient for $N{=}3$ enhancement here --- the resulting $\fusp(2)$-invariant form of the on-shell $N{=}3$ lagrangian being given by \eqref{eq:susy-lag-on-shell-quaternionic2}.

\bibliographystyle{utphys}
\bibliography{AdS,ESYM,Sugra,Geometry,Algebra}

\end{document}